\documentclass{article}
\usepackage[utf8]{inputenc}
\usepackage[english]{babel}
\usepackage{csquotes}
\usepackage{graphicx}
\usepackage[hidelinks]{hyperref}
\usepackage{amsmath}
\usepackage{amsfonts}
\usepackage{algorithm}
\usepackage{algpseudocode}
\usepackage{enumitem}
\usepackage{array}
\usepackage{booktabs}
\usepackage{tikz}
\usetikzlibrary{arrows.meta, fit, positioning, shapes}
\usepackage{subcaption}
\usepackage{nomencl}
\usepackage[left=2cm,right=2cm,top=2cm,bottom=2cm]{geometry}
\usepackage[
    backend=biber,
    style=numeric,
    autocite=footnote,
    sorting=nty,
    sortcase=false,
    url=true,
    date=year,
    eprint=false,
    hyperref=auto,
    maxbibnames=99,
    giveninits=true,
    isbn=false
]{biblatex}
\renewcommand\nomgroup[1]{%
	\item[\bfseries
	\ifstrequal{#1}{G}{Greek symbols}{%
		\ifstrequal{#1}{L}{Latin symbols}{%
			\ifstrequal{#1}{A}{Abbreviations}{%
				\ifstrequal{#1}{X}{Indices}{%
					\ifstrequal{#1}{O}{Symbols}}}}}%
	]\vspace{0.5em}}
    
\hypersetup{%
    pdftitle=An octree-based sampling algorithm for processing big simulation data,
    pdfauthor=Andre Weiner,
    pdfsubject=article,
    unicode=true,
    breaklinks=true,
    colorlinks=false
}

\title{An octree-based sampling algorithm for analyzing big simulation data}
\author{Janis Geise$^{1}$, Sebastian Spinner$^{2}$, Richard Semaan$^{3}$, and Andre Weiner$^{1}$ \\
        \small $^1$Technical University of Dresden, Institute of Fluid Mechanics, 01062 Dresden, Germany,\\
        \small janis.geise@tu-dresden.de, ORCID ID 0009-0009-9537-4294,\\
        \small andre.weiner@tu-dresden.de, ORCID ID 0000-0001-5617-1560,\\
        \small $^2$DLR, German Aerospace Center, 38108 Braunschweig, Germany,\\
        \small sebastian.spinner@dlr.de, ORCID ID  0000-0002-3661-6814\\
        \small $^3$Technical University of Braunschweig, Institute of Fluid Mechanics,\\
        \small 38108 Braunschweig, Germany\footnote{Current affiliation: ZEISS Segment Semiconductor Manufacturing Technology, Carl Zeiss SMT GmbH, 73447 Oberkochen, Germany},\\
        \small richard.semaan@zeiss.com, ORCID ID  0000-0002-3219-0545
}
\date{July 16, 2026}

\providecommand{\keywords}[1]
{
  \small	
  \textbf{\textit{Keywords---}} #1
}

\makenomenclature

\begin{document}

\maketitle

\begin{abstract}
As computational resources continue to increase, the storage and analysis of vast amounts of data will inevitably become a bottleneck in computational fluid dynamics (CFD) and related fields.
Although compression algorithms and efficient data formats can mitigate this issue, they are often insufficient when post-processing large amounts of volume data.
Processing such data may require additional high-performance software and resources, or it may restrict the analysis to shorter time series or smaller regions of interest.
The present work proposes an improved version of the existing \emph{Sparse Spatial Sampling} algorithm ($S^3$) to reduce the data from time-dependent flow simulations. 
The $S^3$ algorithm iteratively generates a time-invariant octree grid based on a user-defined metric, efficiently down-sampling the data while aiming to preserve as much of the metric as possible.
Using the sampled grid allows for more efficient post-processing and enables memory-intensive tasks, such as computing the modal decomposition of flow snapshots.
The enhanced version of $S^3$ is tested and evaluated on the scale-resolving simulations of the flow past a tandem configuration of airfoils in the transonic regime, the incompressible turbulent flow past a circular cylinder, and the flow around an aircraft half-model at high Reynolds and Mach numbers. 
$S^3$ significantly reduces the number of mesh cells by $35 \%$ to $95\%$ for all test cases while accurately preserving the dominant flow dynamics, enabling post-processing of CFD data on a local workstation rather than HPC resources for many cases.
\end{abstract}

\nomenclature[Lx2]{$\mathbf{X}$}{data matrix}
\nomenclature[Lu1]{$\mathbf{U}$}{matrix of left-singular vectors}
\nomenclature[Lv1]{$\mathbf{V}$}{matrix of right-singular vectors}
\nomenclature[Lv0]{$\mathbf{v}_i$}{i$^\mathrm{th}$ right-singular vector}
\nomenclature[Lu0]{$\mathbf{u}_i$}{$i^\mathrm{th}$ component of the velocity field}
\nomenclature[Lm1]{$Ma_\infty$}{free stream Mach number}
\nomenclature[Lu]{$U_\infty$}{free stream velocity}
\nomenclature[Lp]{$p_\infty$}{free stream pressure}
\nomenclature[Gr]{$\rho_\infty$}{free stream density}
\nomenclature[Lr]{$Re$}{Reynolds number}
\nomenclature[Lm2]{$\mathbf{Ma}$}{Mach number field}
\nomenclature[Lm3]{$\widehat{\mathbf{Ma}}$}{interpolated Mach number field}
\nomenclature[Lc]{$c$}{chord length}
\nomenclature[Ln]{$N$}{number of snapshots}
\nomenclature[Lk]{$\mathbf{k}$}{turbulent kinetic energy field}
\nomenclature[Lk1]{$K$}{number of neighbors}
\nomenclature[Lm0]{$M$}{number of mesh cells}
\nomenclature[Lt]{$t$}{time}
\nomenclature[Ld1]{$D$}{number of physical dimensions}
\nomenclature[Ld0]{$d$}{diameter}
\nomenclature[Ll0]{$l$}{cell level}
\nomenclature[Ll1]{$l_0$}{edge length of the initial cell}
\nomenclature[Lv0]{$V_l$}{area (2D) / volume (3D) of a cell of level $l$}
\nomenclature[Lw0]{$w$}{inverse distance weight}
\nomenclature[Lx0]{$\mathbf{x}_n$}{snapshot at time step $n$}
\nomenclature[Lx1]{$\hat{\mathbf{x}}_n$}{interpolated snapshot at time step $n$}
\nomenclature[Lx]{$x$}{coordinate in $x$-direction}
\nomenclature[Ly]{$y$}{coordinate in $y$-direction}
\nomenclature[Lz]{$z$}{coordinate in $z$-direction}
\nomenclature[Ls]{$Sr$}{Strouhal number}
\nomenclature[Lf]{$\mathbf{f}$}{generic flow field}

\nomenclature[Lp0]{$\mathbf{p}_i$}{spatial coordinates of a point $i$}
\nomenclature[Lp1]{$\widehat{\mathbf{p}}_i$}{spatial coordinates of a query point $i$}

\nomenclature[Gt]{$\tau$}{dimensionless time}
\nomenclature[Gn1]{$\boldsymbol{\nu}_t$}{eddy viscosity field}
\nomenclature[Gn]{$\nu$}{kinematic viscosity}
\nomenclature[Gs1]{$\sigma_i$}{$i^\mathrm{th}$ singular value}
\nomenclature[Gs2]{$\mathbf{\Sigma}$}{matrix of singular values}

\nomenclature[Om2]{$\boldsymbol{\mathcal{M}}$}{metric field}
\nomenclature[Om0]{$\mathcal{M}_i$}{metric of cell $i$}
\nomenclature[Om1]{$\widehat{\mathcal{M}}_i$}{interpolated metric of cell $i$}
\nomenclature[Og]{$\mathcal{G}_l$}{gain of a cell at level $l$}
\nomenclature[Om3]{$\widehat{\boldsymbol{\mathcal{M}}}$}{interpolated metric field}
\nomenclature[Om4]{$\mathcal{M}_{\mathrm{approx}}$}{metric approximation quality defined as $\lVert \widehat{\boldsymbol{\mathcal{M}}} \rVert / \lVert \boldsymbol{\mathcal{M}} \rVert$}
\nomenclature[Om5]{$\mathcal{M}_{\mathrm{min}}$}{threshold value for minimum metric approximation quality}

\nomenclature[A]{S$^3$}{Sparse Spatial Sampling}
\nomenclature[A]{AZDES}{Automated Zonal Detached Eddy Simulation}
\nomenclature[A]{DNS}{Direct Numerical Simulation}
\nomenclature[A]{CFD}{Computational Fluid Dynamics}
\nomenclature[A]{HPC}{High Performance Computing}
\nomenclature[A]{SVD}{Singular Value Decomposition}
\nomenclature[A]{TKE}{Turbulent Kinetic Energy}
\nomenclature[A]{DDES}{Delayed Detached Eddy Simulation}
\nomenclature[A]{MAC}{Mean Aerodynamic Chord}

\nomenclature[Xc]{$c$}{child cell}
\nomenclature[Xl]{$\ell$}{leaf cell}
\nomenclature[Xg]{$\mathrm{g}$}{geometry}
\nomenclature[Xa]{$\mathrm{a}$}{adaptive}
\nomenclature[Xu]{$\mathrm{uni}$}{uniform}
\nomenclature[Xr]{$\mathrm{rn}$}{renumber}
\printnomenclature[3cm]

\section{Introduction}
\label{sec:intro}
Increasing computational resources enable the usage of more complex models and scale-resolving simulations in computational fluid dynamics (CFD). 
Moreover, the decreasing cost of high-performance computing (HPC) resources and improved accessibility motivate performing more simulations, e.g., for parameter studies and optimization \cite{duwe_state_2020}.
However, high-fidelity simulations such as large eddy simulations (LES) or detached eddy simulations (DES) yield massive amounts of data that need to be stored and processed.
The progressive usage of machine learning techniques for flow analysis and modeling, e.g., via modal decompositions for dimensionality reduction, requires further availability of highly sampled flow field sequences~\cite{baddoo_kernel_2022,weiner_robust_2023}.
While computational resources were the main limiting factor for CFD for a long time, it is foreseeable that now the available storage capacity and data analysis will become the bottleneck for large-scale simulations \cite{duwe_state_2020}.
Post-processing such amounts of data nowadays often requires HPC resources. 
In 2022, data centers already contributed $2\%$ to the worldwide energy consumption, and a doubling is expected in their energy consumption by 2026 \cite{international_energy_agency_electricity_2024}, emphasizing the need for efficient data management and data reduction.\\ 

Existing data reduction techniques can be classified as lossless or lossy compression algorithms.
While lossless compression algorithms aim to remove redundancies in the data, the achievable compression ratio is limited  \cite{duwe_state_2020}.
Lossless compression is generally applied whenever the exact reconstruction of the data is crucial, e.g., when compressing text files such as source code.
Although lossless compression can reduce the required disk space, it does not reduce the required computational resources for post-processing.
A widely used alternative is lossy compression.
A variety of algorithms already exists that utilize digit rounding, quantization, or similar methods \cite{duwe_state_2020}.
Although lossy compression can reach a significantly higher compression ratio than lossless compression, generally applicable algorithms do not leverage fundamental physical phenomena inherent to fluid flows. \\

Data reduction for post-processing CFD data using flow physics can be accomplished by decreasing the number of snapshots, i.e., the temporal resolution of output files, or by reducing the mesh size of the simulation, i.e., the spatial resolution. 
Decreasing the number of snapshots is rarely an option for tasks such as modal analysis, as a sufficiently high sample frequency is required to avoid undesired effects like aliasing.
Consequently, most of the work conducted so far focuses on utilizing coarse grids to accelerate the simulation in conjunction with coarse-grid corrections applied to improve the obtained solution \cite{hanna_machine-learning_2020, kiener_data-driven_2023, shumilin_self-supervised_2024}.

One possible reduction approach is to optimize the grid topology by removing nodes that are not necessary to represent spatial or temporal variation.
This idea was investigated by Klein et al. \cite{klein_mesh_1996}, aiming to decrease the number of nodes for the surface representation of geometries in medical applications.
More recently, Lorsung and Farimani \cite{lorsung_mesh_2023} utilized deep reinforcement learning to identify mesh nodes that can be removed, accelerating simulations while minimizing the associated accuracy loss.
While the latter approach focused on reducing the runtime of the simulation, Lee et al. \cite{lee_machine_2024} leveraged autoencoders to compress spatiotemporal CFD data, achieving compression by two orders of magnitude.
Deep convolutional autoencoders were also employed by Glaws et al. \cite{glaws_deep_2020} for in-situ data compression of CFD simulations.
For a compression ratio of $64$, the reported reconstruction error was significantly lower than that obtained using singular value decomposition at an equivalent compression ratio.
However, the proposed method relies on a fixed compression ratio and was developed specifically for turbulent flow configurations, which may limit its direct applicability to a wider range of flow scenarios.
Moreover, one should keep in mind that neural network training is usually stochastic and depends on several hyperparameters, so each run yields a slightly different compression and requires tuned training parameters.

An alternative data reduction strategy was investigated by Otero et al. \cite{otero_lossy_2018}.
In their work, the discrete Legendre transform was utilized to first convert the data into the spectral space, followed by a truncation based on modal amplitudes.
Although they were able to achieve compression ratios up to $97\%$ across different flow scenarios, their approach relies on a specific spectral element method and is therefore not easily transferable to other numerical methods or data structures.\\

When it comes to modal analysis tasks, the singular value decomposition (SVD) is the backbone of many state-of-the-art techniques. To deal with large data matrices, incremental and streaming variants of the SVD exist; refer to \cite{kuhl_incremental_2024} for a recent overview.
These methods update the POD basis incrementally by processing individual snapshots or batches of snapshots, thereby reducing memory and storage requirements.
Recent applications to large flow datasets include \cite{li_enhanced_2022, kuhl_incremental_2024, liu_incremental_2026}.
However, the resulting POD modes remain high-dimensional spatial fields, whose visualization and processing still require considerable computational resources.
The same argument holds true for other modal decomposition algorithms.\\

It is important to note that lossless and lossy approaches can be combined to maximize data reduction while minimizing information loss.
For instance, using binary formats such as HDF5 combined with lower precision - e.g., single precision floating point numbers - will lead to a significant decrease in the computational resources required and may already be sufficient in many cases.
If a higher compression ratio is required, lossy algorithms such as mesh reduction algorithms or truncation can be applied additionally.\\

The concept of removing unnecessary grid nodes and associated snapshot data is also the core idea of \emph{Sparse Spatial Sampling} ($S^3$), introduced by Fernex et al. \cite{fernex_sparse_2021}.
$S^3$ consists of three steps.
Initially, a point cloud is generated from the vertices of the original mesh.
Then, a subset of the point cloud is selected based on high values of a user-defined metric.
A sensible metric to investigate temporal fluctuations of turbulent flows is the turbulent kinetic energy (TKE).
Next, a coarse mesh is generated from the reduced point cloud by employing 3D Delaunay triangulation.
One or more Laplacian smoothing iterations are run to improve the mesh quality.
Lastly, the mesh vertices of the coarse mesh are snapped to the closest vertices of the original mesh.
The snapping avoids the need for more sophisticated interpolation methods to map the field data between the meshes.
The new mesh is static (invariant in time).
While demonstrating the potential of mesh reduction based on flow physics, the original $S^3$ algorithm has the following shortcomings:
\begin{itemize}
	\item Point cloud reduction: selecting points based on high metric values, e.g., the temporal variance of a field, leads to large empty regions in some parts of the domain, while most points are clustered in other parts. The derived mesh can display poor quality, and the reduction can be suboptimal in regions where the metric is uniformly high. 
	\item Cell topology: the derived mesh comprises tetrahedral cells. The reduction is less effective if the original mesh comprises hexahedral or polyhedral cells.
	\item Boundary treatment: the algorithm does not handle spurious cells outside the actual domain, which can be created during the remeshing.
	\item Missing interpolation: while snapping points from the derived mesh to the closest ones of the original mesh avoids additional interpolation, the field values of small original mesh cells might be copied to large cells of the derived mesh, potentially introducing a significant interpolation error.
\end{itemize}
In the fully revised version of $S^3$ presented in this contribution, the mesh generation starts with a single cubic (3D) or quadrilateral (2D) cell placed around the numerical domain, which is then refined iteratively based on a user-defined metric field, aiming to approximate the metric field as closely as possible.
The refinement process stops once a user-controlled stopping criterion is reached.
Finally, field values are interpolated to the resulting quadtree/octree mesh.\\

The remainder of this article is structured as follows: section \ref{subsec:scube} presents the updated $S^3$ version.
Section \ref{sec:applications} evaluates $S^3$ on three unsteady flow simulations, namely, the transonic flow past an airfoil tandem configuration (section \ref{subsec:oat}), the fully-resolved turbulent flow past a circular cylinder (section \ref{subsec:cylinder}), and the transonic flow around an aircraft half-model (section \ref{subsec:xrf1}).
We employ a singular value decomposition (SVD) to compare original and reduced data.
The~$S^3$ source code and the examples are publicly available on GitHub \cite{geise_git_2024}.

\section{Sparse Spatial Sampling} \label{subsec:scube}
\begin{figure}[htbp]
		\centering
		\begin{tikzpicture}[
			box/.style={draw, minimum width=4.5cm, minimum height=1cm, align=center},
			dashedboxred/.style={draw=red, dashed, rounded corners=8pt, inner sep=10pt, fit=#1},
			dashedboxblue/.style={draw=blue, dashed, rounded corners=8pt, inner sep=10pt, fit=#1},
			->, >=Stealth
			]
			
			\node[draw=blue, dashed, rounded corners=8pt, minimum width=13cm, minimum height=6cm, line width=1pt] at (-11.5,4.5) {};
			
			\node (metric) at (-15, 7.1) {\color{blue}{1. Computation of the metric field}};
			\node (snapshotsOrig0) at (-15.5, 6) {
				\begin{minipage}{0.2\textwidth}
					\footnotesize $\mathbf{x}_0$
					\centering
					\includegraphics[width=\textwidth]{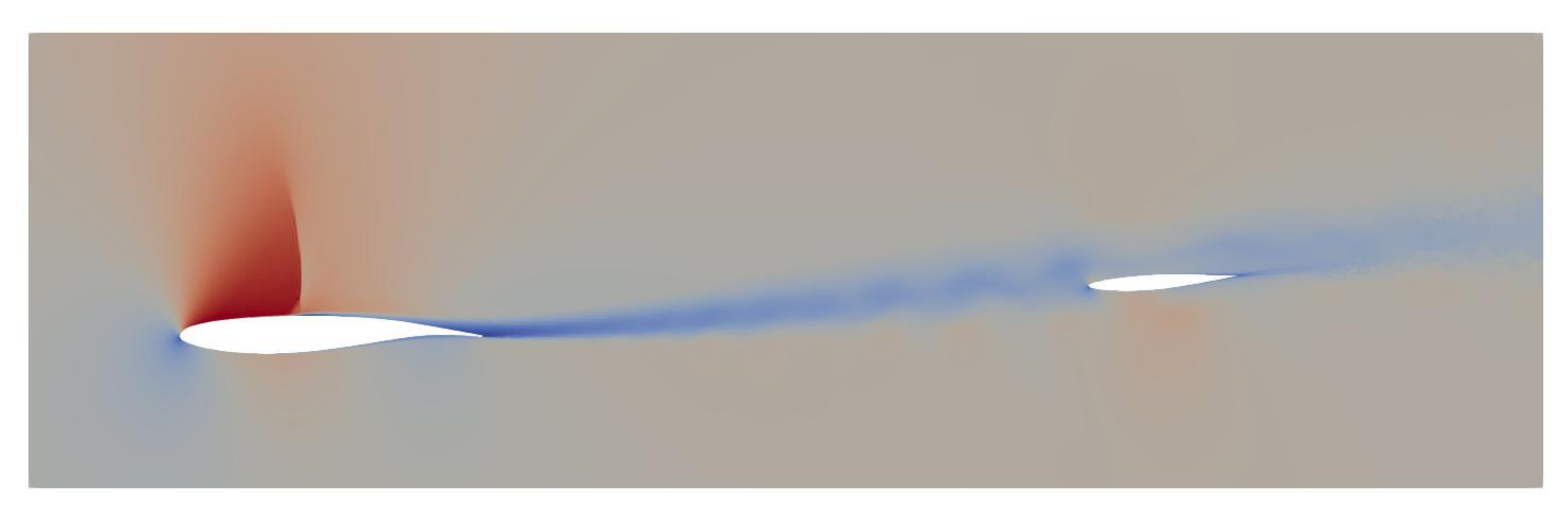} \\
				\end{minipage}
			};
			
			\node (snapshotsOrig1) at (-11.5, 6) {
			\begin{minipage}{0.2\textwidth}
				\footnotesize $\mathbf{x}_1$
				\centering
				\includegraphics[width=\textwidth]{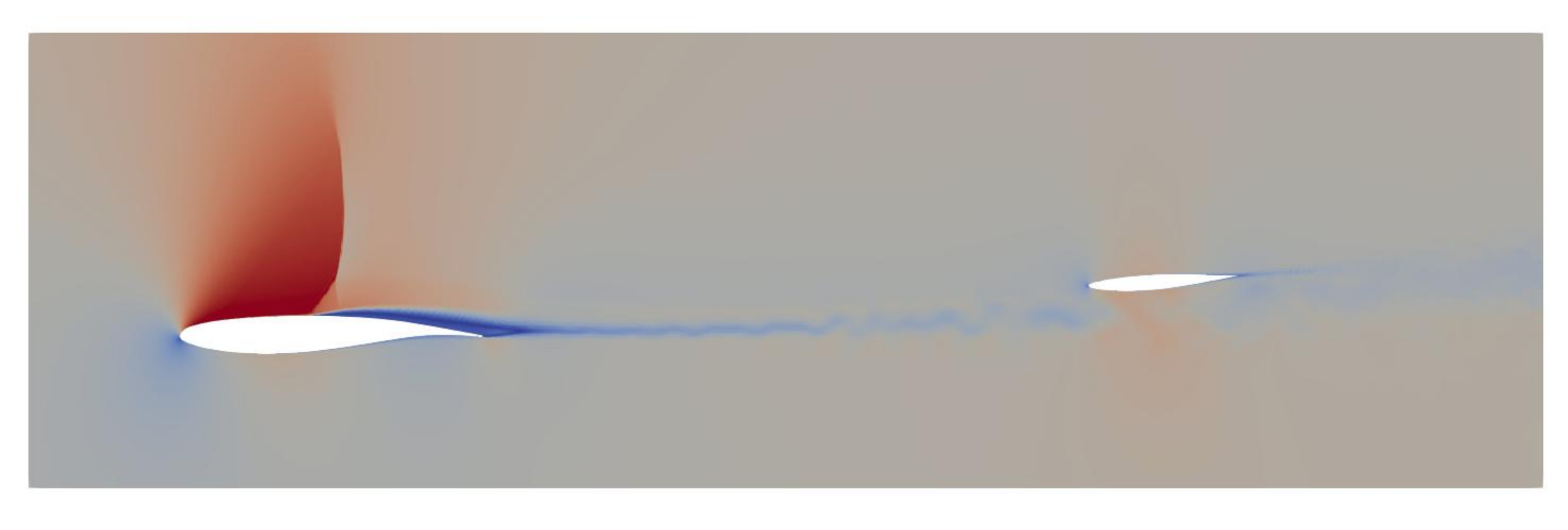} \\
			\end{minipage}
			};
			
			\node (snapshotsOrig2) at (-7.5, 6) {
				\begin{minipage}{0.2\textwidth}
					\footnotesize $\mathbf{x}_N$
					\centering
					\includegraphics[width=\textwidth]{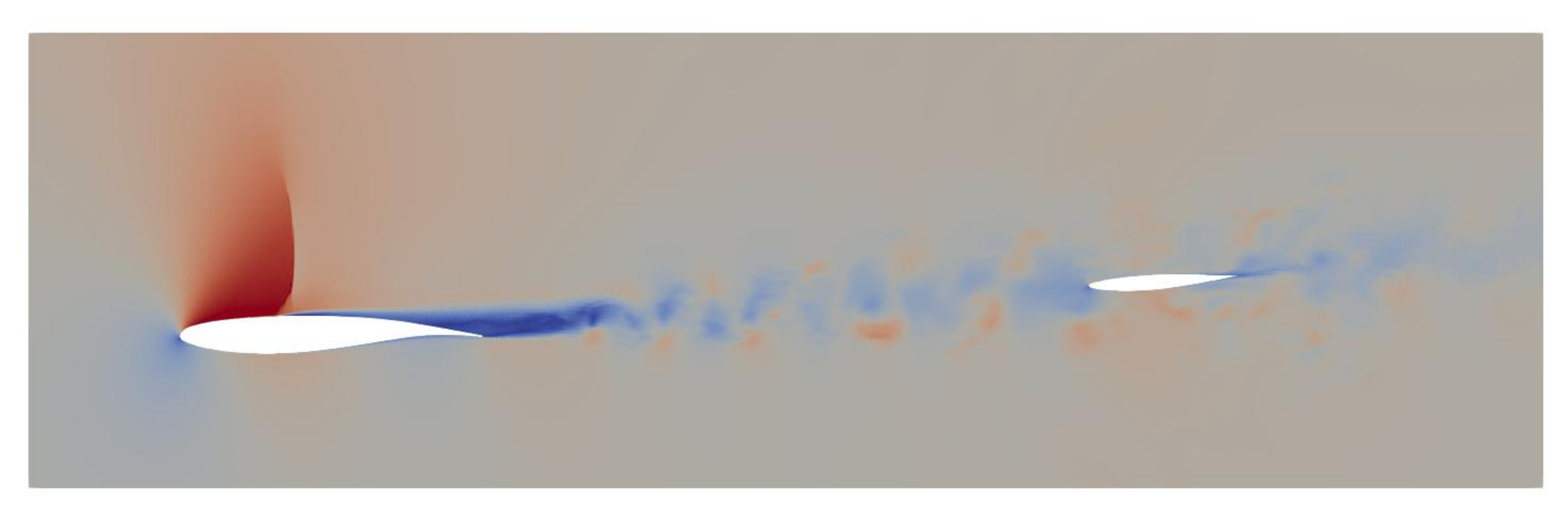} \\
				\end{minipage}
			};
			
			\node (metric) at (-11.5, 3) {
				\begin{minipage}{0.4\textwidth}
					\centering
					\includegraphics[width=\textwidth]{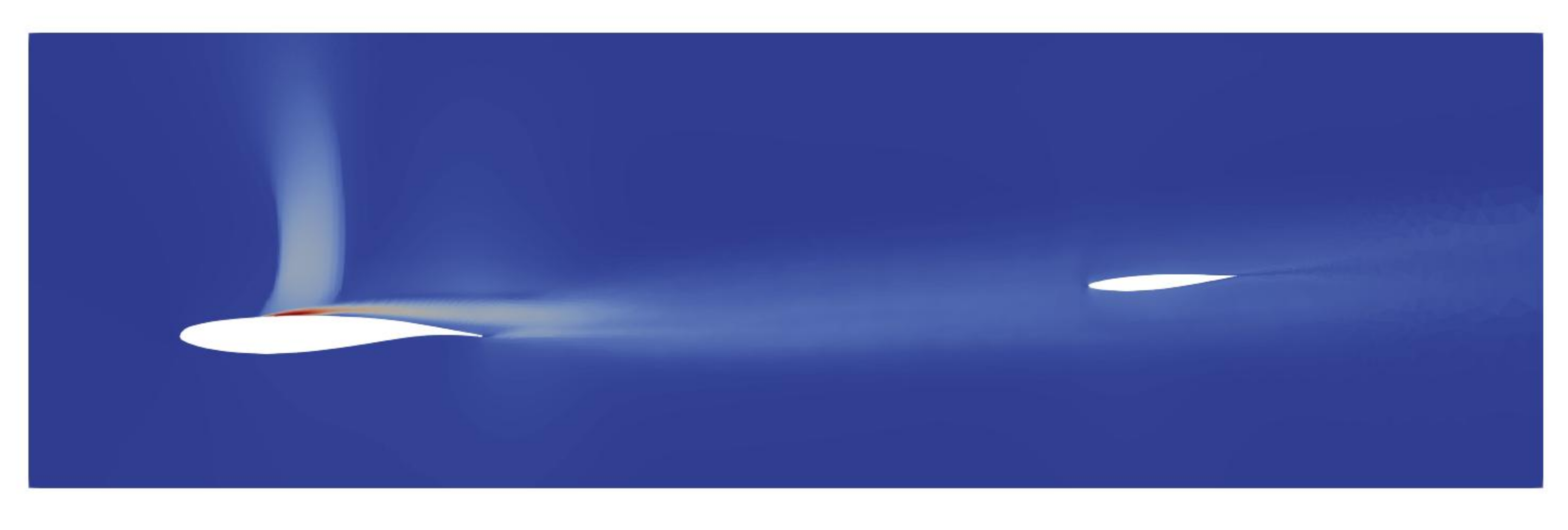} \\
					\footnotesize compute metric field, e.g., $\boldsymbol{\mathcal{M}} = \mathrm{std}(\mathbf{Ma})$
				\end{minipage}
			};
			
			\node[circle, fill=black, inner sep=2pt] (sum) at (-11.5, 4.8) {};
			\draw[->] (-11.5, 4.8) -- (-11.5, 4.25);
			\draw[-] (-15.5, 5.35) -- (-15.5, 4.8);
			\draw[-] (-11.5, 5.35) -- (snapshotsOrig1.south) -- (sum.south);
			\draw[-] (-7.5, 5.35) -- (-7.5, 4.8);
			\draw[-] (-15.5, 4.8) -- (-9.75, 4.8);
			\node (metric) at (-9.5, 6.6) {\footnotesize $...$};
			\node (metric) at (-9.5, 5.75) {\footnotesize $...$};
			\node (metric) at (-9.5, 4.8) {\footnotesize $...$};
			\draw[-] (-9.25, 4.8) -- (-7.5, 4.8);
			
			\node[draw=orange, dashed, rounded corners=8pt, minimum width=13cm, minimum height=13cm, line width=1pt] at (-11.5, -5.5) {};
			
			\node (label2) at (-16.25, 0.65) {\color{orange}{2. Grid refinement}};
			\node (label21) at (-15.9, 0.15) {\color{orange}{2.1 Uniform refinement}};
			
			\node (uniform1) at (-16, -2) {
				\begin{minipage}{0.2\textwidth}
					\centering
					\includegraphics[width=\textwidth]{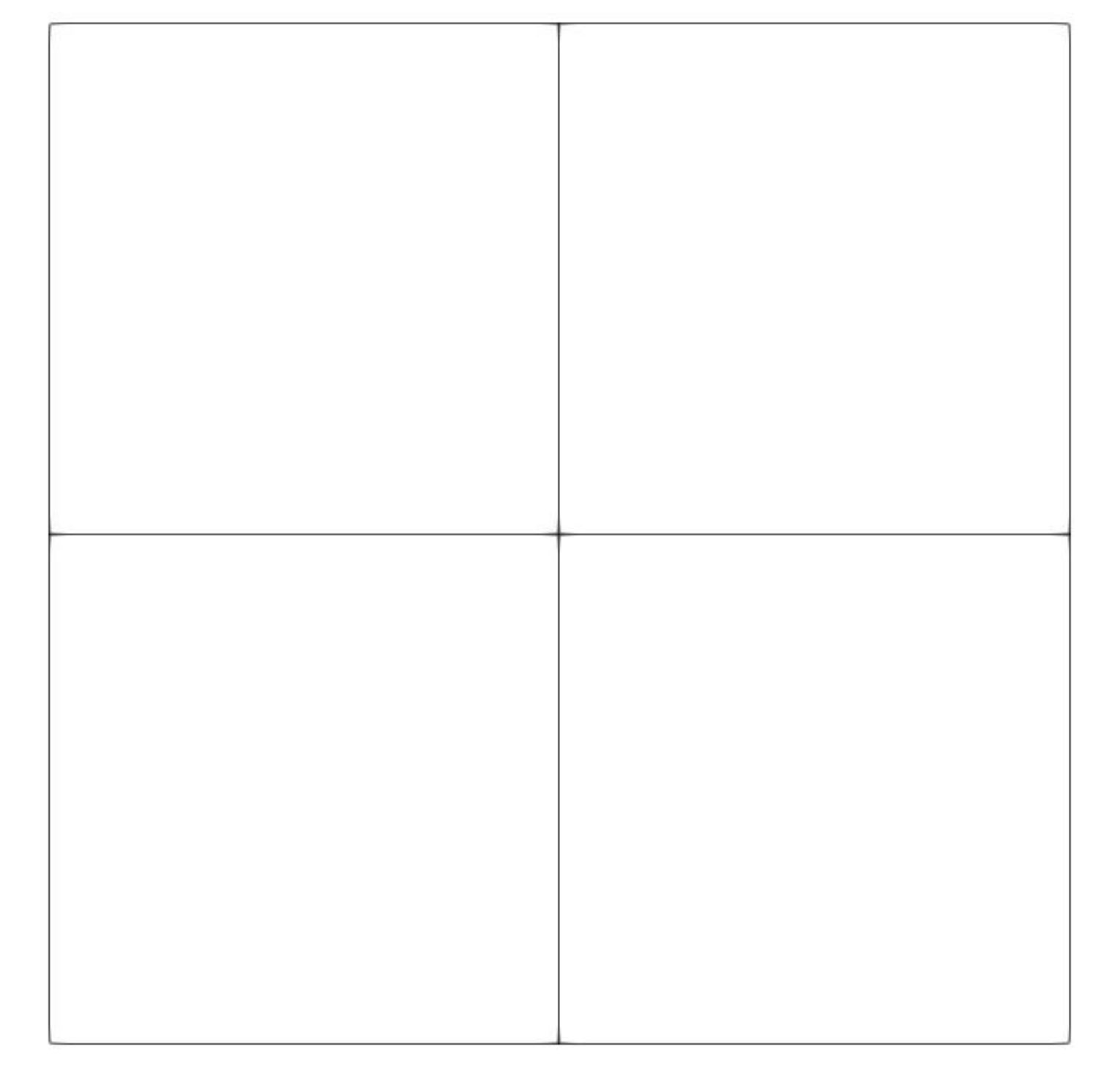} \\
				\end{minipage}
			};
			
			\node (uniform2) at (-11.5, -2) {
				\begin{minipage}{0.2\textwidth}
					\centering
					\includegraphics[width=\textwidth]{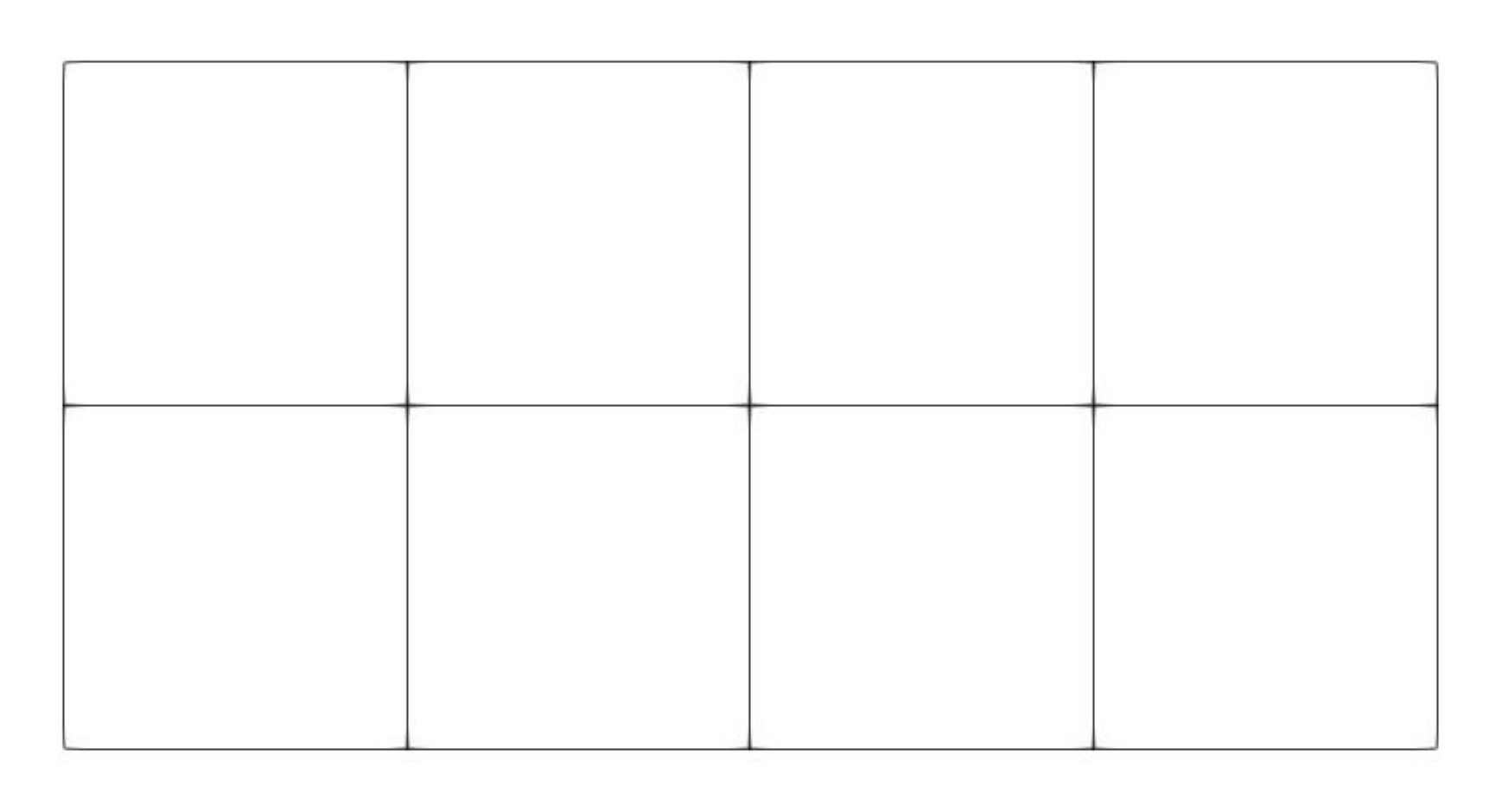} \\
				\end{minipage}
			};
			
			\node (uniform5) at (-7, -2) {
				\begin{minipage}{0.2\textwidth}
					\centering
					\includegraphics[width=\textwidth]{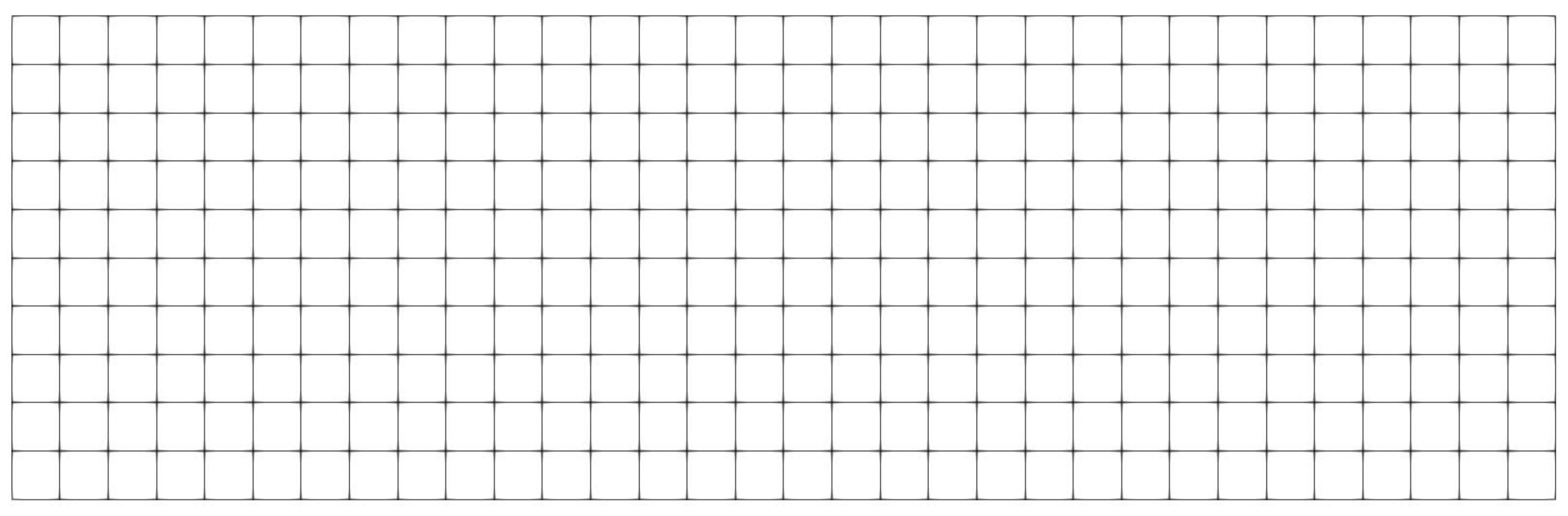} \\
				\end{minipage}
			};
			
			\draw[->] (uniform1.east) -- (uniform2.west);
			\draw[->] (uniform2.east) -- (uniform5.west);
			\node at (-9.25,-1.75) {\footnotesize $...$};
			
			\node[draw=black, dashed, rounded corners=0pt, minimum width=3.2cm, minimum height=1.1cm, line width=1pt] at (-16, -2) {};	
			\node[draw=black, dashed, rounded corners=0pt, minimum width=3.22cm, minimum height=1.1cm, line width=1pt] at (-11.5, -2) {};
			\node[draw=black, dashed, rounded corners=0pt, minimum width=3.48cm, minimum height=1.1cm, line width=1pt] at (-7, -2) {};
			\node(roi) at (-16, -2.8) {region of interest};
			
			\node (metric) at (-15.5, -4.25) {\color{orange}{2.2 Metric-based refinement}};
			
			\node (gain) at (-15, -5.75) {
				\begin{minipage}{0.3\textwidth}
					\centering
					\footnotesize compute / update $\mathcal{G}_l$
					\includegraphics[width=\textwidth]{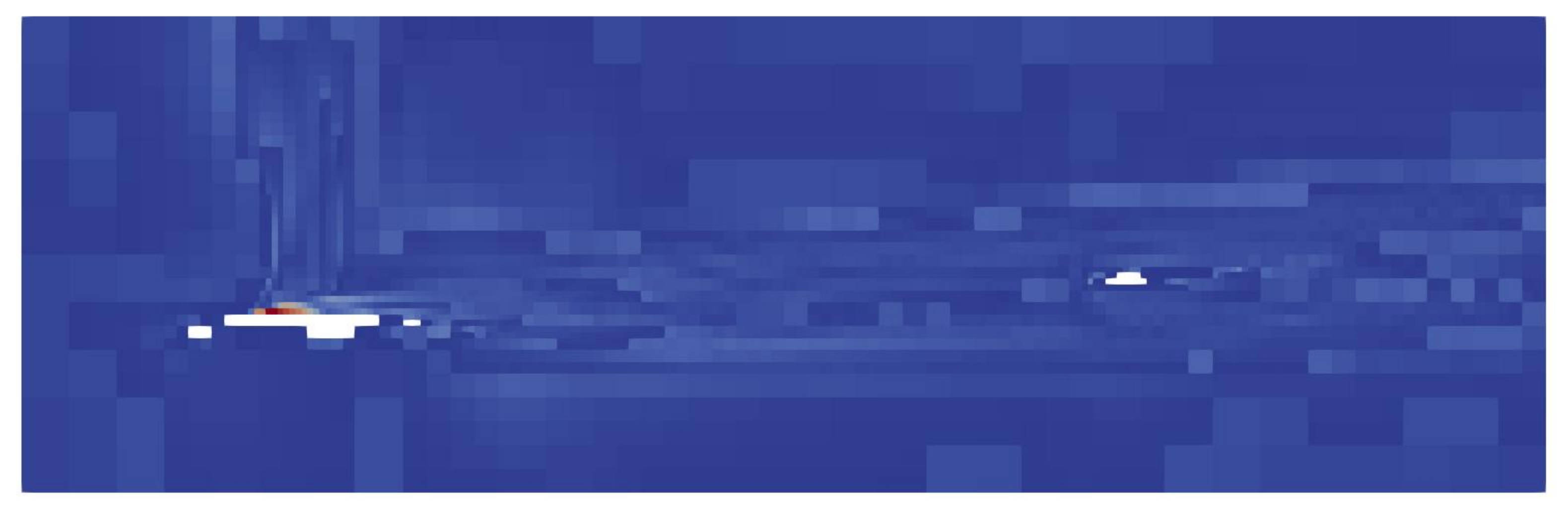} \\
				\end{minipage}
			};
			
			\node (grid) at (-8, -5.75) {
				\begin{minipage}{0.3\textwidth}
					\centering
					\footnotesize perform next refinement step
					\includegraphics[width=\textwidth]{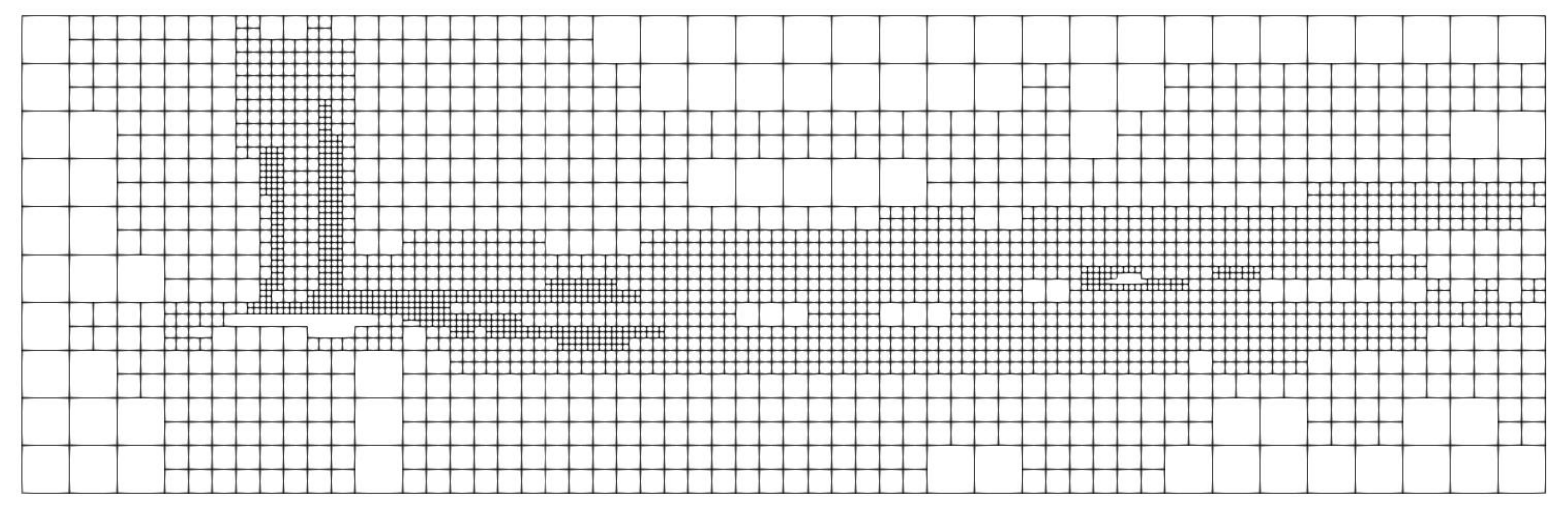} \\
				\end{minipage}
			};
			
			\draw[->] (gain.east) -- (grid.west);
			\node (stop) [diamond, aspect=2, text centered, draw=black, minimum size=0.5cm, inner sep=1pt] at (-11.5,-7.25) {stop?};
			\draw[->] (-15, -7.25) -- (-15, -6.75);
			\draw[-] (-8, -7.25) -- (-8, -6.75);
			\draw[->] (-8, -7.25) -- (stop);
			\draw[-] (stop) -- (-15, -7.25);
			\node (yes) at (-11.15,-8.25) {yes};
			\node (no) at (-13.5,-7.05) {no};
			
			\node (metric) at (-14.9, -7.75) {\color{orange}{2.3 Geometry refinement (optional)}};
			
			\node[circle, fill=black, inner sep=2pt] (gref) at (-11.5,-9) {};
			\draw[->, dashed] (stop.south) -- (gref.north);
			\draw[->, dashed] (gref.west) -- (-12.25, -9);
			\draw[->, dashed] (gref.east) -- (-10.75, -9);
			
			\node (snapshots) at (-15, -9) {
				\begin{minipage}{0.3\textwidth}
					\centering
					\includegraphics[width=\textwidth]{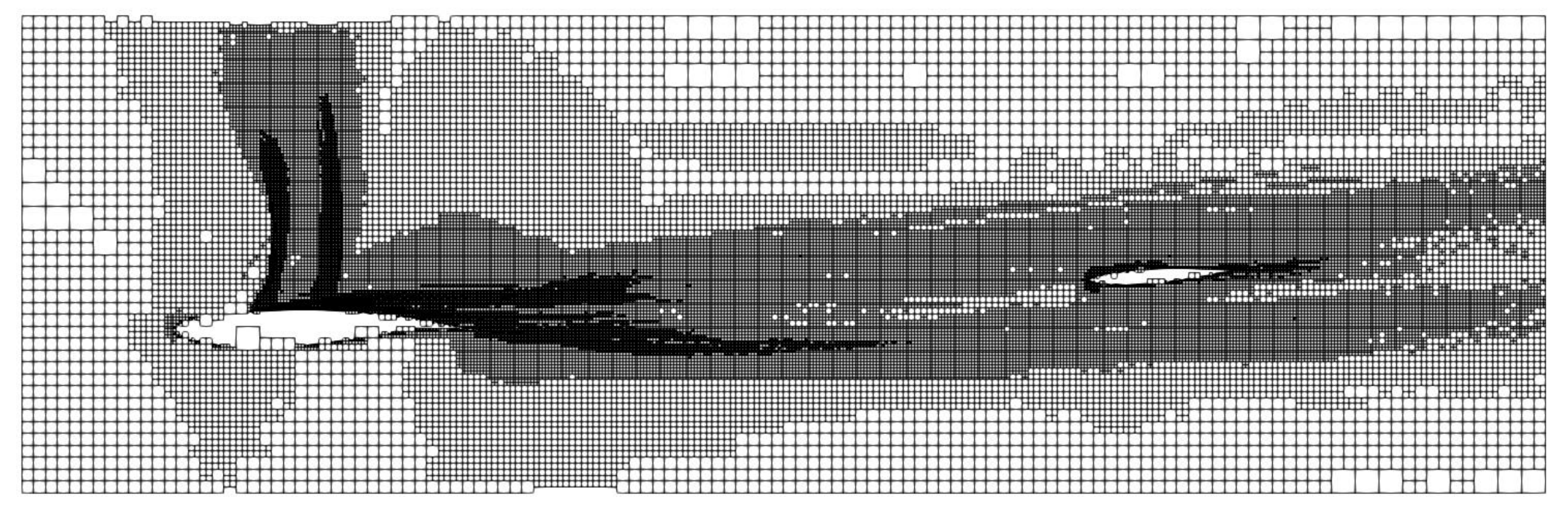}
				\end{minipage}
			};
			
			\node (snapshots) at (-8, -9) {
				\begin{minipage}{0.3\textwidth}
					\centering
					\includegraphics[width=\textwidth]{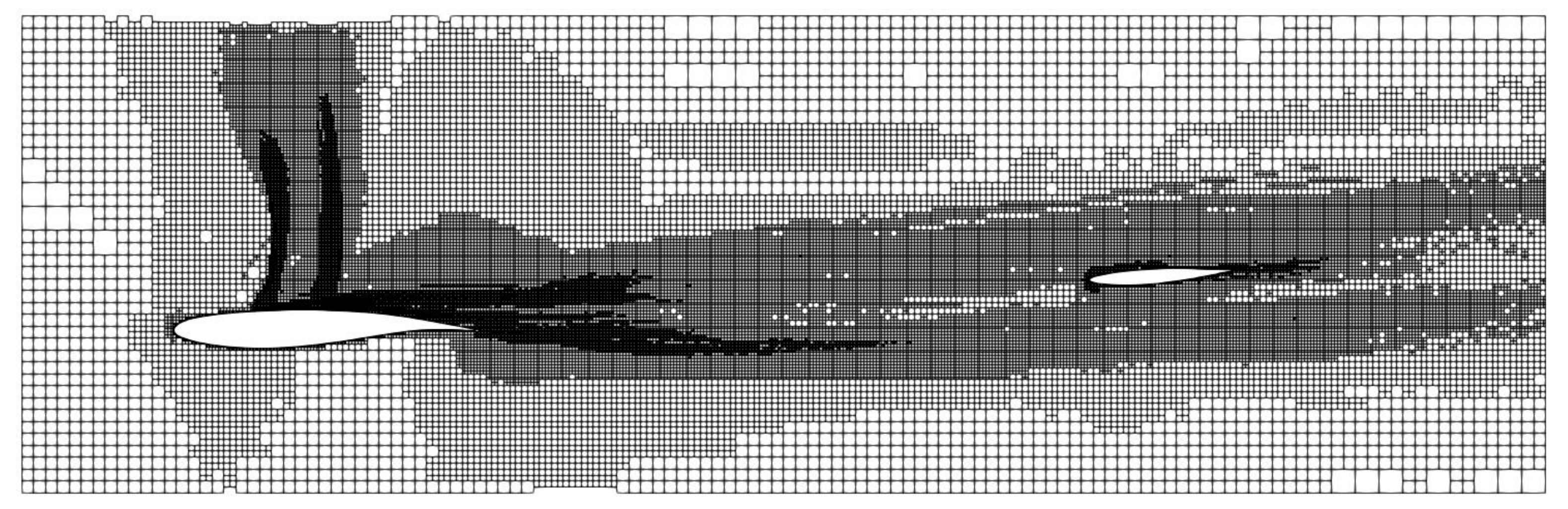} \\
				\end{minipage}
			};
			
			\node (snapshots) at (-14, -11) {
				\begin{minipage}{0.25\textwidth}
					\centering
					\includegraphics[width=\textwidth]{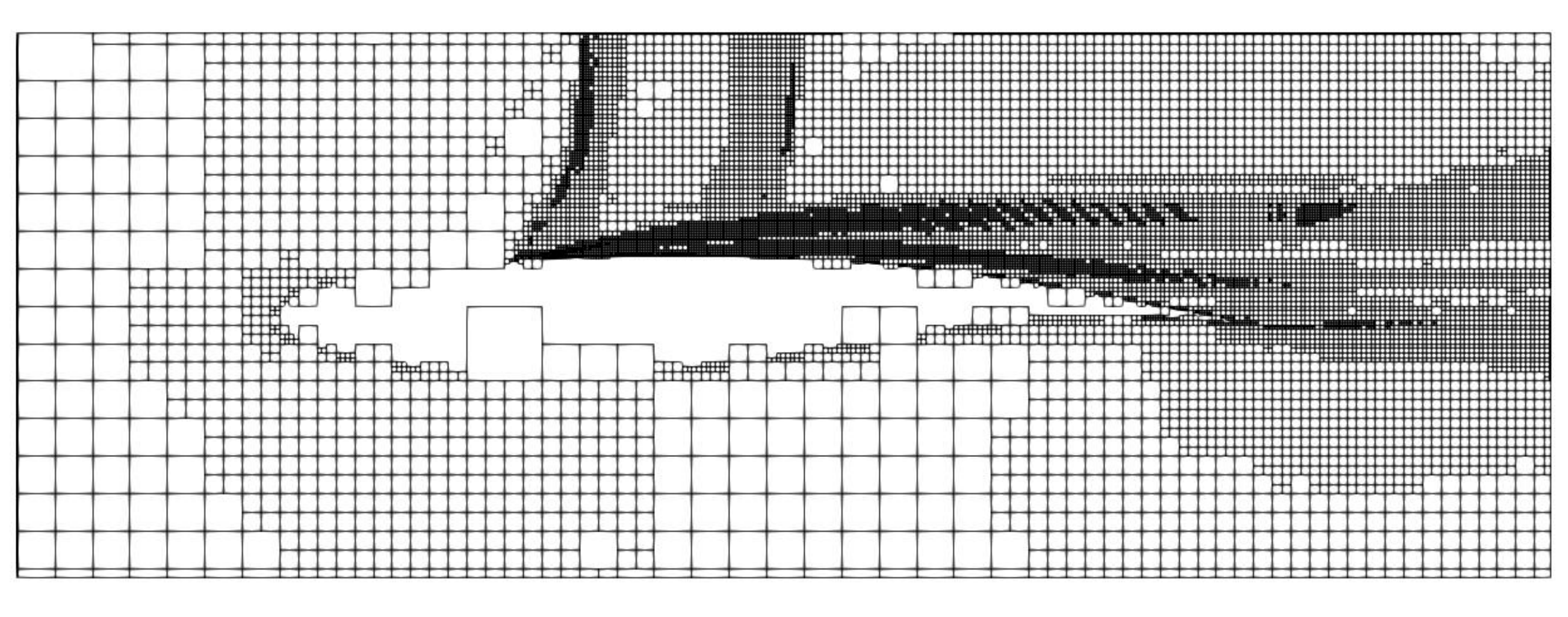}
				\end{minipage}
			};
			
			\node (snapshots) at (-8, -11) {
				\begin{minipage}{0.25\textwidth}
					\centering
					\includegraphics[width=\textwidth]{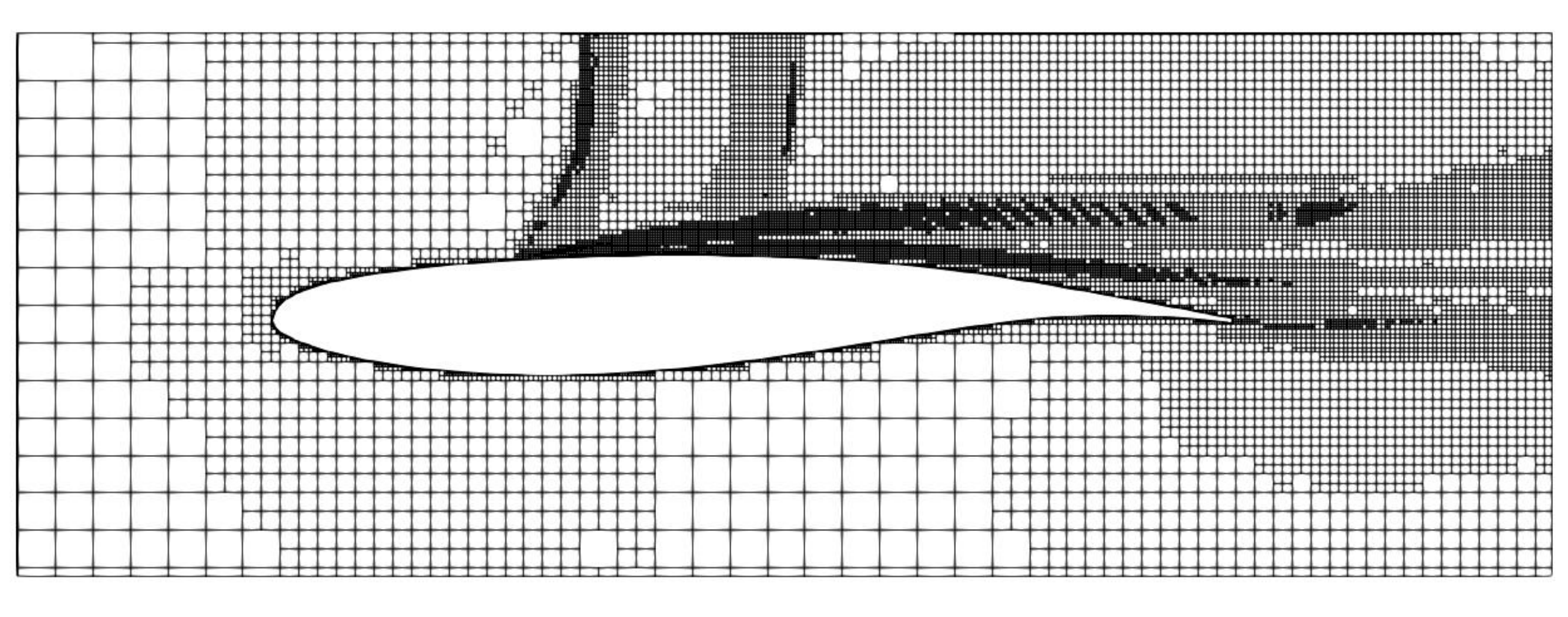}
				\end{minipage}
			};
			
			\node[draw=red, rounded corners=0pt, minimum width=1.2cm, minimum height=0.25cm, line width=1pt] at (-9.6, -9.25) {};
			\node[draw=red, rounded corners=0pt, minimum width=1.2cm, minimum height=0.25cm, line width=1pt] at (-16.6, -9.25) {};
			
			\node[draw=red, rounded corners=0pt, minimum width=4.3cm, minimum height=1.55cm, line width=1pt] at (-8, -11) {};
			\node[draw=red, rounded corners=0pt, minimum width=4.3cm, minimum height=1.55cm, line width=1pt] at (-14, -11) {};
			
			\draw[draw=red, line width=1pt, -] (-10.21, -9.37) -- (-10.15, -10.23);
			\draw[draw=red, line width=1pt, -] (-9, -9.37) -- (-5.86, -10.23);
			
			\draw[draw=red, line width=1pt, -] (-17.21, -9.37) -- (-16.12, -10.23);
			\draw[draw=red, line width=1pt, -] (-16, -9.37) -- (-11.85, -10.23);
			
			\node[draw=red, dashed, rounded corners=8pt, minimum width=13cm, minimum height=2.5cm, line width=1pt] at (-11.5, -13.75) {};
			
			\node (inter) at (-16.4, -12.85) {\color{red}{3. Interpolation}};
			
			\node (snapshotsInter0) at (-15.5, -14) {
				\begin{minipage}{0.2\textwidth}
					\centering
					\includegraphics[width=\textwidth]{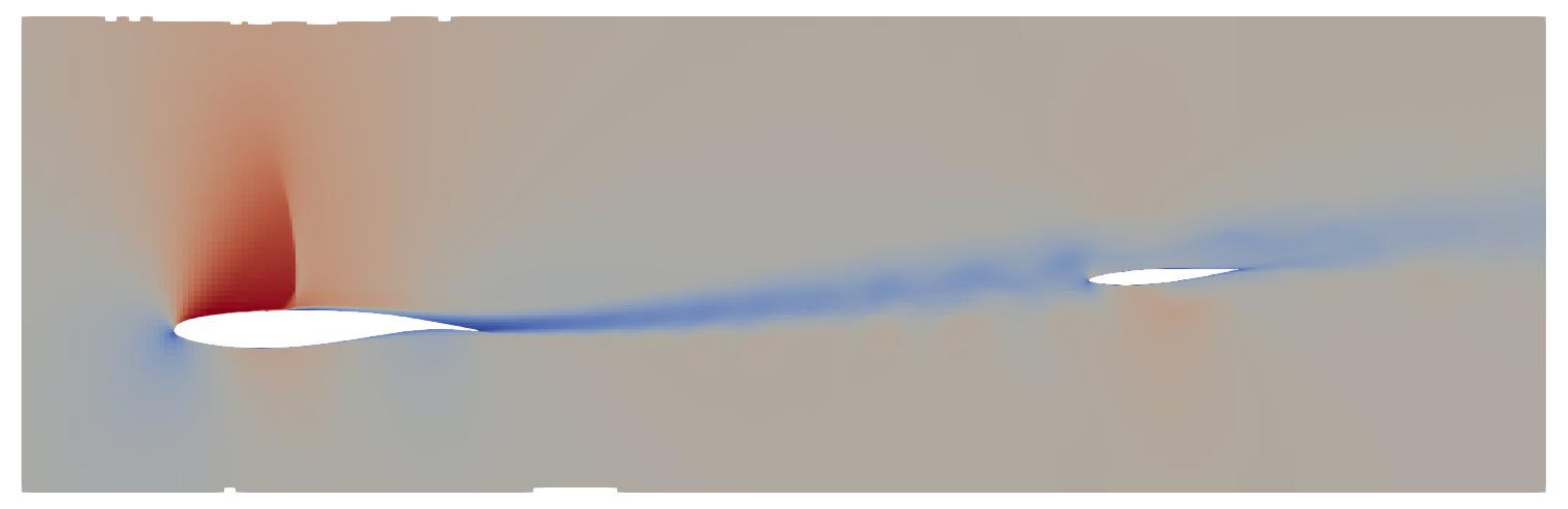} \\
					\footnotesize $\hat{\mathbf{x}}_0$
				\end{minipage}
			};
			
			\node (snapshotsInter1) at (-11.5, -14) {
				\begin{minipage}{0.2\textwidth}
					\centering
					\includegraphics[width=\textwidth]{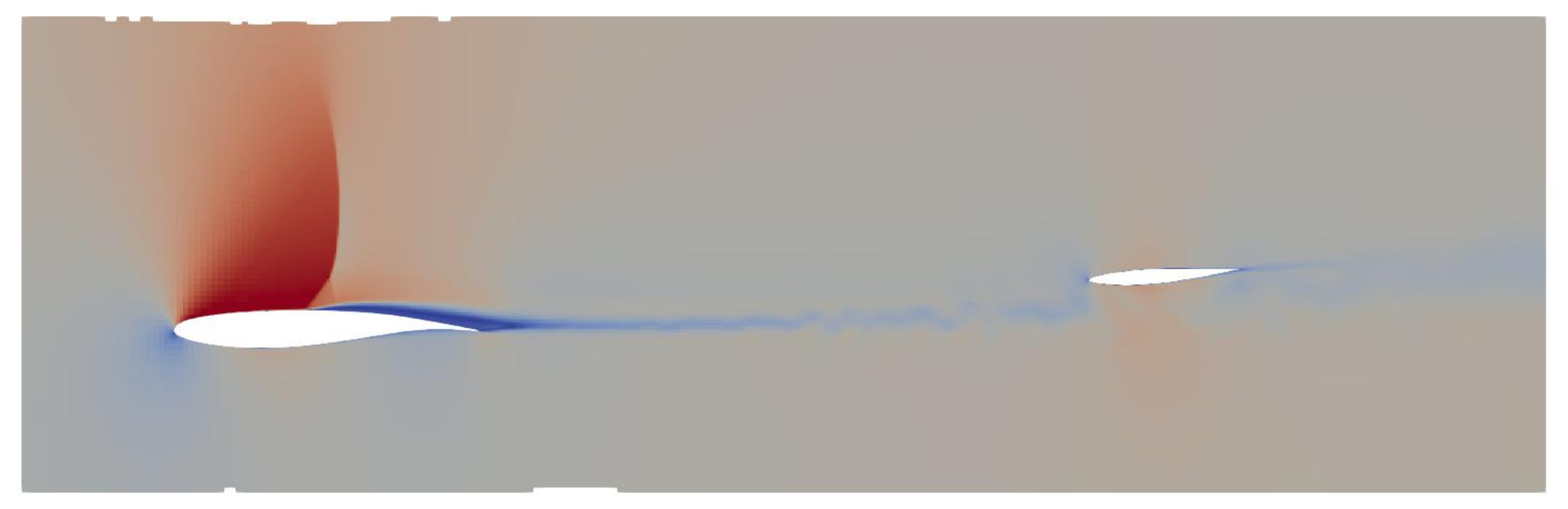} \\
					\footnotesize $\hat{\mathbf{x}}_1$
				\end{minipage}
			};
			
			\node (metric) at (-9.5, -13.8) {\footnotesize $...$};
			\node (metric) at (-9.5, -14.6) {\footnotesize $...$};
			
			\node (snapshotsInter2) at (-7.5, -14) {
				\begin{minipage}{0.2\textwidth}
					\centering
					\includegraphics[width=\textwidth]{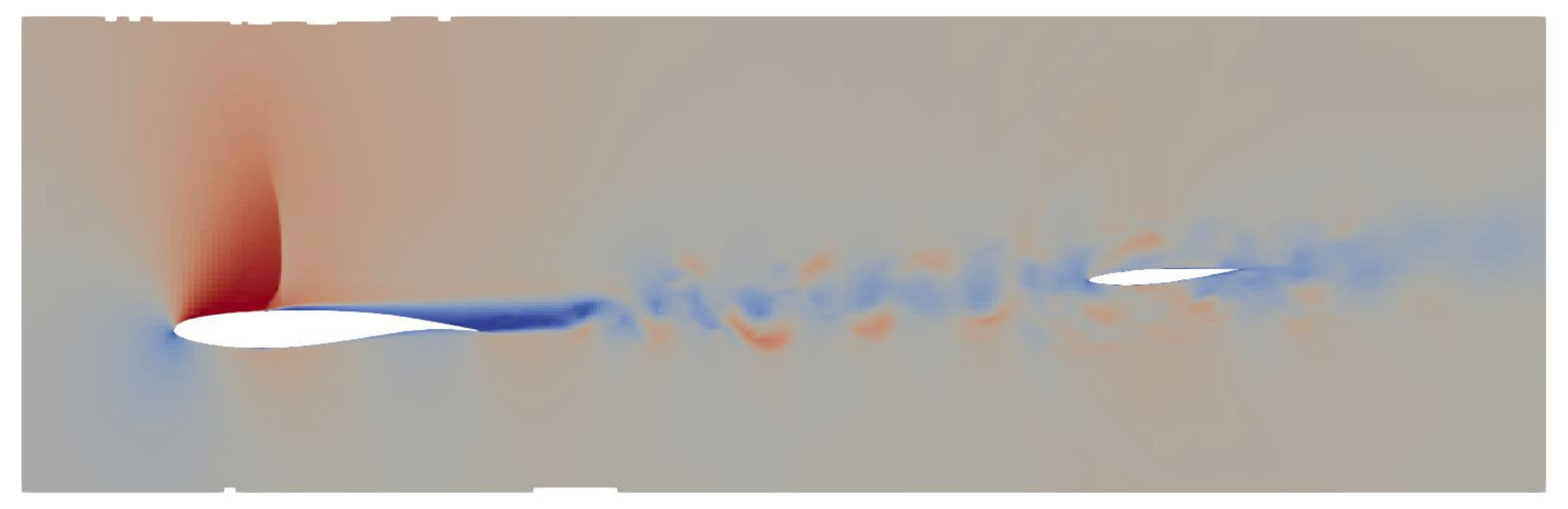} \\
					\footnotesize $\hat{\mathbf{x}}_N$
				\end{minipage}
			};
			
		\end{tikzpicture}
		\caption{The three main steps of $S^3$. The stopping criterion is either the maximum number of leaf cells $N_{\ell,\max}$ or the minimum percentage of the original metric that must be captured. The depicted test case represents a generic tandem configuration of an \emph{ONERA OAT15A} airfoil (front) and a \emph{NACA64A110} airfoil (rear).}
		\label{fig:scube_scheme}
	\end{figure}

The new version of $S^3$ follows the same principle idea as the original version presented in \cite{fernex_sparse_2021}, namely generating a time-invariant coarse grid based on a user-defined metric field.
This coarse grid can then be used to post-process CFD data more efficiently.
Rather than successively coarsening the original grid, the revised version of $S^3$ starts by creating a single large cell, which is refined iteratively.
This procedure enables the use of efficient recursive trees to store and update the mesh with each iteration.\\
As already mentioned in section (\ref{sec:intro}), $S^3$ consists of three main steps: (i) the computation of a metric field, (ii)  the adaptive, metric-based grid generation, and (iii) the interpolation of the original data onto the new grid. This process is illustrated in fig. (\ref{fig:scube_scheme}) and will be discussed thoroughly below.

Suppose that scalar field values are given at $M$ spatial points $\mathbf{p}_i = [x_i, y_i, z_i]^T$ of the original mesh, with $i \in \{1, 2, \ldots, M\}$.
For a co-located finite volume mesh, these points correspond to the cell centers, so $M$ is equivalent to the number of cells composing the mesh.
Since the example data used in section (\ref{sec:applications}) are generated by finite volume codes, we stick to this interpretation of $M$ for the remainder of this manuscript.
However, in principle, $S^3$ does not rely on the mesh topology or the underlying numerical method.
The first step is the calculation of a metric field $\boldsymbol{\mathcal{M}} \in \mathbb{R}^{M}$ based on the time series of $N$ snapshots at discrete times $t_n$ of one or more scalar fields $\mathbf{f}_n \in \mathbb{R}^{M}$ with $n \in \{1, 2, \ldots, N\}$.
The metric is a user-defined scalar field that should express the importance of each location $\mathbf{p}_i$.
For example, a simple but effective metric to accurately capture the shock oscillations of transonic shock buffet is the temporal standard deviation of the local Mach number, i.e., $\boldsymbol{\mathcal{M}} = \mathrm{std}( \mathbf{Ma})$ with
\begin{equation}
    \mathrm{std}( \mathbf{f})^2 = \frac{1}{N-1} \sum_{n=1}^{N} \left( \mathbf{f}_n - \bar{\mathbf{f}} \right)^2,
\quad
\bar{\mathbf{f}} = \frac{1}{N}\sum_{n=1}^{N} \mathbf{f}_n.
\end{equation}
There is virtually no constraint on $\boldsymbol{\mathcal{M}}$ other than it has to contain a scalar value for each point $\mathbf{p}_i$.
Users can easily tailor it to their specific target application and embed physical knowledge, e.g., to focus the refinement on regions with strong temporal fluctuations or production of turbulent kinetic energy.
The target-specific metric enables higher compression rates of the data while maintaining flexibility.

\noindent
As shown in fig. (\ref{fig:scube_scheme}), the grid generation consists of three steps.
First, a uniform background grid is created (step $2.1$), followed by an adaptive, metric-based refinement (step $2.2$), and an optional geometry refinement (step $2.3$).
The mesh structure is a quadtree for 2D or an octree for 3D cases.
Each cell has the same edge length in all coordinate directions.
The uniformly refined background mesh serves as a starting point for the metric-based refinement, which accelerates the overall process because the adaptive refinement is more computationally expensive.
The mesh generation starts by creating an initial cell with edge length $l_0$ that contains all points $\mathbf{p}_i$ (not shown in fig (\ref{fig:scube_scheme})).
Although the uniform refinement is straightforward, it serves here to explain the terminology.
Depending on the number of spatial dimensions, $D\in \left\{2, 3\right\}$, each cell marked for refinement is split into four or eight child-cells for 2D and 3D cases, respectively.
The cell that has been split is referred to as the root or parent cell.
All cells without child cells are called leaf cells and represent the current grid.
During each uniform refinement iteration, all available leaf cells are marked for refinement.
Since the original numerical domain is seldom a cube, an increasing refinement level inevitably leads to child cells being located outside the original numerical domain.
As shown in fig. (\ref{fig:scube_scheme}), these invalid cells are removed from the tree to avoid refining cells in regions that are not part of the domain.
A child cell is only considered outside the original domain if all $2^D$ corner points are located outside.
The uniform refinement stops after a predefined number of iterations.\\

Next, the metric-based refinement (step 2.2) follows, where in each refinement iteration, only a subset of the leaf cells is split.
Suppose the cell centers of a leaf cell are denoted as $\widehat{\mathbf{p}}_j$, $j\in \left\{ 1, 2, \ldots, 2^D\right\}$.
The cell volume of a cell that results from $l$ subdivisions of the initial cube is:
\begin{center}
	\begin{equation}
		V_l = \frac{1}{2^D} \left(\frac{l_0}{2^{l-1}}\right)^D.
	\end{equation}
\end{center}
The adaptive refinement process requires an estimate of the metric at the cell centers $\widehat{\boldsymbol{p}}_j$.
We employ an off-the-shelf K-nearest neighbors (KNN) algorithm to interpolate between the original point cloud and the octree mesh (\emph{scikit-learn} library \cite{pedregosa_scikit-learn_2012, cunningham_k-nearest_2022}).
Let $\left\{ (\mathbf{p}_k, \mathcal{M}_k)\right\}_{k=1}^K$ be the set of the $K$ nearest neighbor points of a query point $\widehat{\mathbf{p}}_j$ and their corresponding metric values.
Then the metric value at the query point can be approximated as:
\begin{equation}
	\widehat{\mathcal{M}}_{j} = \dfrac{\sum_{k = 1}^{K} \mathcal{M}_k w_k}{\sum_{k = 1}^{K} w_k},\qquad w_k = \frac{1}{||\mathbf{p}_k - \widehat{\mathbf{p}}_j ||},
\end{equation}
where we employ inverse distance weighting by default.
Motivated by the topology of Cartesian grids, we set $K=8$ and $K=26$ for 2D and 3D cases, respectively.
Since the original mesh is typically much denser than the sub-sampled mesh, the precise value of $K$ played no significant role in the performed numerical tests.

\begin{algorithm}[b!]
	\begin{algorithmic}[1]
		\Require Metric field $\boldsymbol{\mathcal{M}}$, points $\mathbf{p}_i$, tolerance $\mathcal{M}_{\min} \;\lor\;$ max. number of leaf cells $N_{\ell,\max}$
		\State Initialize KNN, sampling tree
		\State Create root cell covering the computational domain
		\State Compute reference norm $\lVert \boldsymbol{\mathcal{M}} \rVert$
		\State Perform $N_\mathrm{uni}$ uniform refinement cycles	\Comment{creates uniform background mesh}
        \State Evaluate gain $\mathcal{G}$ for all leaf cells using eq.~\eqref{eq:gain}
		\While{$||\widehat{\boldsymbol{\mathcal{M}}}|| / ||\boldsymbol{\mathcal{M}}|| \leq \mathcal{M}_{\min} \;\lor\; N_\ell \leq N_{\ell,\max}$}
		\State Select $N_{c}$ cells with largest $\mathcal{G}$ \Comment{avoid too large updates of $\widehat{\boldsymbol{\mathcal{M}}}$}
		\State Include neighboring cells with $\Delta l > 1$ \Comment{optional constraint}
		\State Refine selected cells
        \State Update gain $\mathcal{G}$ for all new leaf cells using eq.~\eqref{eq:gain}
		\State Remove invalid cells from tree
		\State Update $||\widehat{\boldsymbol{\mathcal{M}}}||$ and $N_\ell$
		\EndWhile
		\If{geometry refinement is enabled} 
			\For{each geometry $g$}
				\State Identify cells intersecting or near $g$
				\State Refine cells until prescribed target level $l_{\max}$ is reached \Comment{no re-evaluation of $\mathcal{M}_{\mathrm{approx}}$}
			\EndFor
		\EndIf
		\State Re-number leaf cells for export and interpolation
		\State \textbf{Output} Generated grid
	\end{algorithmic}
	\caption{Improved Sparse Spatial Sampling pseudo algorithm}
	\label{algo:s_cube}
\end{algorithm}
In contrast to the original $S^3$ algorithm, the metric value is not used directly to mark leaf cells for refinement.
Instead, we estimate by how much the approximation of the original metric field improves when splitting a leaf cell.
The improvement of the approximation when splitting a leaf cell at refinement level $l$ is denoted as gain $\mathcal{G}_l$ and defined as:
\begin{center}
	\begin{equation}
		\mathcal{G}_l = \frac{V_{l+1}}{\mathcal{G}_0} \sum_{j = 1}^{2^D} | \widehat{\mathcal{M}}_l - \widehat{\mathcal{M}}_{l+1,j}|,
		\label{eq:gain}
	\end{equation}
\end{center}
where we normalize by the gain of the initial cell, $\mathcal{G}_0$, for numerical stability.
The summation in eq. \eqref{eq:gain} measures the error reduction resulting from a potential split of the leaf cell and promotes refinement in regions where the metric field undergoes strong spatial variations.
In contrast to the original version of $S^3$, regions with uniformly high metric are not excessively refined.
To quantify how much the local approximation error contributes to the global approximation error in terms of a volume integral over the domain, the error is weighted by the child cells' volume $V_{l+1}$.
The volume weighting also avoids large jumps in the refinement level between neighboring leaf cells, which in turn tends to improve interpolation errors when mapping fields between original and octree mesh.
Optionally, the implementation allows limiting the difference in refinement level between neighboring cells to one, which is useful, e.g., when computing gradients on the octree mesh.

Refining only the leaf cell with the highest gain value in each adaptive refinement results in the most efficient octree mesh, but it also leads to an impractically high iteration count.
Therefore, the current leaf cells are first sorted from the highest to the lowest gain value, and then the first $N_c$ cells are marked for refinement.
$N_c$ may be varied linearly between $N_{c,\mathrm{start}}$ and $N_{c,\mathrm{end}}$, with $N_{c,\mathrm{start}} \ge N_{c,\mathrm{end}}$, to promote a more fine-grained refinement towards the end.
By default, we set $N_c = N_{c,\mathrm{start}} = N_{c,\mathrm{end}} = 0.01 \, M$, which appeared as a reasonable compromise between effective cell placement and computational speed in a variety of numerical tests.
As in the uniform refinement, leaf cells that fall outside the domain are removed at the end of each iteration.

Before starting the next iteration, the user-defined stopping criterion is evaluated.
The stopping criterion is either the maximum number of leaf cells $N_{\ell,\max}$ or the minimum percentage of the original metric that must be captured.
While the first criterion is straightforward to evaluate, the latter is not directly available, since the current number of leaf cells differs from the number of cells in the original mesh.
Suppose that $\widehat{\boldsymbol{\mathcal{M}}}\in \mathbb{R}^{N_\ell}$ is a vector holding the approximated metric of all $N_\ell$ leaf cells.
To stop the refinement, we measure the captured metric as the ratio of the vector norms and compare it to a user-defined threshold value, i.e., stop if $ \mathcal{M}_\mathrm{approx} = || \widehat{\boldsymbol{\mathcal{M}}}|| / ||\boldsymbol{\mathcal{M}}|| \ge \mathcal{M}_{\min}$.
Note that approximation errors in $\widehat{\boldsymbol{\mathcal{M}}}$ may cancel out when comparing the norms.
However, we found the impact of this cancellation to be negligible.

Once the metric-based refinement is completed, an optional mesh refinement can be performed near geometries to improve their representation (step 2.3).
This additional refinement is advantageous if the metric has only small gradients near geometries, but further post-processing requires a good approximation of the geometry.

Once the refinement is complete, the final leaf cells are extracted from the tree and converted into a suitable CFD mesh description.
We denote the extraction as the renumbering (rn) step in the remainder of this manuscript.

Interpolating the original data onto the generated grid marks the final step (step $3$).
We employ the same KNN interpolation approach as for the metric field.
Although not observed in the investigated test cases, the lack of exception handling for missing neighbors near domain boundaries or thin geometries, such as the trailing edge of an airfoil, may lead to spurious values and may necessitate a more careful neighbor selection.
Depending on the size of the snapshots and the available memory, the snapshots are either loaded and interpolated all at once, in batches, or snapshot by snapshot.
The interpolated snapshots are exported to an HDF5 binary file format.
A summary of all steps is shown in pseudo-algorithm (\ref{algo:s_cube}).

\noindent
It was generally found that the resulting grid topology is only slightly influenced by the hyperparameters of $S^3$, provided they are chosen within reasonable bounds; therefore, results of conducted hyperparameter studies are not included here.
A summary of all hyperparameters and their default values can be found in table (\ref{table:def_values}).

\begin{table}[htb]
	\begin{center}
		\caption{Hyperparameters and associated default values for $S^3$. The parameters $N_{c,  \mathrm{start}}$ and $N_{c,\mathrm{end}}$ denote the number of cells selected for refinement at the beginning and at the end of the adaptive refinement, respectively, while $M$ represents the number of cells of the original grid. The default stopping criterion is activated when the current approximation of the metric field $||\boldsymbol{\widehat{\mathcal{M}}}||$ relative to the original metric field $||\boldsymbol{\mathcal{M}}||$ exceeds the given threshold $\mathcal{M}_{\min}$.}
		\begin{tabular}{@{}lcccc@{}}
			\toprule
			& $N_\mathrm{uni}$ & $N_{\mathrm{c, start}}$ & $N_{\mathrm{c, end}}$ & stopping criterion \\
			\midrule
			default value & $5$ & $0.01 \, M$ & $N_{c,\mathrm{start}}$ & $||\widehat{\boldsymbol{\mathcal{M}}}|| / ||\boldsymbol{\mathcal{M}}|| \ge \mathcal{M}_{\min}$ \\ 
			\bottomrule
		\end{tabular}
		\label{table:def_values}
	\end{center}
\end{table}

\section{Results}
\label{sec:applications}
\subsection{Airfoil tandem configuration}\label{subsec:oat}
The first test case is a tandem airfoil configuration in high-speed stall conditions.
The data are kindly provided by research partners within the research unit FOR 2895 \cite{for_2895_unsteady_nodate}.
A detailed description of the numerical setup is presented in \cite{kleinert_wake_2023, kleinert_numerical_2023}.
The leading airfoil within the tandem configuration is a supercritical \emph{ONERA OAT15A} with chord length $c_\mathrm{front}~=~0.15 \, m$ at an angle of attack of $5^\circ$, whereas the rear airfoil is a \emph{NACA64A110} with chord length $c_\mathrm{rear}~=~0.075 \,m$ placed at a horizontal distance of $2c_{\mathrm{front}}$ behind the leading airfoil's trailing edge.
The inflow Mach number is $Ma_{\infty}~=~0.72$, leading to a freestream Reynolds number of $Re_\infty~=~2 \cdot 10^6$ based on the chord length of the leading airfoil.
Under these conditions, the leading airfoil exhibits periodic shock-boundary layer interactions known as shock buffet.
Vortices are generated at the leading airfoil's trailing edge and in the separating boundary layer.
These vortices are propagated through the wake and impact on the rear airfoil.

\begin{figure}[htbp]
	\begin{center}
		\subfloat[original grid]{\includegraphics[width=0.42\textwidth]{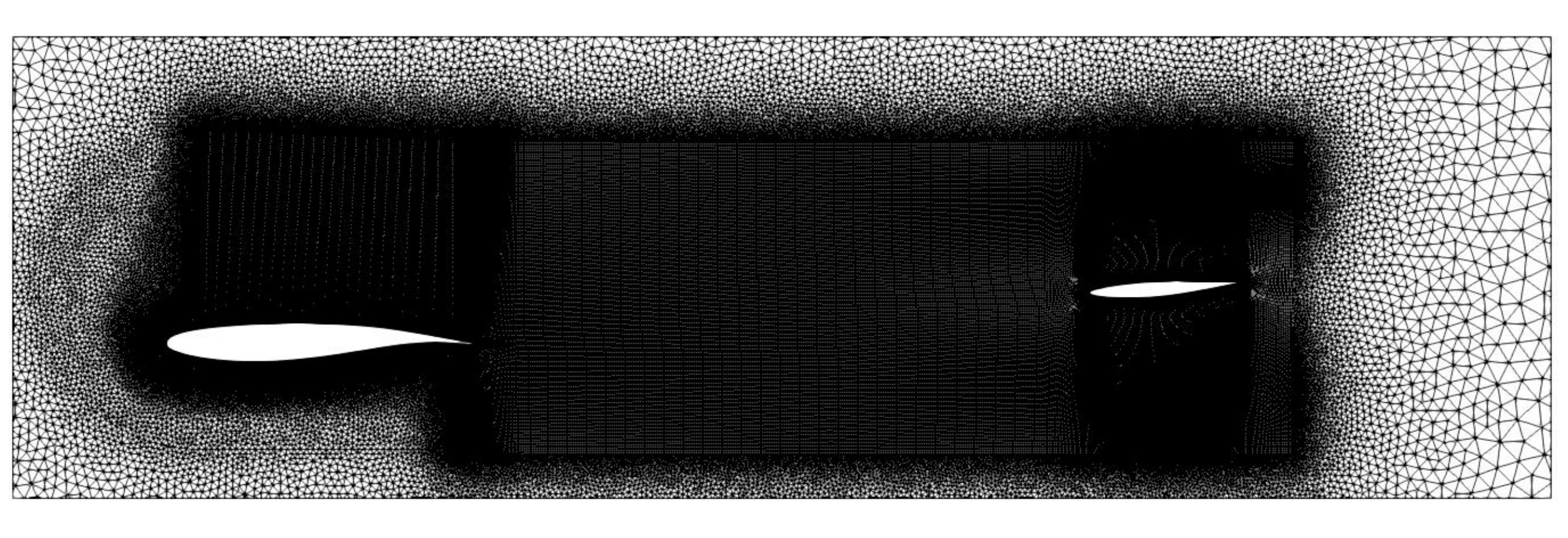}\label{fig:original_grid_oat}}
		\hspace*{5mm}
		\subfloat[original metric field $\boldsymbol{\mathcal{M}}_1$]{%
			\begin{tikzpicture}[baseline]
				\node[anchor=south west, inner sep=0] (image) at (0,0)
				{\includegraphics[width=0.42\textwidth]{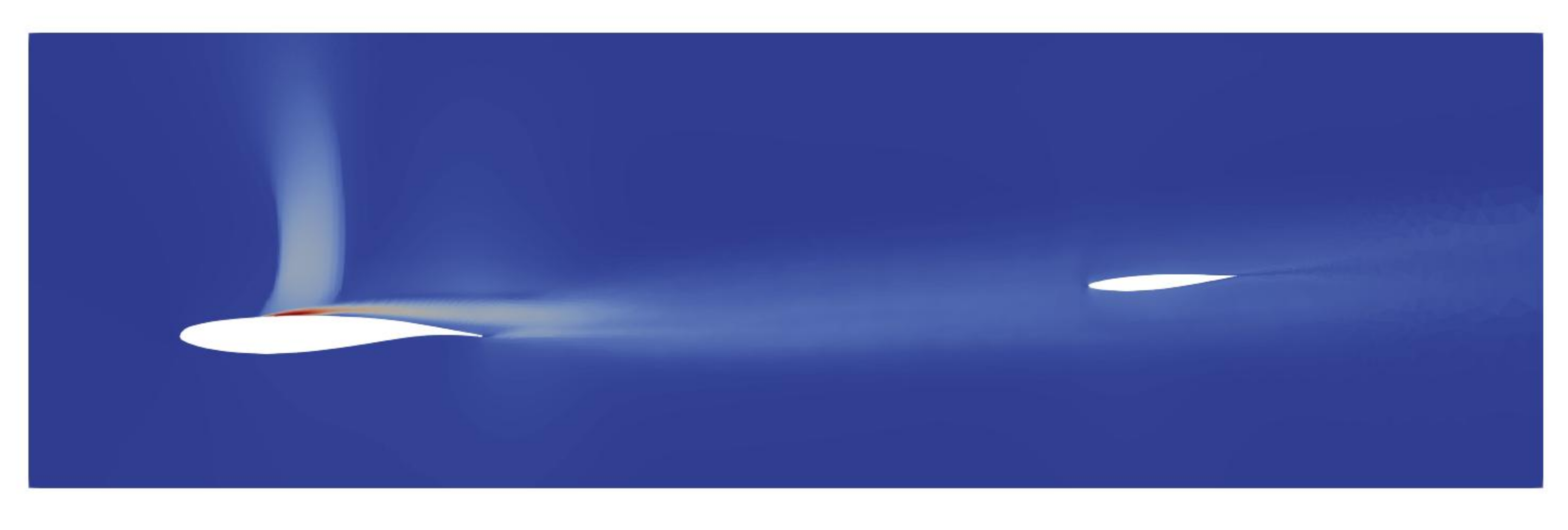}};
				\begin{scope}[x={(image.south east)}, y={(image.north west)}]
					\draw[-{Stealth}, thick, black] (0.115, 0.36) -- ++(0.075, 0) node[below=2pt, midway] {\small $x$};
					\draw[-{Stealth}, thick, black] (0.115, 0.36) -- ++(0, 0.225) node[left=2pt, midway] {\small $z$};
				\end{scope}
			\end{tikzpicture}
			\label{fig:original_metric_oat}}
		\\
		\subfloat[grid generated by $S^3$]{\includegraphics[width=0.42\textwidth]{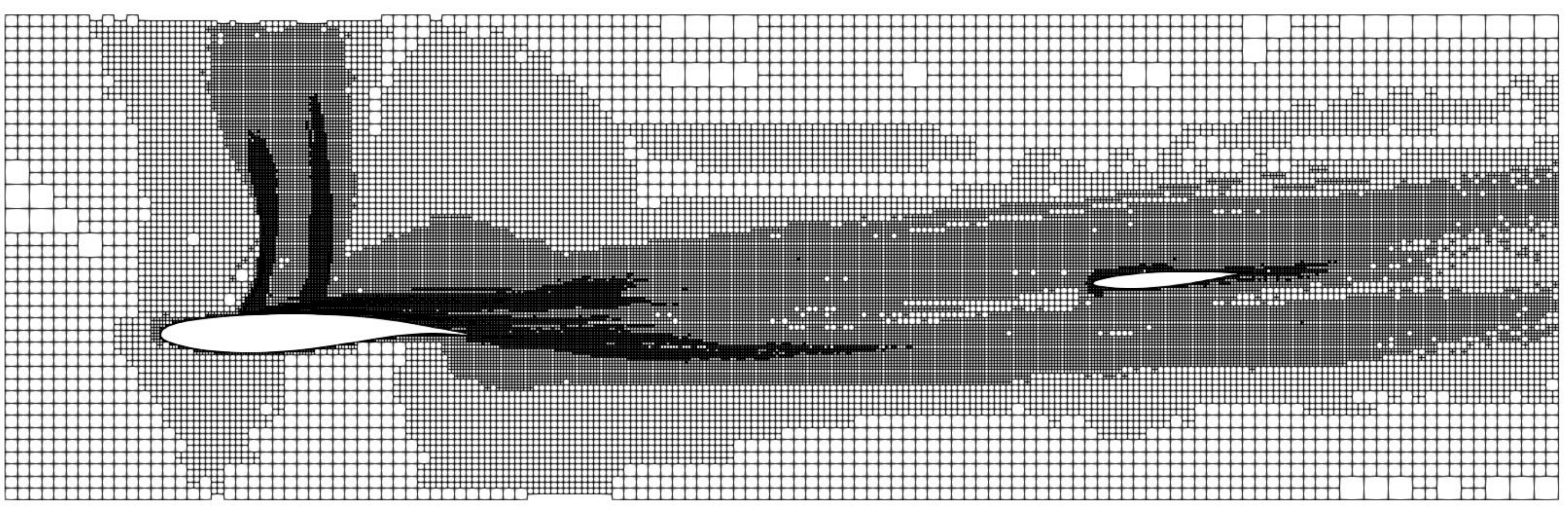}\label{fig:interpolated_grid_oat}}
		\hspace*{5mm}
		\subfloat[interpolated metric field $\widehat{\boldsymbol{\mathcal{M}}}_1$]{\includegraphics[width=0.42\textwidth]{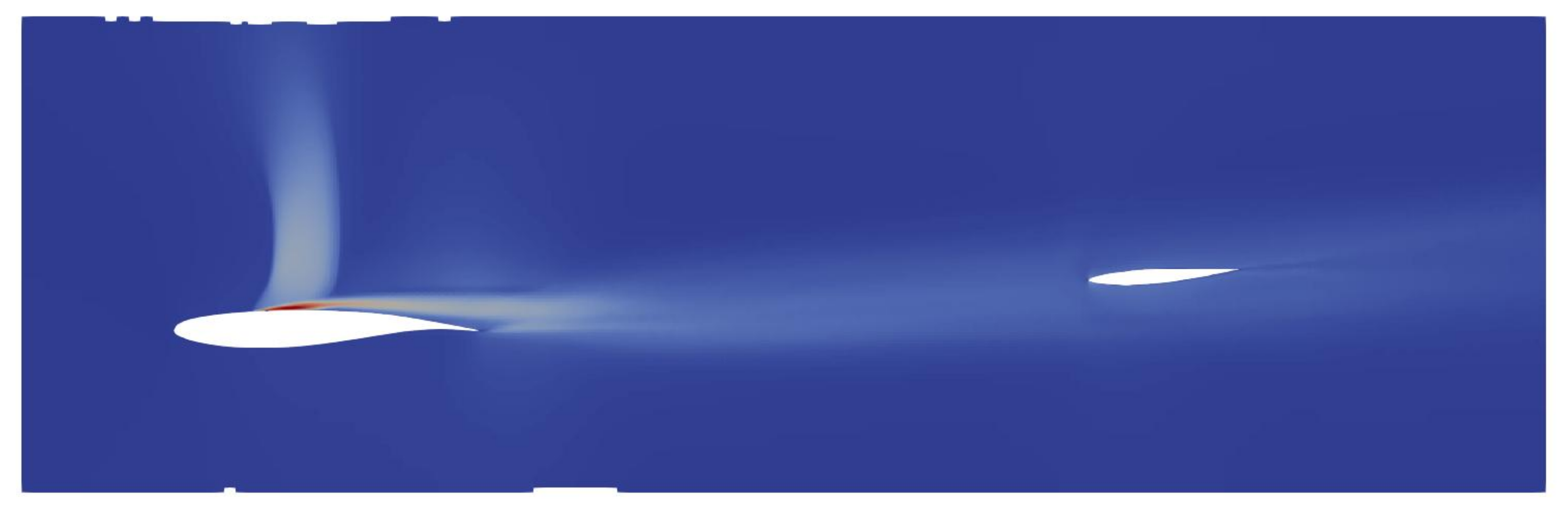}\label{fig:interpolated_metric_oat}}
	\end{center}
	\caption{The left column shows a comparison of the original grid (\ref{fig:original_grid_oat}) and the grid generated by $S^3$ for a captured metric of $ \mathcal{M}_\mathrm{approx}~=~|| \widehat{\boldsymbol{\mathcal{M}}}_1|| / ||\boldsymbol{\mathcal{M}}_1||~=~0.75$ (\ref{fig:interpolated_grid_oat}). The right column depicts the metric on the original grid (\ref{fig:original_metric_oat}) and the metric interpolated onto the $S^3$-grid (\ref{fig:interpolated_metric_oat}). Both contour plots are scaled from zero to $||\boldsymbol{\mathcal{M}}_1||_\infty$.}
\end{figure}

The original finite volume mesh consists of $16 \cdot 10^6$ grid points in total.
The simulation was executed with the DLR TAU code, employing automated zonal detached eddy simulation and an SSG/LRR-$\omega$ Reynolds stress model.
In contrast to \cite{kleinert_wake_2023}, only a two-dimensional plane at a fixed spanwise location is analyzed here.
The plane consists of $2.68 \cdot 10^5$ cells.
Since the majority of the mesh cells are accumulated in the vicinity of the two airfoils, the region of interest is restricted to $x/c_{\mathrm{front}} \in [-0.5, 4.5]$, $z/c_{\mathrm{front}} \in [-0.5, 1.0]$, where $x$ is the coordinate in streamwise direction, and $z$ is the second plane coordinate normal to $x$.
The origin is located at the leading airfoil's nose; refer to fig.~(\ref{fig:original_metric_oat}).
The region of interest contains $2.456 \cdot 10^5$ cells in total.	
The dataset comprises $559$ snapshots taken at an equidistant time interval of $\Delta t~=~4.4 \cdot 10^{-5}\,s$.
Storing the time sequence of a scalar field occupies approximately $550$ MB of disk space (single-precision floating-point numbers).

Since shock oscillations and periodic vortex shedding characterize the shock buffet, the first metric field for this test case is defined as the temporal standard deviation of the local Mach number computed over all snapshots, $\boldsymbol{\mathcal{M}}_1~=~\mathrm{std}(\mathbf{Ma})$.
We provide results obtained with more generic metric definitions at the end of this section.

\noindent
Fig.~(\ref{fig:original_grid_oat}) shows the original grid within the defined region of interest, along with the metric depicted in fig.~(\ref{fig:original_metric_oat}).
The increased metric in the shock region and the wake of the leading airfoil are clearly visible.
The metric reaches its maximum in the region where the boundary layer detaches.
The stopping criterion is set to a minimum approximation of the metric, $\mathcal{M}_{\min}~=~0.75$.
Due to the refinement of multiple cells within a single iteration, discussed in section (\ref{subsec:scube}), the mesh generated by $S^3$ captures $\mathcal{M}_{\mathrm{approx}}~=~0.752$.
The final mesh, shown in fig.~(\ref{fig:interpolated_grid_oat}), consists of $5.24 \cdot 10^4$ cells corresponding to a reduction of $78.67\%$. 
The progression of $\mathcal{M}_{\mathrm{approx}}$ with respect to the adaptive refinement iterations is shown in the appendix, section \ref{sec:appendix:cylinder} (fig.~(\ref{fig:progress_refinement})).
We also report detailed timings for the individual refinement steps in section (\ref{subsec:timings}).
It is important to note that for this test case, a subsequent geometry refinement is employed to demonstrate the ability of $S^3$ to resolve complex geometries properly, which is not included in the captured metric.
However, this geometry refinement is not always important, e.g., when computing the SVD.
By omitting the geometry refinement, a total reduction of $81.54\%$ can be achieved without further loss of information.	

\begin{figure}[htbp]
	\begin{center}
		\includegraphics[width=0.9\textwidth]{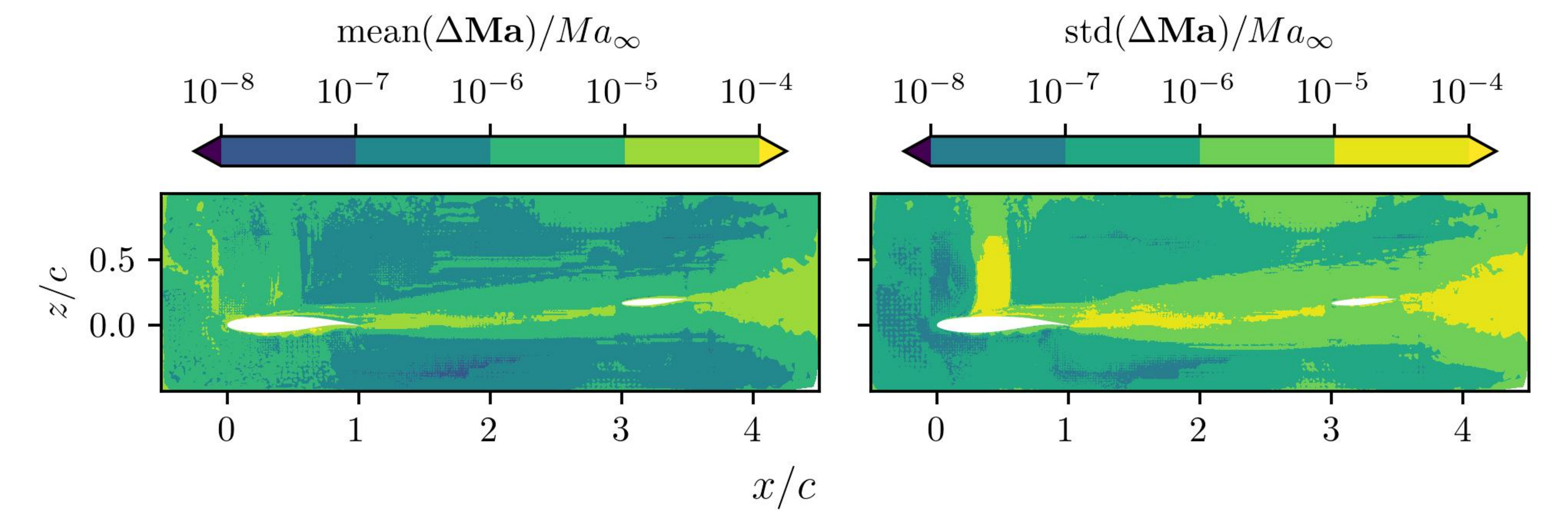}
		\caption{Temporal mean (left) and standard deviation (right) of the absolute spatial error $\Delta \mathbf{Ma}_n~=~|\mathbf{Ma}^\ast_n - \mathbf{Ma}_n|$, weighted with the square-root of the cell volume and scaled with the free stream Mach number $Ma_{\infty}$ for $\mathcal{M}_{\mathrm{approx}}~=~0.75$.}
		\label{fig:spatial_error_oat}
	\end{center}
\end{figure}
\noindent
To assess the quality of the interpolated Mach number snapshots $\widehat{\mathbf{Ma}}$, we map the interpolated field back to the original grid using KNN interpolation, denoted as $\mathbf{Ma}^\ast$, and compare to the original field $\mathbf{Ma}$.
The overall local interpolation error (original grid $\rightarrow$ $S^3$ $\rightarrow$ original grid) of the $n$th snapshot is $\Delta \mathbf{Ma}_n~=~|\mathbf{Ma}^\ast_n - \mathbf{Ma}_n|$.
The temporal mean and standard deviation of the weighted interpolation error over all snapshots are shown in fig.~(\ref{fig:spatial_error_oat}).
Both the normalized mean error and the standard deviation are very small, $O(10^{-5})$, with their maximum values located in the wake region aft of the rear airfoil.

As we show next, the approximation error is sufficiently small for most post-processing and analysis tasks.
The SVD is a matrix factorization that provides an optimal low-rank representation of data arranged in matrix form.
If the sequence of flow states is arranged into a data matrix as a sequence of column vectors, the matrix of left-singular vectors, $\mathbf{U}$, represents spatial structures (equivalent to the so-called POD modes), while the matrix of right-singular vectors, $\mathbf{V}$, represents temporal importance.
Both sets of vectors are sorted according to the singular values, $\mathbf{\Sigma}$, from most to least important.
We provide a compact mathematical description in section (\ref{subsec:svd}).
Comparing the SVDs computed with the original and reduced data matrices poses a challenging test that validates the suitability of $S^3$ data for analyzing dominant coherent structures.
The SVD is employed here as a representative example of a memory-intensive post-processing task, serving as a benchmark throughout all test cases considered in this work.

\begin{figure}[htbp]
	\begin{center}
		\includegraphics[width=0.75\textwidth]{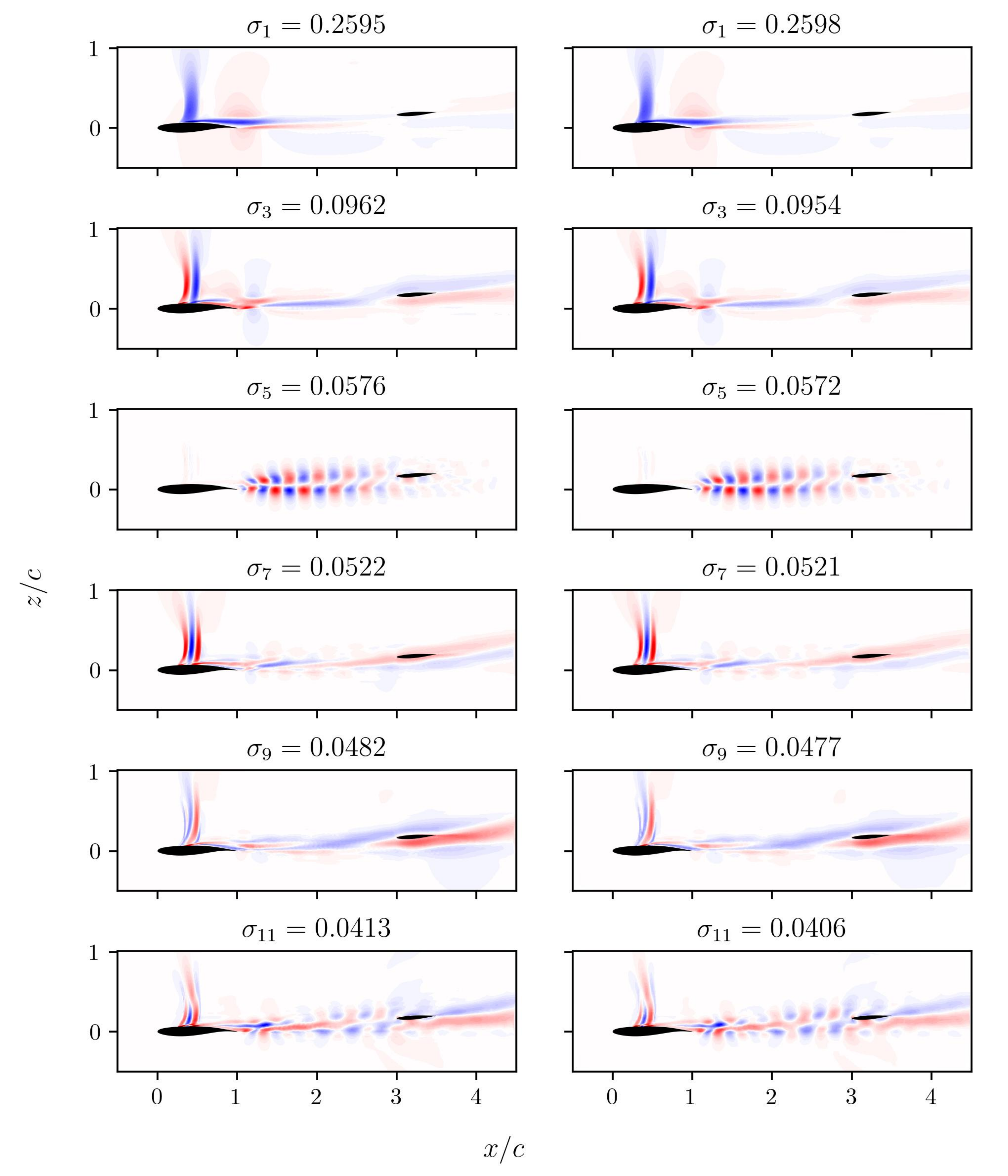}
		\caption{Comparison of the leading POD modes and the associated singular values for the tandem configuration. The left column shows the POD modes on the original grid, while the right column depicts the POD modes on the grid generated with $S^3$ and $\mathcal{M}_{\mathrm{approx}}~=~0.27$. The colorscale is identical for all contours and bounded by $\pm||\mathbf{U}||_\infty$.}
		\label{fig:modes_oat}
	\end{center}
\end{figure}

\begin{figure}[htbp]
	\begin{center}
		\includegraphics[width=0.8\textwidth]{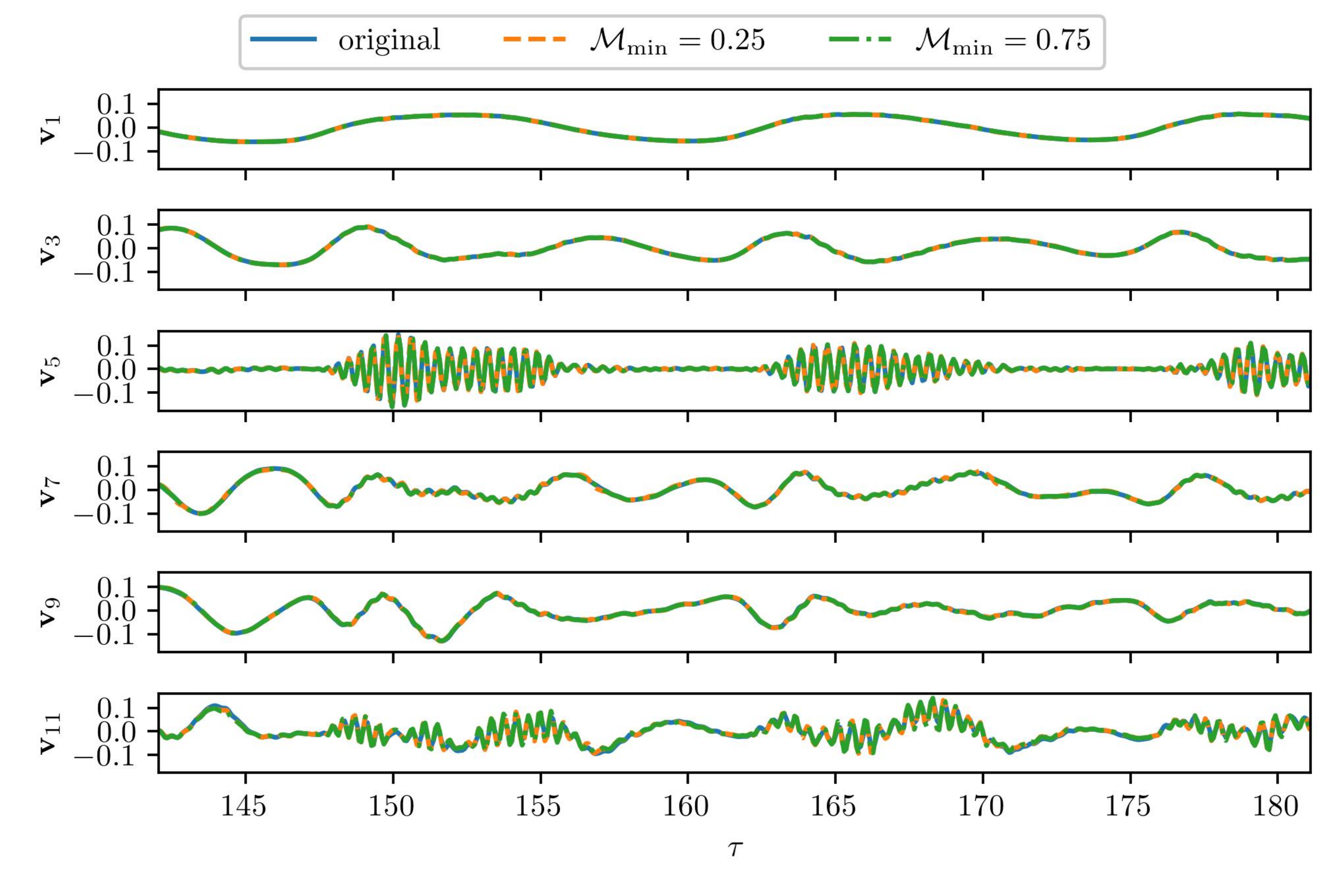}
		\caption{Comparison of the leading right-singular vectors for the tandem configuration. As for the modes, only every second mode coefficient is shown.}
		\label{fig:mode_coefficients_oat}
	\end{center}
\end{figure}

For the tandem configuration, the data matrix is constructed based on the local Mach number.
We tested a variety of threshold values in the range $0.25 \leq \mathcal{M}_\mathrm{min} \leq 0.75$ and found a strong robustness of the modal decomposition with respect to the stopping criterion.
For brevity, we only present POD modes (left-singular vectors) for the lowest threshold value.
Fig.~(\ref{fig:modes_oat}) compares every second of the first $12$ modes and the corresponding singular values $\sigma_i$ for $\mathcal{M}_{\mathrm{approx}}~=~0.27$ (note that $\mathcal{M}_{\mathrm{approx}}$ can be slightly larger than $\mathcal{M}_\mathrm{min}$).
We show every second mode, since spatially periodic flow structures give rise to pairs of similar modes.
Modes one to four encode the main shock motion and the associated boundary layer separation typical for buffet.
Modes four and five represent periodic vortex shedding at the leading airfoil's trailing edge.
Higher-order modes show a mixture of these two phenomena.
For more details on the physical interpretation, we refer to the literature \cite{weiner_robust_2023,kleinert_wake_2023}.
Despite the low threshold value, the modes are visually nearly indistinguishable, and the first six singular values differ only in the fourth significant digit.
More precisely, the error is around $0.1\%$ for $\sigma_{1}$ and increases to $1.7\%$ for $\sigma_{11}$.
We included a comparison for the first $100$ singular values in the appendix (\ref{fig:eigenvalues_oat}) for both $\mathcal{M}_\mathrm{min} = 0.25$ and $0.75$.
The absolute as well as the cumulative error remain low for both threshold values.
The low error in the modes and singular values is also reflected in the mode coefficients $\mathbf{v}_i$ (right-singular vectors), shown in fig.~(\ref{fig:mode_coefficients_oat}).
The time is non-dimensionalized with the free stream velocity $U_{\infty}$ and the chord length of the leading airfoil, yielding $\tau~=~tU_{\infty} / c_{\mathrm{front}}$.
Both phase and amplitude are in good agreement for the tested threshold values, even for higher-order modes, depicted in the appendix, fig. (\ref{fig:v_temporal_error_oat}), for first $100$ mode coefficients.
The error remains in the order of $O(10^{-5} \dots 10^{-2})$ across all time steps and mode coefficients.

So far, the error as well as the SVD results were assessed on the same quantity that was used to generate the grid, namely, the local Mach number field.
To demonstrate the influence of the chosen metric on the resulting grid and interpolation error, we generate two new grids and assess the interpolation error on fields that have not been part of the metric field.
It should be noted that the choice of the metric field generally depends on the quantities of interest, because the adaptive grid generation process is driven by the metric field, as discussed in Section (\ref{subsec:scube}).
Since the quantity of interest is not always known a priori, we define the next metric field more generically as the superposition of temporal mean flow quantities, i.e., $\boldsymbol{\mathcal{M}}_2~=~\mathrm{mean}(\mathbf{p}) / p_\infty + \mathrm{mean}(\boldsymbol{\rho}) / \rho_\infty + \mathrm{mean}(\sqrt{\mathbf{u}_1^2+\mathbf{u}_3^2}) / U_\infty$, where $\mathbf{p}$,
$\boldsymbol{ \rho }$, $\mathbf{u}_1$, and $\mathbf{u}_3$ are the pressure, density, and velocity component fields in $x$ and $z$, respectively. The subscript $\infty$ indicates the respective free stream values.
Similarly, the third metric is defined as the superposition of temporal standard deviations, i.e., $\boldsymbol{\mathcal{M}}_3~=~\mathrm{std}(\mathbf{p}) / p_\infty + \mathrm{std}(\boldsymbol{\rho}) / \mathbf{\rho}_\infty + \mathrm{std}(\sqrt{\mathbf{u}_1^2+\mathbf{u}_3^2}) / U_\infty$.
$\boldsymbol{\mathcal{M}}_1~=~\mathrm{std}(\mathbf{Ma})$ remains the same as before.
All cases use $\mathcal{M}_{\min}~=~0.75$ as stopping criterion.

\begin{figure}[htbp]
	\begin{center}
		\subfloat[resulting grid for $\boldsymbol{\mathcal{M}}_2$]{\includegraphics[width=0.42\textwidth]{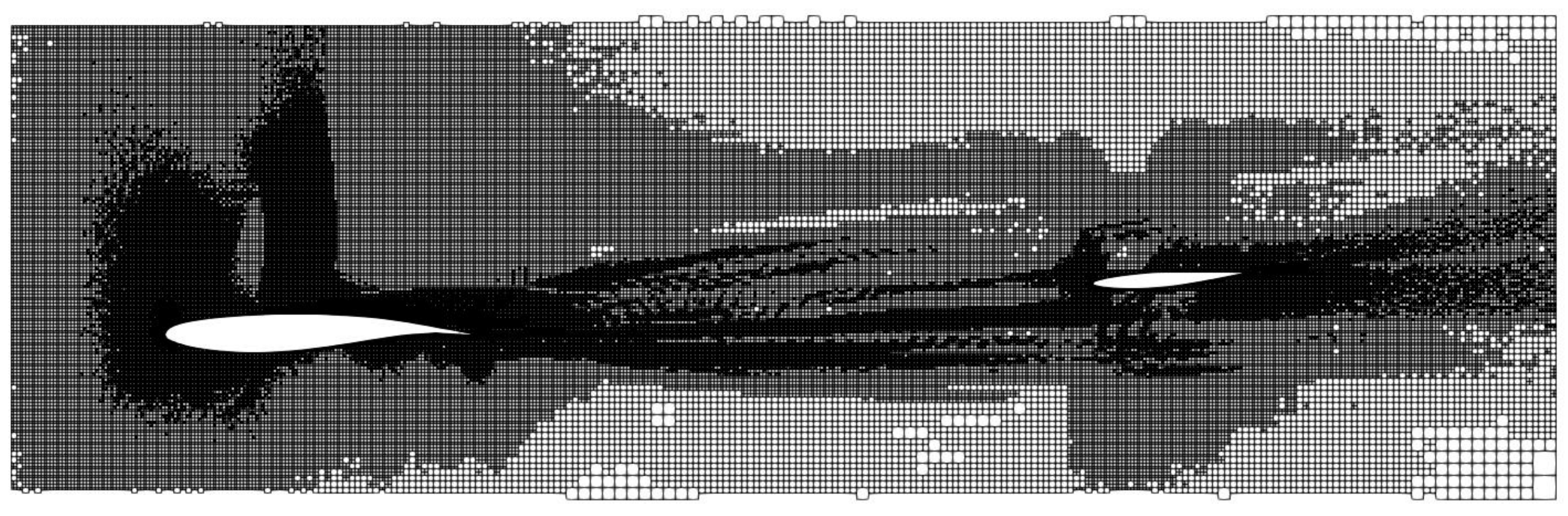}\label{fig:oat_grid_m2}}
		\hspace*{5mm}
		\subfloat[interpolated metric field $\widehat{\boldsymbol{\mathcal{M}}}_2$]{\includegraphics[width=0.42\textwidth]{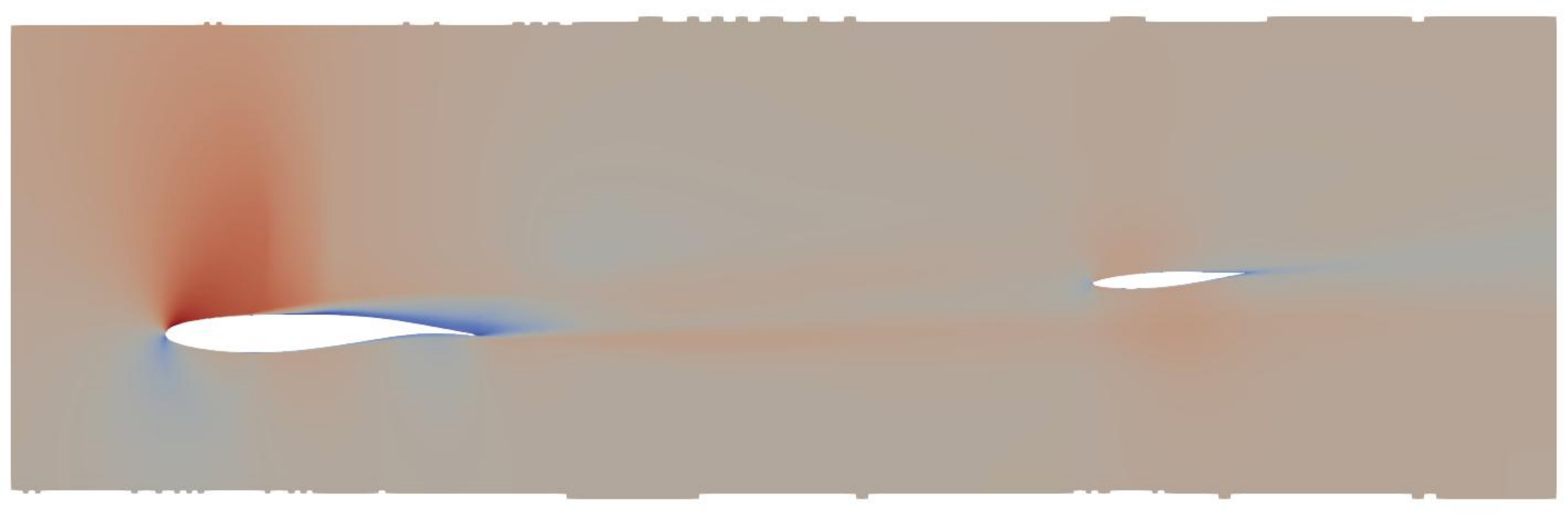}
			\label{fig:oat_metric_m2}}
		\\
		\subfloat[resulting grid for $\boldsymbol{\mathcal{M}}_3$]{\includegraphics[width=0.42\textwidth]{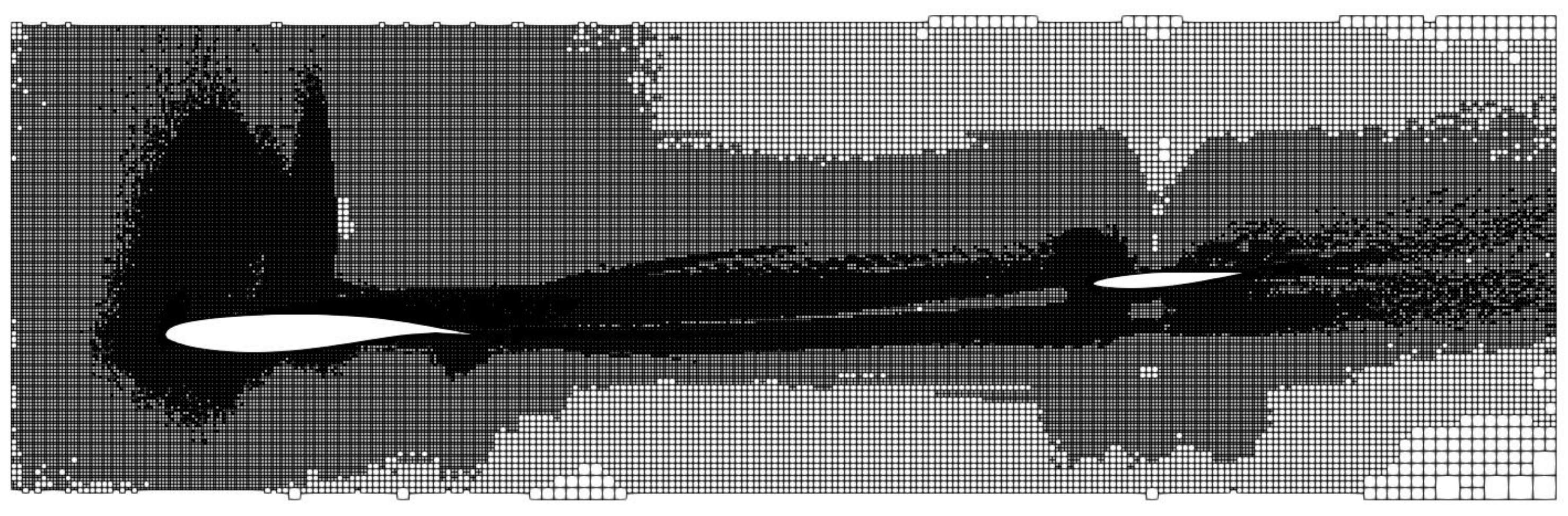}\label{fig:oat_grid_m3}}
		\hspace*{5mm}
		\subfloat[interpolated metric field $\widehat{\boldsymbol{\mathcal{M}}}_3$]{\includegraphics[width=0.42\textwidth]{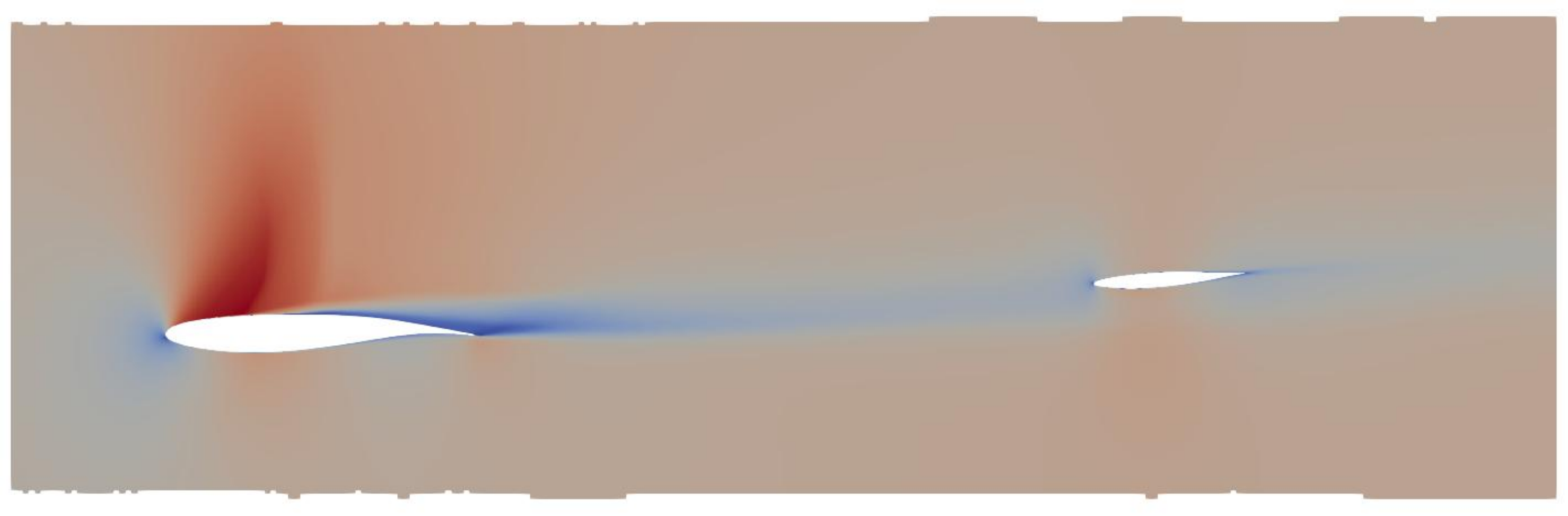}\label{fig:oat_metric_m3}}
	\end{center}
	\caption{The left column shows the grids generated by $S^3$ for $\boldsymbol{\mathcal{M}}_2$ (\ref{fig:oat_grid_m2}) and $\boldsymbol{\mathcal{M}}_3$ (\ref{fig:oat_grid_m3}). Both grids capture a metric of $ \mathcal{M}_\mathrm{approx}~=~|| \widehat{\boldsymbol{\mathcal{M}}}|| / ||\boldsymbol{\mathcal{M}}||~=~0.75$. The right column depicts the interpolated metric fields $\widehat{\boldsymbol{\mathcal{M}}}_2$ (\ref{fig:oat_metric_m2}) and $\widehat{\boldsymbol{\mathcal{M}}}_3$ (\ref{fig:oat_metric_m3}), respectively. Both contour plots are scaled from zero to $||\boldsymbol{\mathcal{M}}||_\infty$.}
	\label{fig:oat_generic}
\end{figure}
Figure (\ref{fig:oat_generic}) shows the resulting grids as well as the metric fields $\boldsymbol{\mathcal{M}}_2$ and $\boldsymbol{\mathcal{M}}_3$.
While the metric fields show significant differences, e.g., in the wake regions of the front airfoil, the resulting grids generated by $S^3$ are of a similar shape.
The cell counts for both metrics are nearly the same, leading to reductions of $41.41\%$ for $\boldsymbol{\mathcal{M}}_2$ and $40.58\%$ for $\boldsymbol{\mathcal{M}}_3$.
The increase in the number of cells compared to $\mathcal{M}_1$ originates from a finer mesh in the front part of the domain as well as a more refined wake region.
Since regions with strong variations of the metric field are less localized for $\boldsymbol{\mathcal{M}}_2$ and $\boldsymbol{\mathcal{M}}_3$ compared to $\boldsymbol{\mathcal{M}}_1$, larger portions of the generated meshes exhibit high mesh density ($S^3$ operates on the spatial change of the metric field).

\begin{figure}[htbp]
	\begin{center}
		\includegraphics[width=0.75\textwidth]{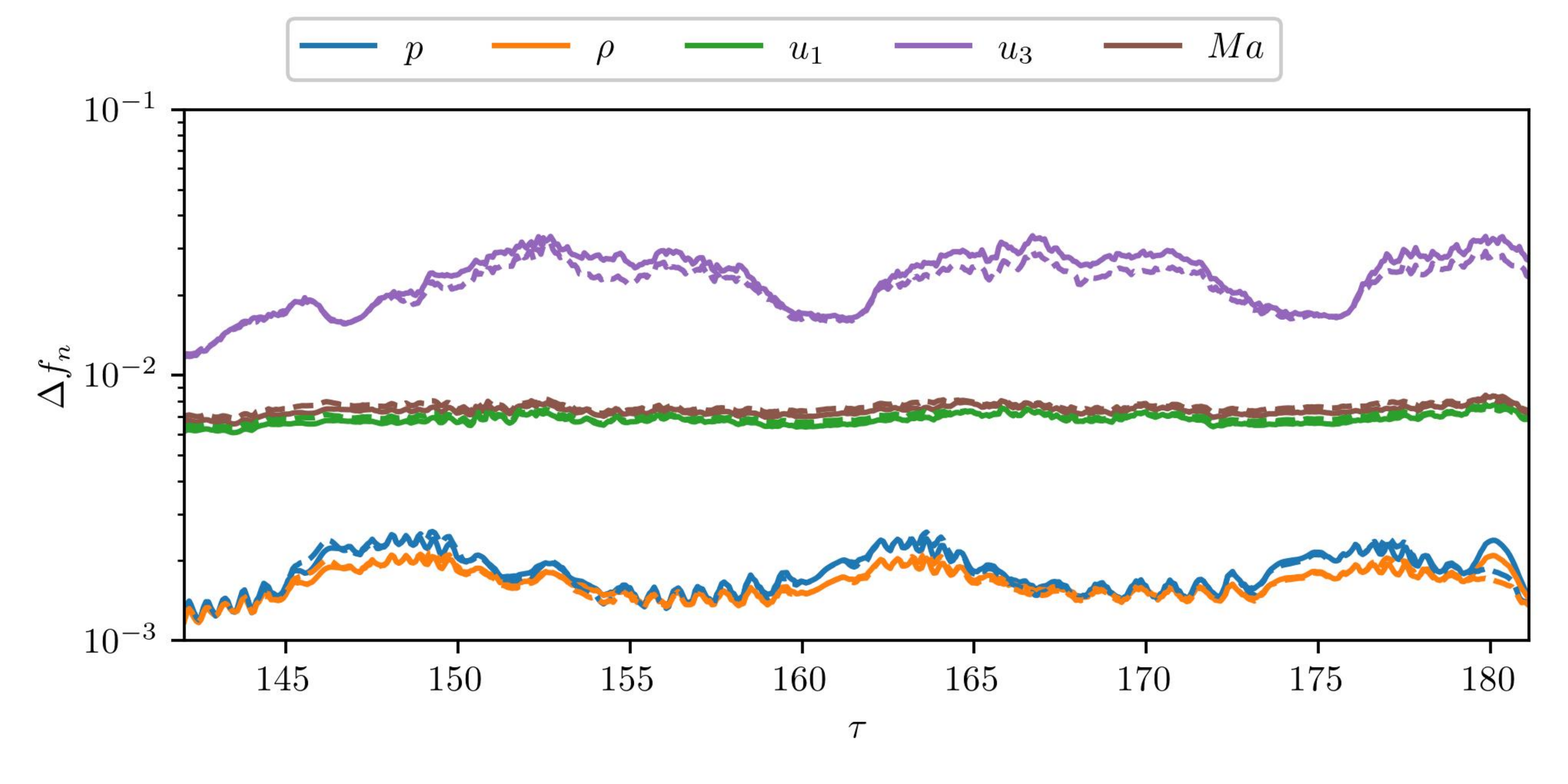}
		\caption{Absolute temporal error $\Delta f_n~=~||\mathbf{f}^\ast_n - \mathbf{f}_n||_2 / ||\mathbf{f}_n||_2$, weighted with the square-root of the cell volume and scaled with the free stream Mach number $Ma_{\infty}$. The solid line marks $\boldsymbol{\mathcal{M}}_2$ while the dashed line belongs to $\boldsymbol{\mathcal{M}}_3$.}
		\label{fig:temporal_error_oat}
	\end{center}
\end{figure}
To quantify the information loss of the different grids, we compare the absolute temporal error defined as \linebreak$\Delta f_n=||\mathbf{f}_n^\ast~-~\mathbf{f}_n||_2 / ||\mathbf{f}_n||_2$ and shown in fig.~(\ref{fig:temporal_error_oat}).
The differences between the errors for $\boldsymbol{\mathcal{M}}_2$ (solid line) and $\boldsymbol{\mathcal{M}}_3$ (dashed line) remain small over the course of the simulation.
The approximated pressure and density exhibit the lowest errors, $O(10^{-3})$. The error in the local Mach number and the streamwise velocity component is approximately six times larger.
For the velocity component $u_3$, the error is consistently higher than for all other fields, $O(10^{-2})$.
In summary, the error remains at similar levels over time and below $3\%$ for all quantities.
Note that the error for $u_2$ is omitted here because the flow is statistically 2D.

The behavior of the temporal error can be explained by looking at the metric field $\boldsymbol{\mathcal{M}}_2$ and $\boldsymbol{\mathcal{M}}_3$.
The pressure as well as density are scalar fields, while the velocity vector is only considered via its norm.
The velocity component $u_3$ is significantly smaller than $u_1$.
Therefore, small interpolation errors yield larger relative errors, as shown in fig. (\ref{fig:temporal_error_oat}).
Even though the local Mach number field was not part of the metric field, the overall interpolation error is similar to all other fields.

We also investigated the absolute spatial error of all fields as done for $\boldsymbol{\mathcal{M}}_1$.
The trends of spatial and temporal errors are similar, which is why they are omitted here.
However, appendix, fig. (\ref{fig:spatial_error_mu_only}) depicts the spatial errors of the local Mach number and the velocity component $u_3$, which constitute the worst-case scenario for the presented test case.
Both the normalized mean error and the standard deviation remain in the order of $O(10^{-5})$, again with their maximum values located in the wake region aft of the rear airfoil. 
It should also be noted that all errors shown here include the interpolation error made by $S^3$, as well as the interpolation error made when interpolating these fields back to the original grid, which is required to compute the spatial error.

Finally, we report the absolute global error for all three metric fields tested in Table (\ref{table:global_error}).
The global error is defined as $\Delta f~=~\sqrt{\sum_n ||\mathbf{f}^\ast_n - \mathbf{f}_n||_2^2/\sum_n ||\mathbf{f}_n||_2^2}$.
For $\boldsymbol{\mathcal{M}}_1$, all quantities except for the Mach number field were not part of the metric definition.
Since the grid created with $\boldsymbol{\mathcal{M}}_1$ has significantly fewer cells than the ones created with $\boldsymbol{\mathcal{M}}_2$ and $\boldsymbol{\mathcal{M}}_3$, the global interpolation error is consistently higher across all fields.
All errors are in the range $O(10^{-4}) - O(10^{-2})$, mostly caused by interpolation errors near the airfoil geometries and in the wake of the rear airfoil.

\begin{table}[htb]
	\begin{center}
		\caption{Absolute global error for all metrics. All cases use $\mathcal{M}_{\min}~=~0.75$ as a stopping criterion. All values are scaled by $10^{-2}$.}
		\begin{tabular}{@{}lccccc@{}}
			\toprule
			& \multicolumn{5}{c}{$\times 10^{-2}$} \\
			\cmidrule(l){2-6}
			& $Ma$ & $p$ & $\rho$ & $u_1$ & $u_3$ \\
			\midrule
			$\mathcal{M}_1$ & $1.36$ & $0.120$ & $0.183$ & $1.290$ & $1.926$ \\ 
			$\mathcal{M}_2$ & $0.577$ & $0.038$ & $0.066$ & $0.543$ & $0.931$ \\ 
			$\mathcal{M}_3$ & $0.623$ & $0.037$ & $0.070$ & $0.579$ & $0.976$\\ 
			\bottomrule
		\end{tabular}
		\label{table:global_error}
	\end{center}
\end{table}

We conclude that the resulting grid is robust against the choice of the metric.
However, a metric tailored to a specific task (or group of tasks) will yield higher reduction rates.
The choice of the metric is therefore a compromise between generalization and compression ratio.

\subsection{Flow past a circular cylinder}\label{subsec:cylinder}
A direct numerical simulation (DNS) of the flow past a circular cylinder, in the following legends abbreviated as \emph{cylinder}, was chosen to demonstrate the data reduction abilities of $S^3$ on a more challenging 3D test case.
The simulation setup is based on the work of Lehmkuhl et al. \cite{lehmkuhl_low-frequency_2013}.
The Reynolds number based on the cylinder diameter $d=0.1\, m$, the uniform inflow velocity $U_{\infty}~=~39 \, m/s$, and the kinematic viscosity $\nu~=~10^{-3}\, m^2/s$ is $Re~=~3900$.
The extent of the numerical domain along the streamwise, lateral, and axial directions is $x / d \in [0, 24]$, $y / d \in [0, 20]$, and $z / d \in [0, \pi]$, respectively.
The cylinder is placed at $x/d~=~8$ and $y/d~=~10$.
The block-structured mesh consists of $9.416 \cdot 10^6$ hexahedral cells.
The incompressible simulation is executed using \emph{OpenFOAM-v2206} \cite{esi_group_openfoam_2206}. 
For more information on the numerical setup and the flow physics, we refer to \cite{lehmkuhl_low-frequency_2013}.
The numerical setup is publicly available on GitHub \cite{weiner_common_nodate}.

The time interval between two subsequent snapshots is set to $\Delta t~=~2.5~\cdot~10^{-4}\,s$, corresponding to $\Delta \tau~=~\Delta t U_{\infty} / d~=~0.0975$ convective time units (CTU), and a total number of $500$ snapshots in the range $ 75\leq \tau \le 124$ is collected, corresponding to roughly $50$ CTU.
The selected time interval only extends over $11$ vortex shedding cycles.
This limitation stems from the available memory on a single node of our HPC cluster when computing the SVD of the original CFD data, which is needed for validation.
We define the state vector using the instantaneous velocity fluctuations in all three directions.
The full data matrix occupies $120$ GB on the hard drive when stored in double-precision binary format.

For this test case, we use the TKE as the metric field, which is defined as
\begin{center}
	\begin{equation}
		\mathbf{k}~=~\frac{1}{2} \left(\overline{(\mathbf{u}_x^\prime)^2} + \overline{(\mathbf{u}_y^\prime)^2} + \overline{(\mathbf{u}_z^\prime)^2}\right),\quad \overline{(\mathbf{u}_i^\prime)^2}~=~\overline{
        \left(\mathbf{u}_i - \overline{\mathbf{u}}_i\right)^2
        },
	\end{equation}
\end{center}
where the overbar denotes temporal averaging and $\mathbf{u}_i$ is the $i$th component of the velocity vector field.
This choice is based on the goal to accurately capture the wake dynamics, which is a key property of the presented flow configuration.

\begin{figure}[htbp]
	\begin{center}
		\subfloat[original grid]{\includegraphics[width=0.42\textwidth,trim={0 7cm 0 7cm},clip]{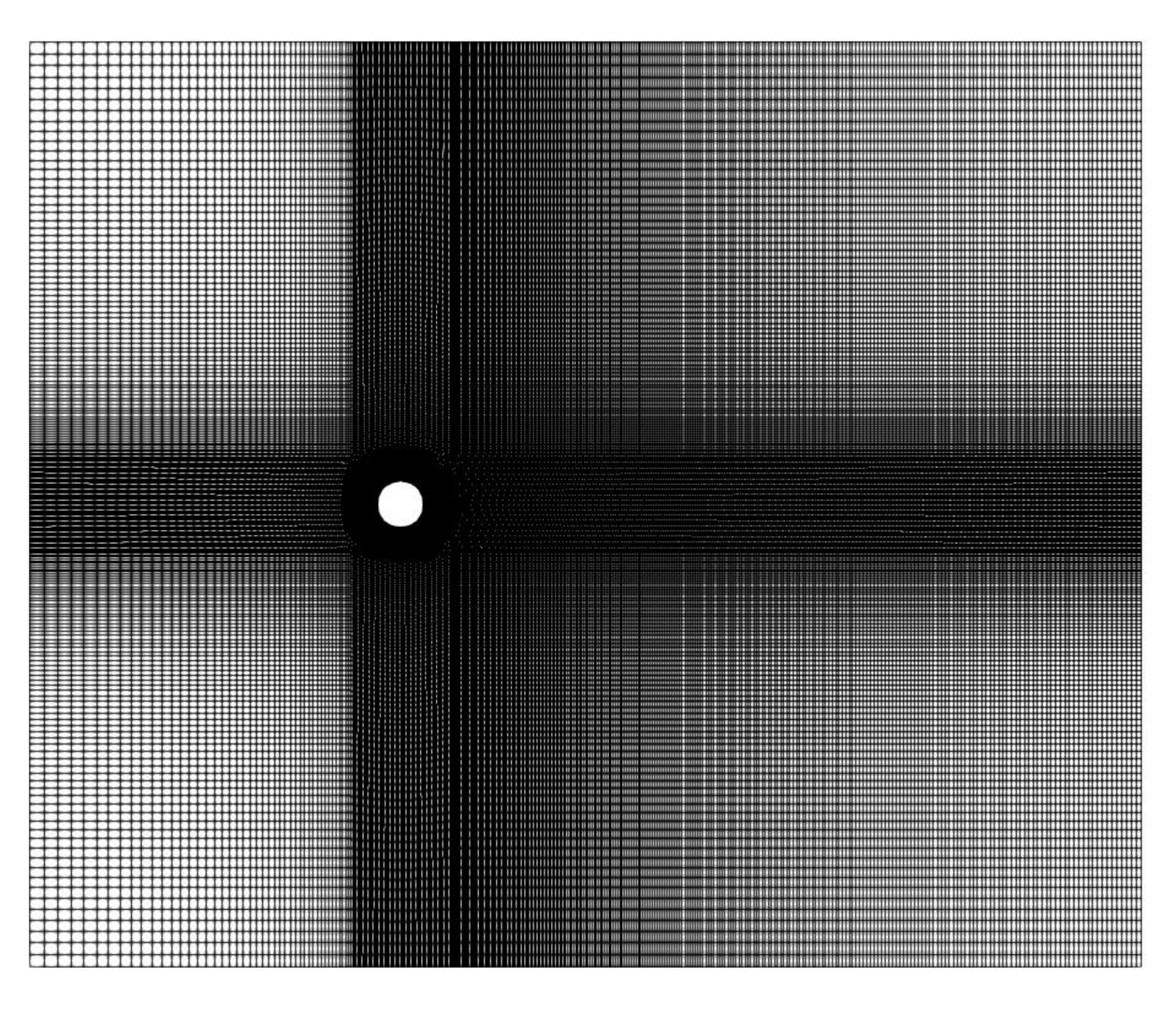}\label{fig:original_grid_cylinder_xy}}
		\subfloat[original metric field $\boldsymbol{\mathcal{M}}$]{%
			\raisebox{0pt}[0pt][0pt]{%
				\begin{tikzpicture}[baseline]
					\node[anchor=south west, inner sep=0] (image) at (0,0)
					{\includegraphics[width=0.42\textwidth,trim={0 7cm 0 7cm},clip]{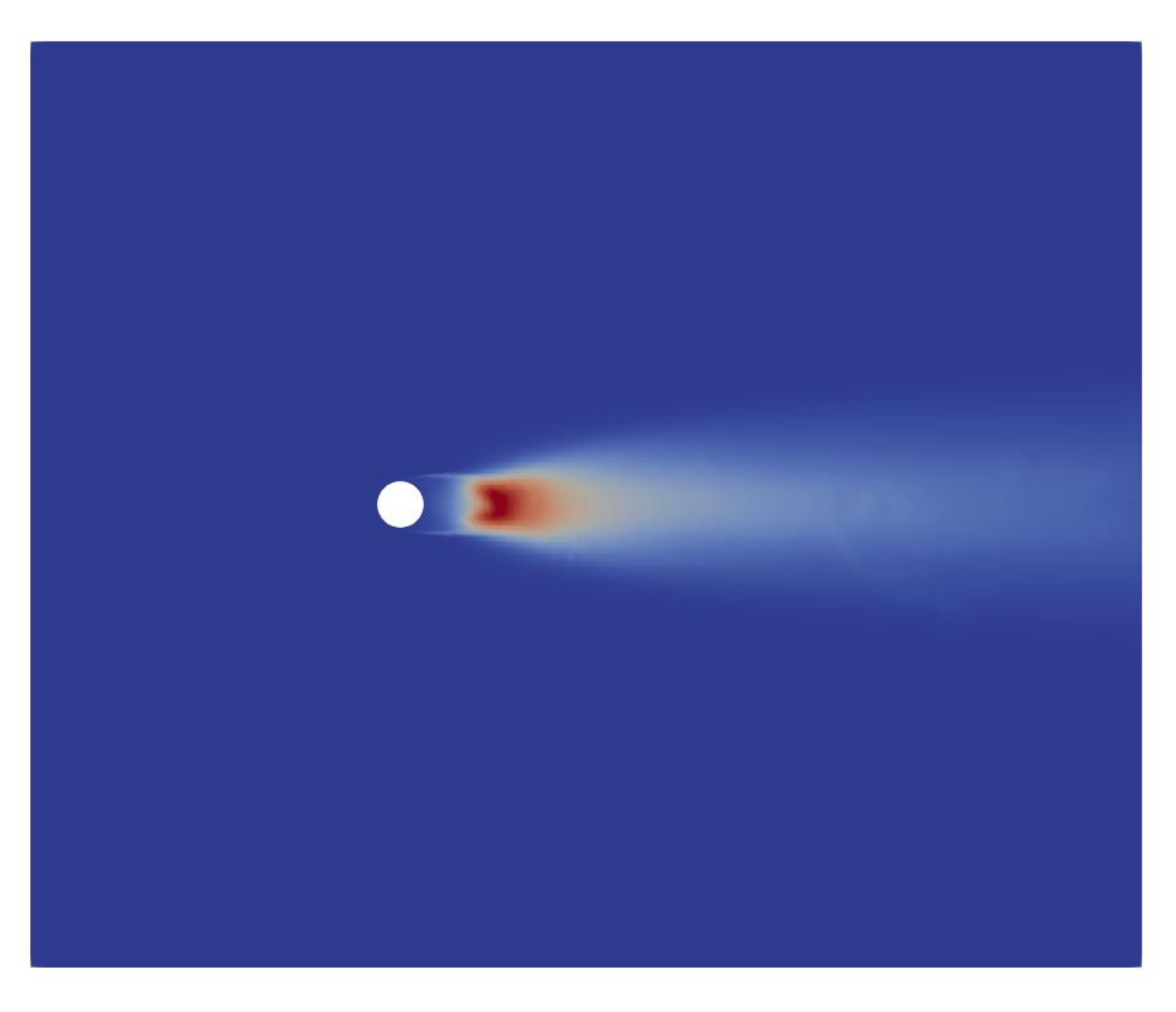}};
					\begin{scope}[x={(image.south east)}, y={(image.north west)}]
						\draw[-{Stealth}, thick, black] (0., -0.05) -- ++(0.1, 0) node[below=2pt, midway] {\small $x$};
						\draw[-{Stealth}, thick, black] (0.0, -0.05) -- ++(0, 0.2) node[above] {\small $y$};
					\end{scope}
				\end{tikzpicture}%
			}
			\label{fig:original_metric_cylinder_xy}}
		\\
		\subfloat[grid generated by $S^3$]{\includegraphics[width=0.42\textwidth,trim={0 7cm 0 7cm},clip]{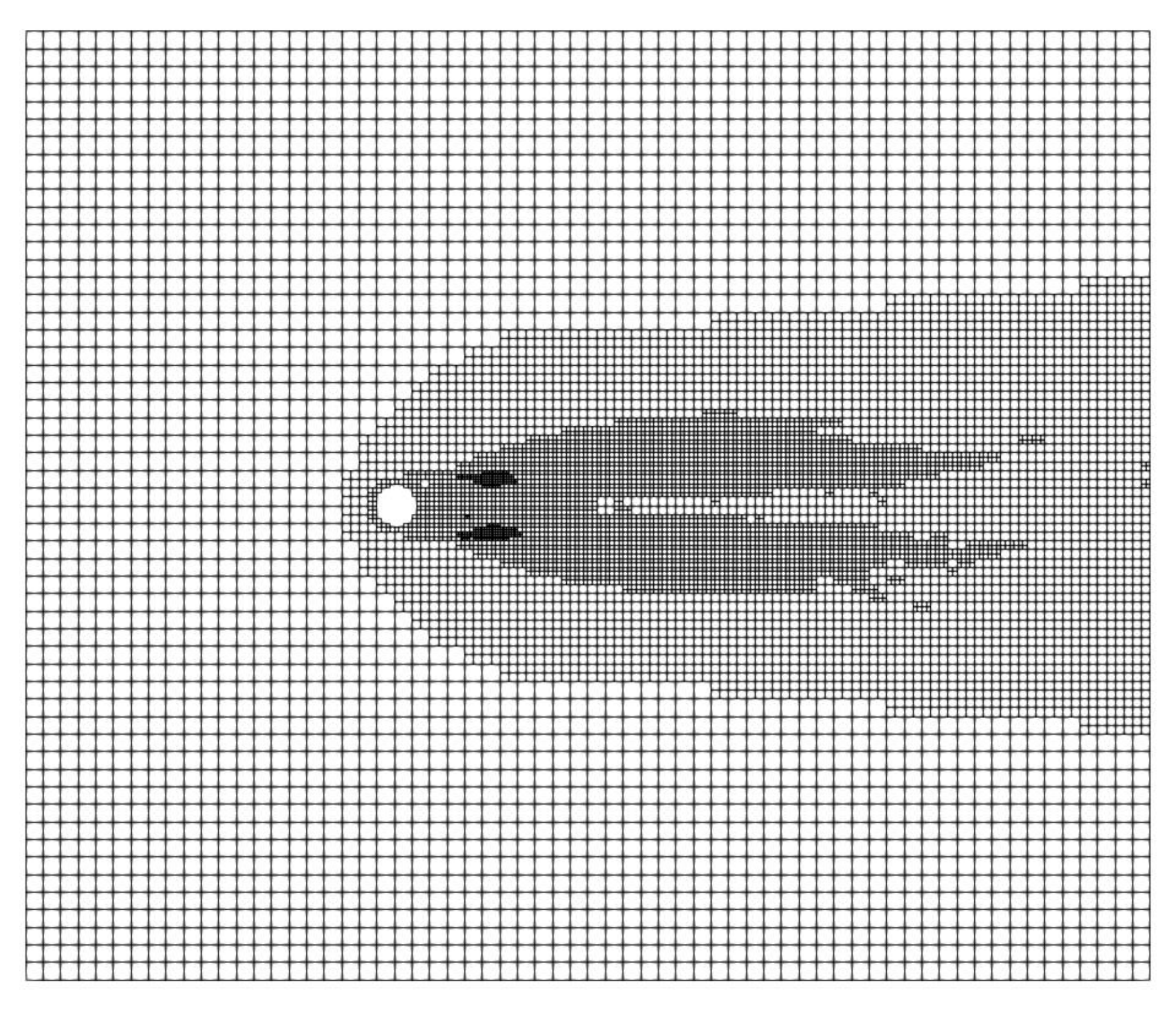}\label{fig:interpolated_grid_cylinder_xy}}
		\hspace*{1.25mm}
		\subfloat[interpolated metric field $\widehat{\boldsymbol{\mathcal{M}}}$]{\includegraphics[width=0.42\textwidth,trim={0 7cm 0 7cm},clip]{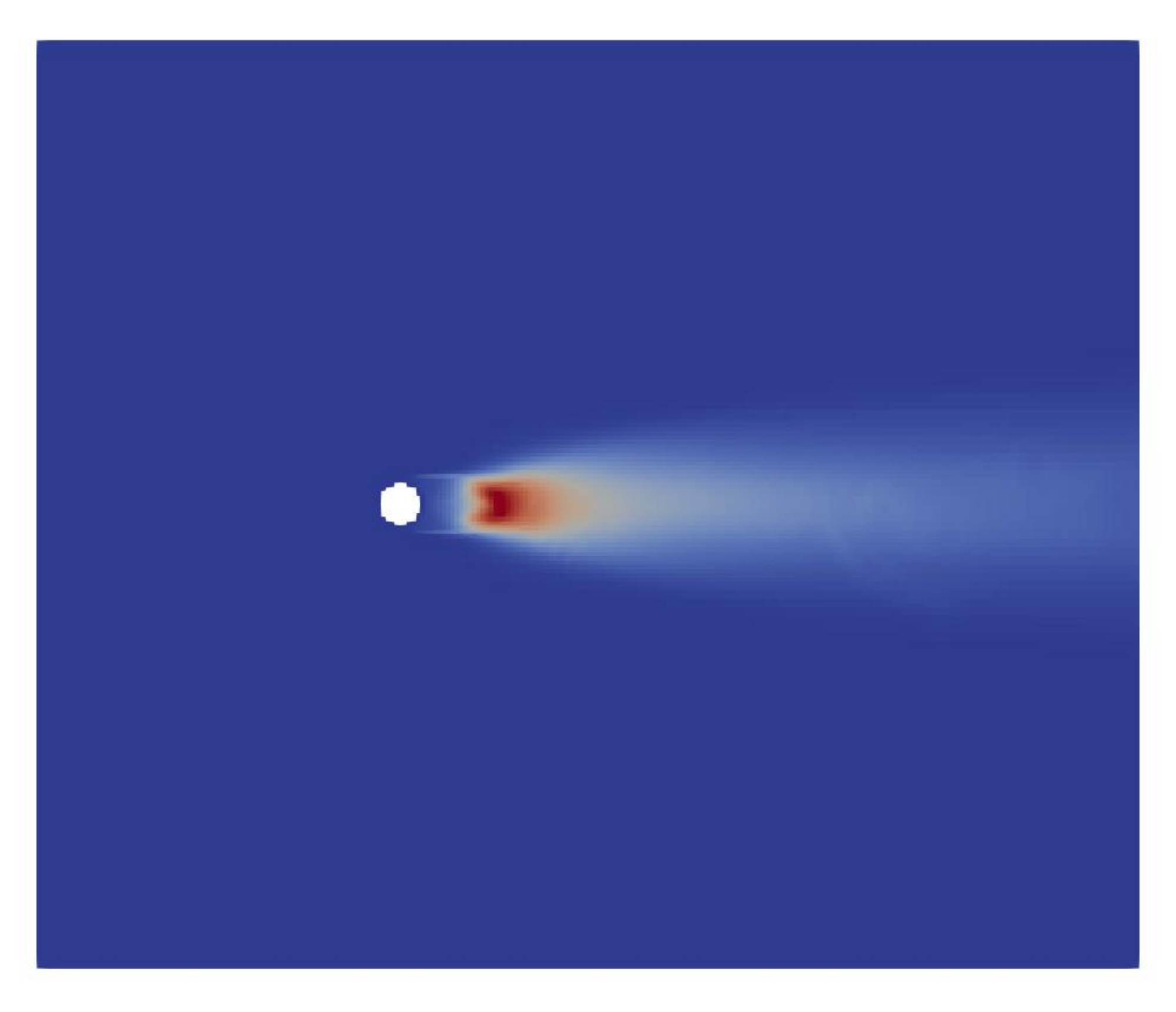}\label{fig:interpolated_metric_cylinder_xy}}
	\end{center}
	\caption{The left column shows a comparison of the original grid (\ref{fig:original_grid_cylinder_xy}) with the grid generated by $S^3$ for a captured metric of $\mathcal{M}_{\mathrm{approx}}~=~0.27$ (\ref{fig:interpolated_grid_cylinder_xy}). The right column depicts the metric on the original grid (\ref{fig:original_metric_cylinder_xy}), and the metric interpolated onto the grid created by $S^3$ (\ref{fig:interpolated_metric_cylinder_xy}). All figures depict the $x$-$y$-plane at $z / d~=~\pi/2$. Both contour plots are scaled from zero to $||\boldsymbol{\mathcal{M}}||_\infty$.}
\end{figure}

\begin{figure}[htbp]
	\begin{center}
		\subfloat[original grid]{\includegraphics[width=0.42\textwidth]{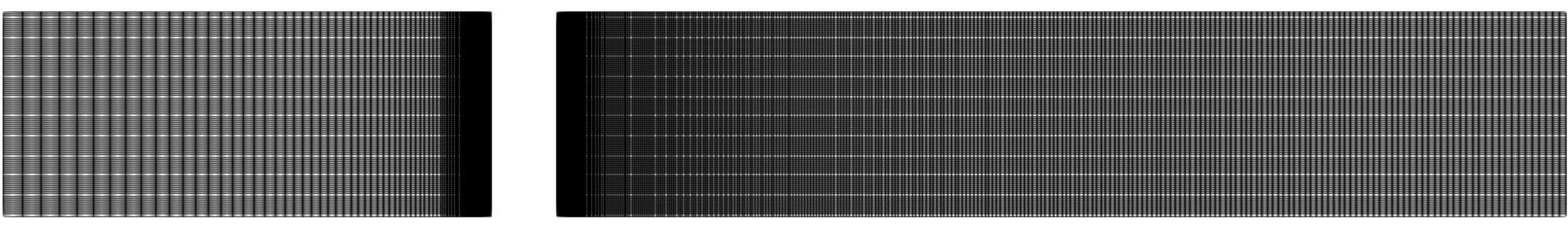}\label{fig:original_grid_cylinder_xz}}
		\hspace*{0.5cm}
		\subfloat[original metric field $\boldsymbol{\mathcal{M}}$]{%
			\raisebox{0pt}[0pt][0pt]{%
				\begin{tikzpicture}[baseline]
					\node[anchor=south west, inner sep=0] (image) at (0,0)
					{\includegraphics[width=0.42\textwidth]{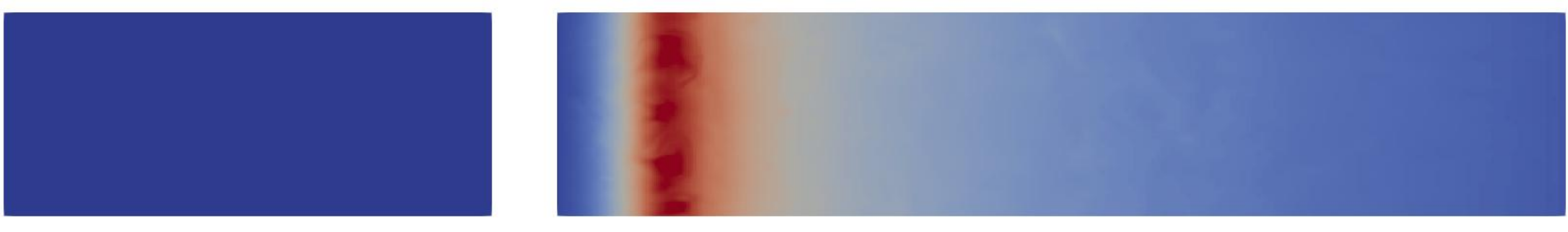}};
					\begin{scope}[x={(image.south east)}, y={(image.north west)}]
						\draw[-{Stealth}, thick, black] (-0.02, -0.1) -- ++(0.1, 0) node[below=2pt, midway] {\small $x$};
						\draw[-{Stealth}, thick, black] (-0.02, -0.1) -- ++(0, 0.6) node[above] {\small $z$};
					\end{scope}
				\end{tikzpicture}%
			}
			\label{fig:original_metric_cylinder_xz}}
		\\
		\subfloat[grid generated by $S^3$]{\includegraphics[width=0.435\textwidth]{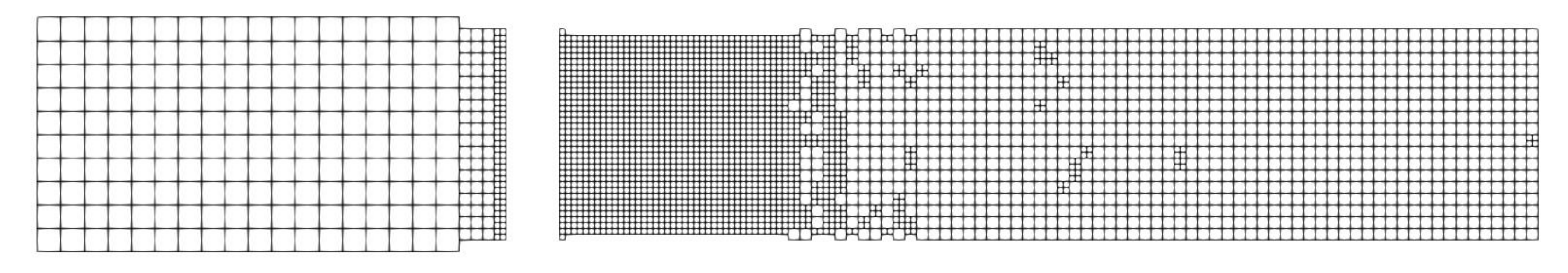}\label{fig:interpolated_grid_cylinder_xz}}
		\hspace*{0.65cm}
		\subfloat[interpolated metric field $\widehat{\boldsymbol{\mathcal{M}}}$]{\includegraphics[width=0.435\textwidth]{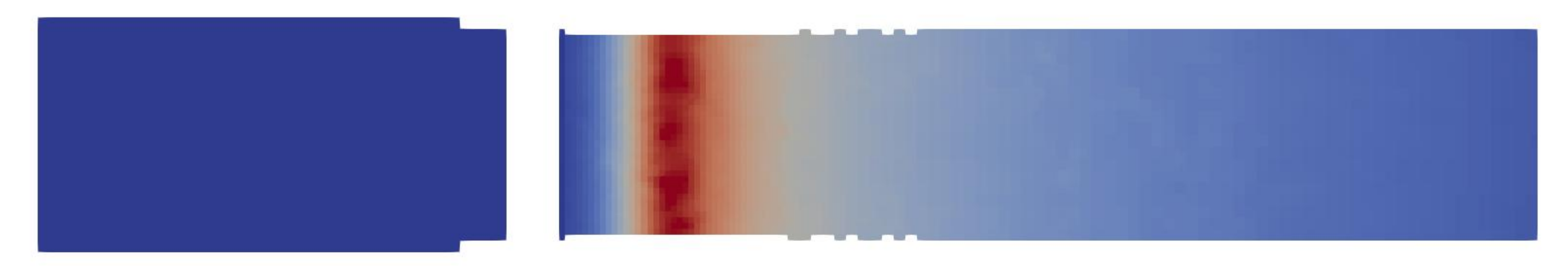}\label{fig:interpolated_metric_cylinder_xz}}
	\end{center}
	\caption{The left column shows a comparison of the original grid (\ref{fig:original_grid_cylinder_xz}) with the grid generated by $S^3$ for a captured metric of $\mathcal{M}_{\mathrm{approx}}~=~0.27$ (\ref{fig:interpolated_grid_cylinder_xz}). The right column depicts the metric on the original grid (\ref{fig:original_grid_cylinder_xz}), and the metric interpolated onto the grid created by $S^3$ (\ref{fig:interpolated_metric_cylinder_xz}). All figures depict the $x$-$z$-plane at $y / d~=~8$. Both contour plots are scaled from zero to $||\boldsymbol{\mathcal{M}}||_\infty$.}
\end{figure}

Fig.~(\ref{fig:original_grid_cylinder_xy}) shows the original grid in the $x$-$y$-plane at $z/d~=~\pi/2$ along with $\boldsymbol{\mathcal{M}}$ in fig.~(\ref{fig:original_metric_cylinder_xy}).
It is clearly visible that the major part of the TKE is located in the wake of the cylinder, while its magnitude decreases in a streamwise direction.
The generated coarse grid consists of $2.188 \cdot 10^5$ cells, corresponding to a reduction rate of $97.68\%$ when using $\mathcal{M}_{\mathrm{min}}=0.25$ as the stopping criterion ($\mathcal{M}_{\mathrm{approx}}~=~0.27$).
A slice through the resulting grid is depicted in fig.~(\ref{fig:interpolated_grid_cylinder_xy}).
Clearly, $S^3$ strongly refines the near wake regions where gradients of the target metric are high, while creating a coarse grid in regions with low TKE.
Regions with uniformly high metric values are not excessively refined, e.g., at a distance of one to three diameters downstream of the cylinder.
As expected, the new grid is roughly symmetric with respect to the horizontal dividing line through the cylinder.
Small asymmetries may occur because of the limited accuracy of the sampled TKE and the empirically chosen number of leave cells marked for refinement in each iteration.
$S^3$ is able to capture the TKE without a qualitative deterioration of its approximation, as shown in fig.~(\ref{fig:interpolated_metric_cylinder_xy}).
Fig.~(\ref{fig:interpolated_grid_cylinder_xz}) shows a second $x$-$z$-slice at $y/d=8$.
The grid is approximately homogeneous in $z$-direction, and the TKE is captured well on the coarse grid.
It is also visible that the domain boundaries in the axial direction are not captured uniformly well due to differences in the refinement level.
This behavior could be avoided by an additional geometry refinement step for the outer boundary, which we did include here.
The same applies to the cylinder boundary, which is only captured roughly.
Comparable results for $\mathcal{M}_{\mathrm{min}}=0.75$ are available in section (\ref{sec:appendix:cylinder}).


\begin{figure}[htbp]
	\begin{center}
		\subfloat[$x$-component]{\includegraphics[width=0.45\textwidth]{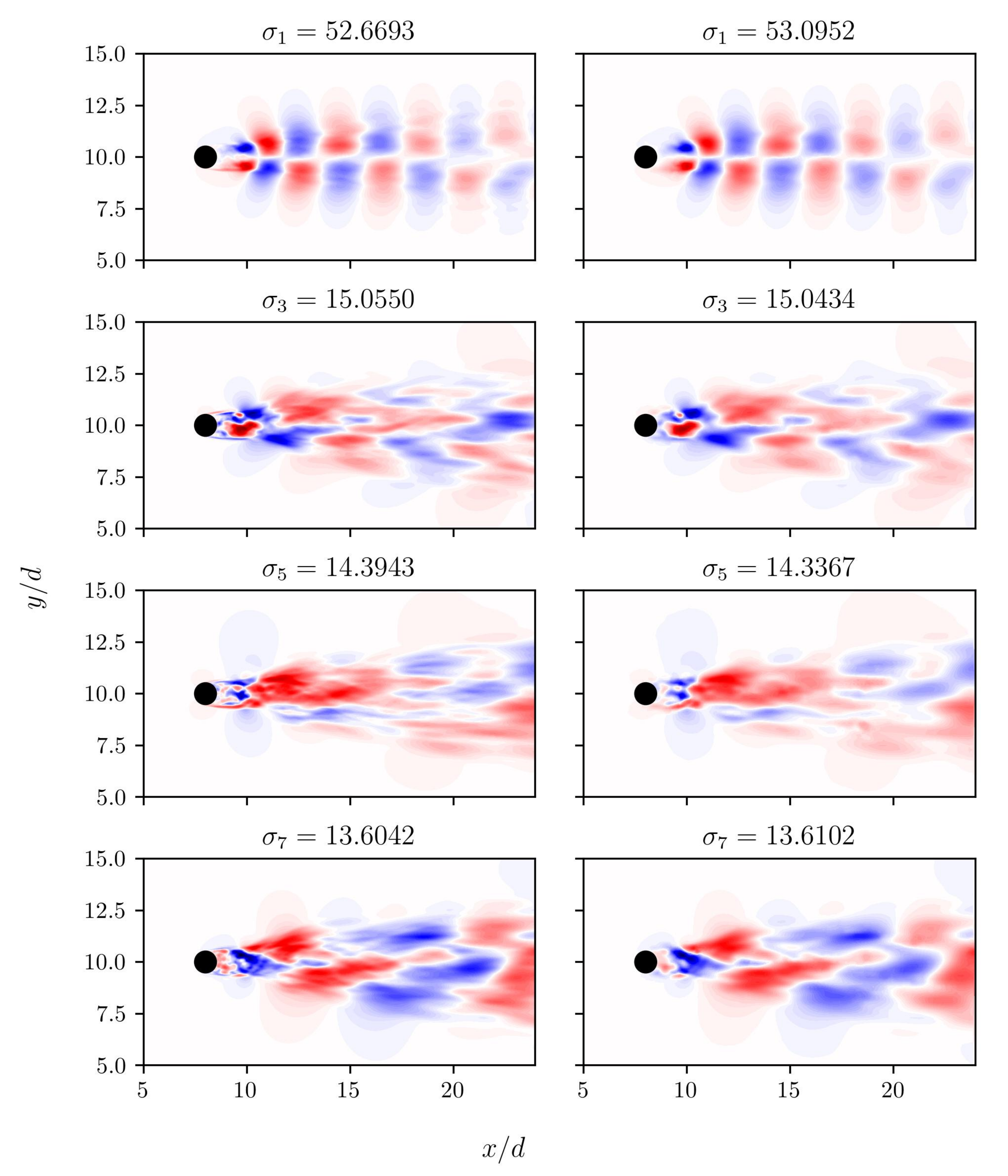}}
		\hspace*{5mm}
		\subfloat[$y$-component]{\includegraphics[width=0.45\textwidth]{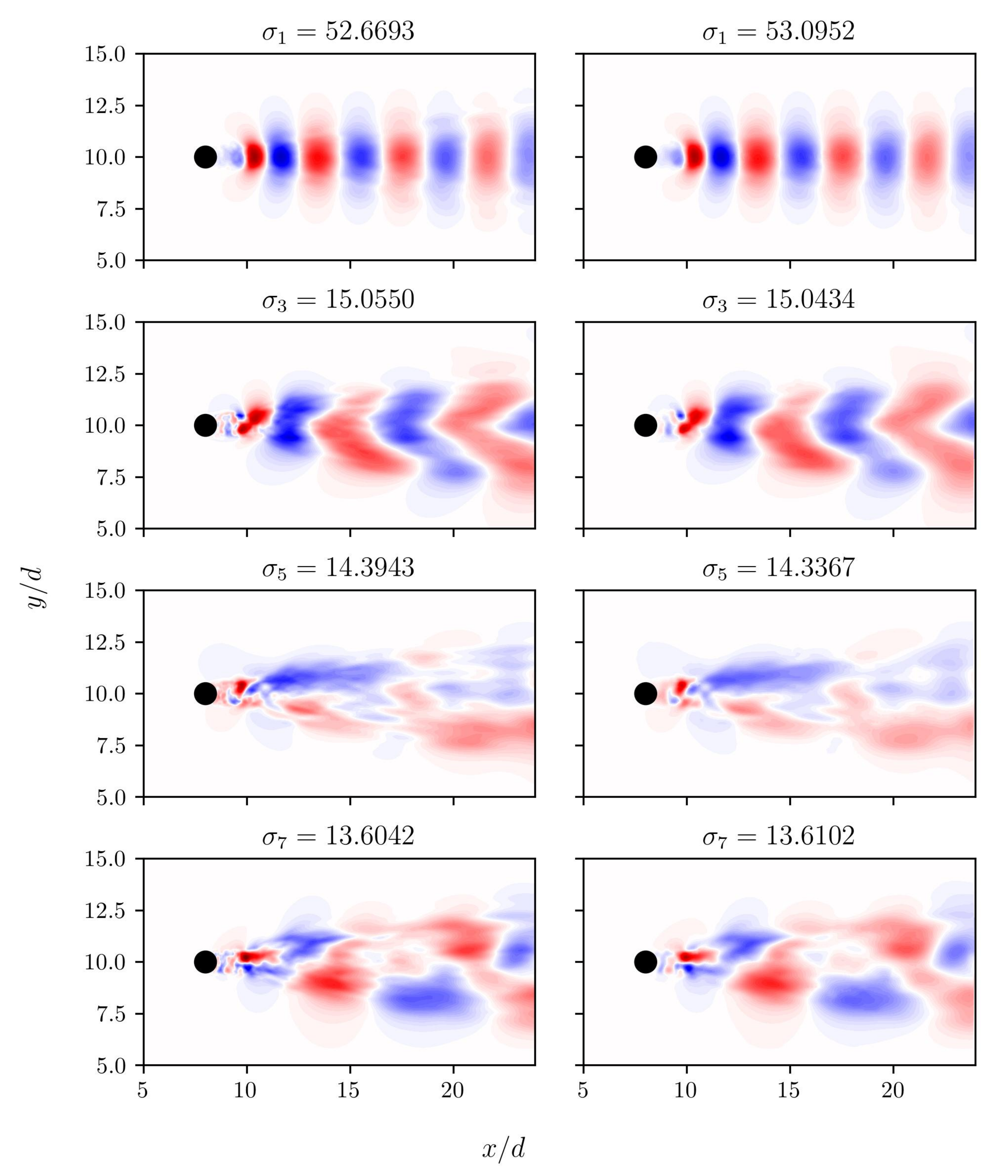}}
	\end{center}
	\caption{Comparison of the first four uneven POD modes and associated singular values for the original data (left columns) and the $S^3$-interpolated data with $\mathcal{M}_{\mathrm{approx}}~=~0.27$  (right columns) in the $x$-$y$-plane at $z/d~=~\pi / 2$ for the cylinder test case. The colorscale is identical for all contours and bounded by $\pm||\mathbf{U}||_\infty$.}
	\label{fig:comparison_pod_mode_cylinder_xy}
\end{figure}
For a quantitative assessment of the reduced data, we build the data matrix using the velocity fluctuations in all three directions and compute the SVD.
Fig.~(\ref{fig:comparison_pod_mode_cylinder_xy}) shows the first four uneven left-singular vectors and the associated singular values.
The modes' $z$-component is omitted for compactness, but the agreement is equally good as for the other two components.
The difference in the singular values is slightly larger than for the tandem configuration, but the deviation for the depicted modes remains below $1\%$.

\begin{figure}[htbp]
	\begin{center}
		\includegraphics[width=0.8\textwidth]{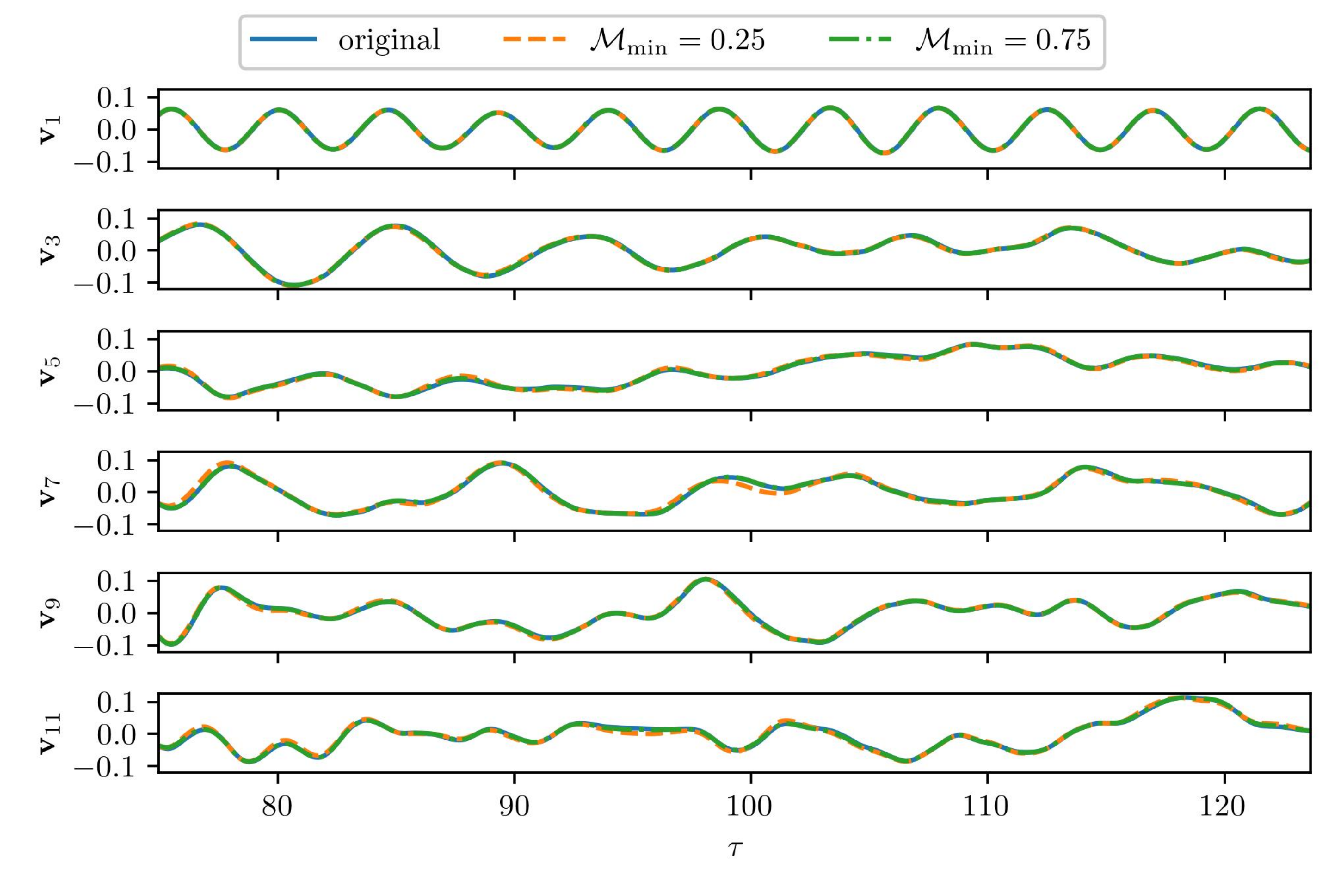}
		\caption{Comparison of the first four uneven right-singular vectors for the cylinder test case.}
		\label{fig:mode_coefficients_cylinder}
	\end{center}
\end{figure}

The good agreement between the original and reduced data is also reflected in the right-singular vectors in fig.~(\ref{fig:mode_coefficients_cylinder}).
While minor deviations are visible for $\mathcal{M}_\mathrm{min}=0.25$, the results are visually indistinguishable when capturing roughly two-thirds of the metric.
The higher threshold value still leads to an $85\%$ reduction in the cell count.
The influence of the threshold value on the cell count and the associated computational cost is discussed in section (\ref{subsec:timings}).
Overall, it is fair to say that the same conclusions can be drawn from the SVDs of the original and reduced data.
\newpage

\subsection{Flow around an aircraft}\label{subsec:xrf1}

We finally consider the flow around an aircraft half-model to demonstrate the potential of $S^3$ to efficiently reduce extremely large amounts of data.
A delayed detached eddy simulation (DDES) of the XRF-1 wind tunnel model was carried out by Spinner et al. \cite{spinner_modal_2026} at $Re~=~3.3\cdot 10^6, Ma_\infty~=~0.84$, and an angle of attack of $\alpha~=~-4^\circ$.
The simulation was executed using the DLR TAU code with an SSG/LRR ln-$\omega$ turbulence model acting as background RANS model in the hybrid RANS-LES simulation.
A detailed description of the numerical setup and methods is provided in \cite{spinner_modal_2026}.\\

A total number of $104$ snapshots are used for the analysis, representing approximately $49.5$ CTU with $
\mathrm{CTU}~=~\mathrm{MAC}/U_\infty$ (MAC - mean aerodynamic chord).   
The computational mesh consists of $M~=~1.65 \cdot 10^8$ cells, occupying $0.761$ GB of disk space per snapshot and field, and leading to a size of roughly $80$ GB for the data matrix based on the local Mach number.
The metric field is composed of the temporal standard deviation of the Mach number field and eddy viscosity, respectively, and is computed using all of the $104$ snapshots.
Both parts are weighted, leading to a final metric field of $\boldsymbol{\mathcal{M}}~=~0.6~\cdot~\mathrm{std}(\mathbf{Ma})~+~0.4~\cdot~\mathrm{std}(\boldsymbol{\nu}_t)$.
As for the tandem configuration, the Mach number variation ensures a high resolution in regions of shock motion.
The addition of the eddy viscosity to the metric field accounts for large separated, turbulent structures convected into the far field, where temporal changes of the local Mach number are small.
The weighting factor expresses a mild prioritization of shock over far-field unsteadiness.
We did not study the sensitivity of the resulting mesh to the weighting factor, but we assume its influence to be minor.
The region of interest for the analysis was reduced to $x/\mathrm{MAC} \in [-2.5445, 12.7226]$, $y/\mathrm{MAC} \in [0.0509, 5.089]$
and $z/\mathrm{MAC} \in [-2.5445, 2.5445]$ with $\mathrm{MAC}~=~0.1965 \, m$ containing a total number of $M~=~1.021 \cdot 10^8$ cells.
The nose of the aircraft is located at $x/\mathrm{MAC} \approx 0.9$, $y/\mathrm{MAC}~=~0$ and $z/\mathrm{MAC}~=~0$.

\begin{figure}[htbp]
	\begin{center}
    \subfloat[original grid in the $x$-$z$-plane]{\includegraphics[width=0.45\textwidth]{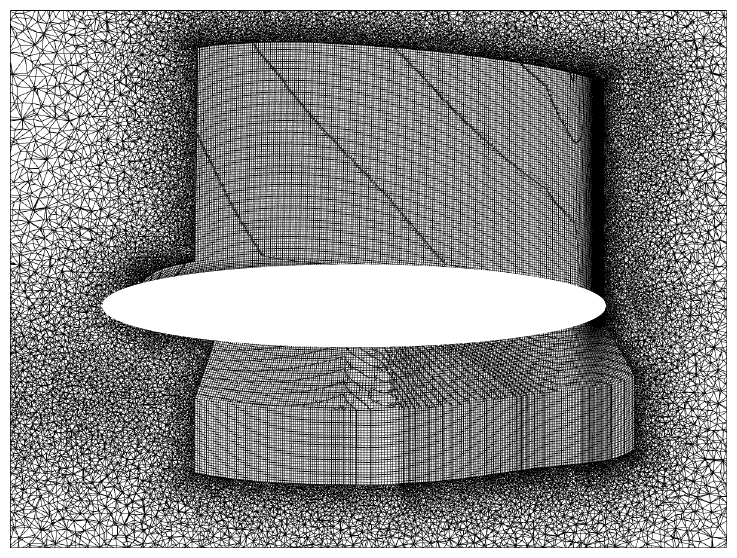}\label{fig:xrf1:grid_orig_x}}
		\hspace*{1cm}
		\subfloat[original grid in the $y$-$z$-plane]{\includegraphics[width=0.45\textwidth]{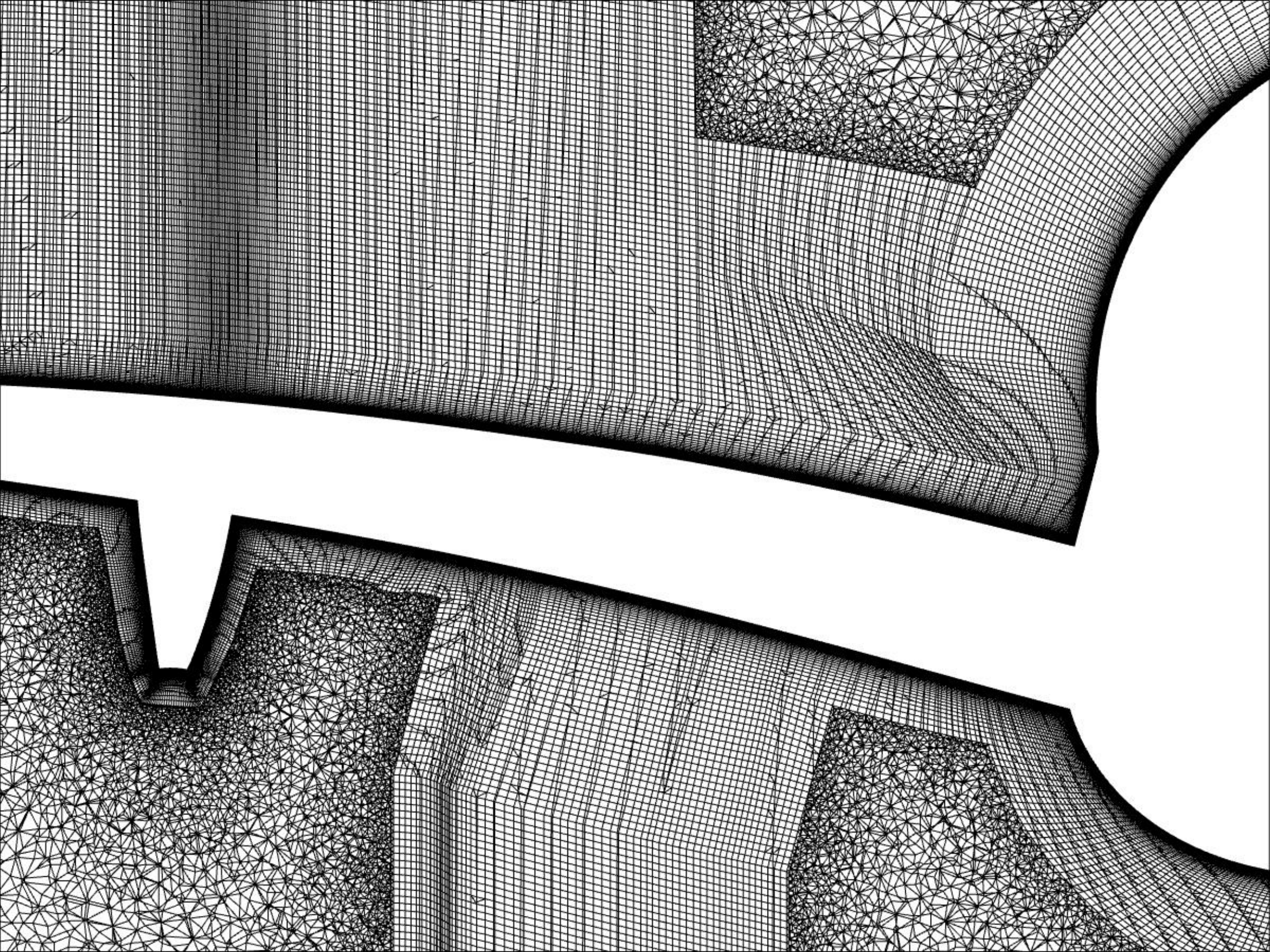}\label{fig:xrf1:grid_orig_y}}
	\end{center}
	\caption{Original grid in the $x$-$z$-plane at $x/\mathrm{MAC} \approx 4.9$ (\ref{fig:xrf1:grid_orig_x}) and in the $y$-$z$-plane at $y/\mathrm{MAC} \approx 0.9$ (\ref{fig:xrf1:grid_orig_y}). The airfoil geometry is proprietary to Airbus and therefore redacted in fig.~(\ref{fig:xrf1:grid_orig_x}).}
	\label{fig:xrf1_grid_orig}
\end{figure}
Fig.~(\ref{fig:xrf1:grid_orig_x}) shows an enlarged view of the original grid in the $x$-$z$-plane at $y/\mathrm{MAC}~=~0.9274$ while fig.~(\ref{fig:xrf1:grid_orig_y}) depicts the original grid in the $y$-$z$-plane at $x/\mathrm{MAC} \approx 4.9$.
Due to the enormous snapshot size, we computed the metric field incrementally using Welford's algorithm \cite{welford_note_1962}.

\begin{figure}[htbp]
	\begin{center}
        \subfloat[grid generated by $S^3$ in the $x$-$z$-plane]{\includegraphics[width=0.45\textwidth]{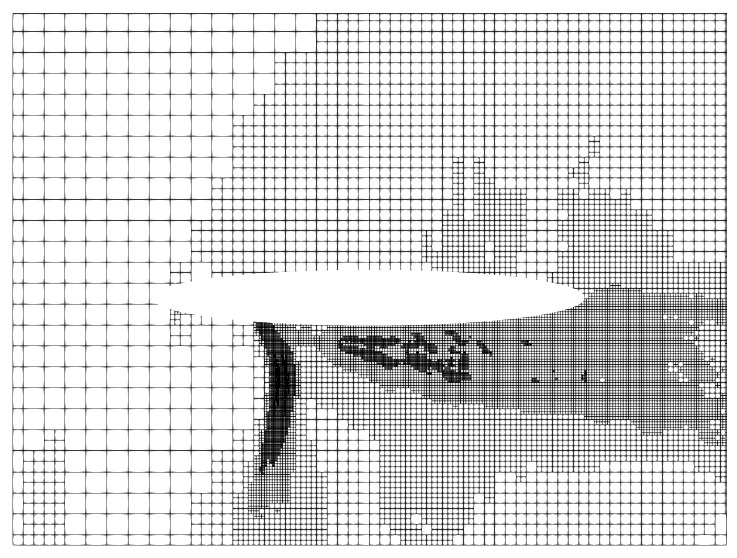}\label{fig:xrf1:grid_x}}        
		\hspace*{1cm}
		\subfloat[grid generated by $S^3$ in the $y$-$z$-plane]{\includegraphics[width=0.45\textwidth]{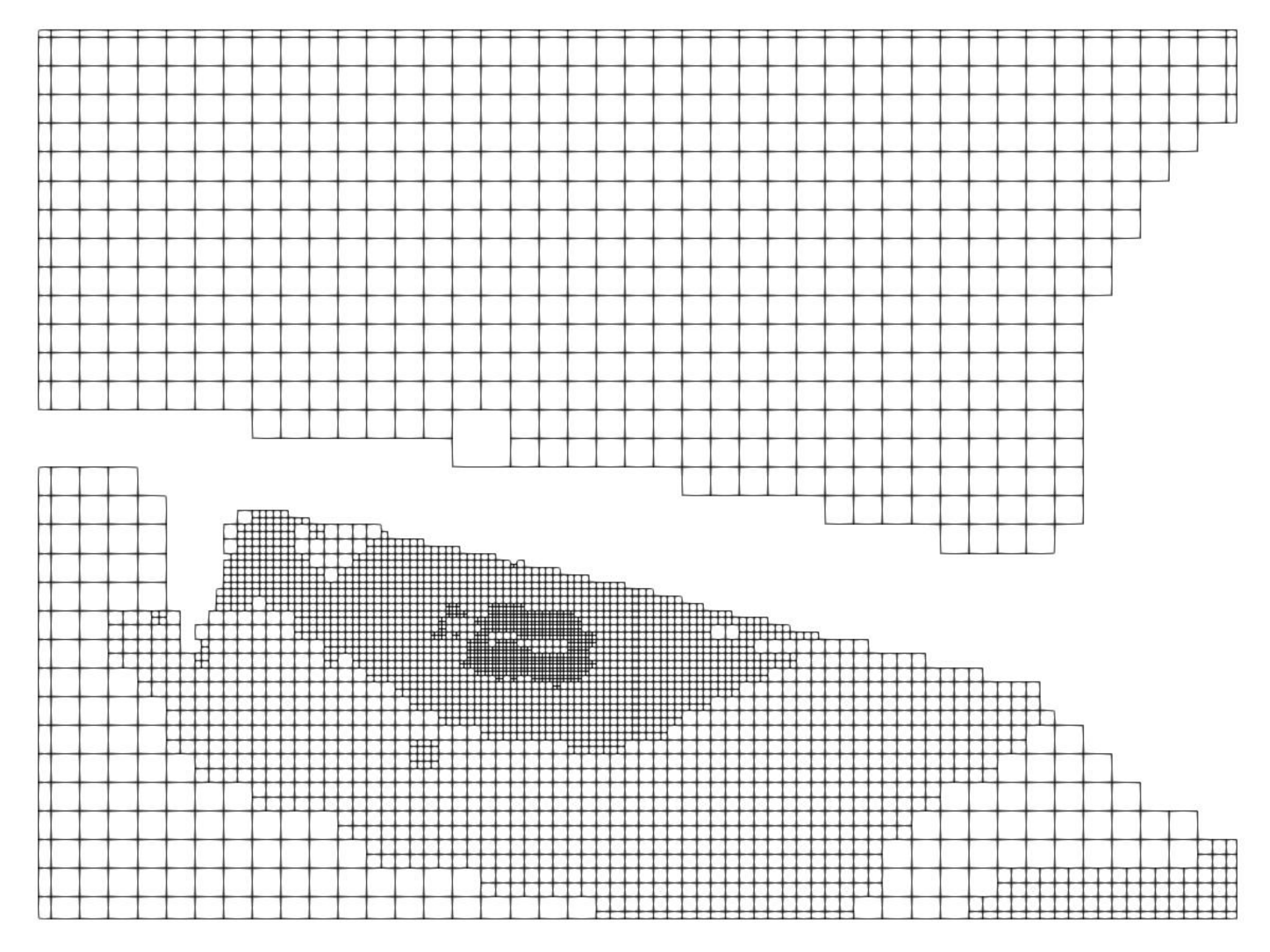}\label{fig:xrf1:grid_y}}
		\newline
      	\subfloat[interpolated metric field in the $x$-$z$-plane]{\includegraphics[width=0.45\textwidth]{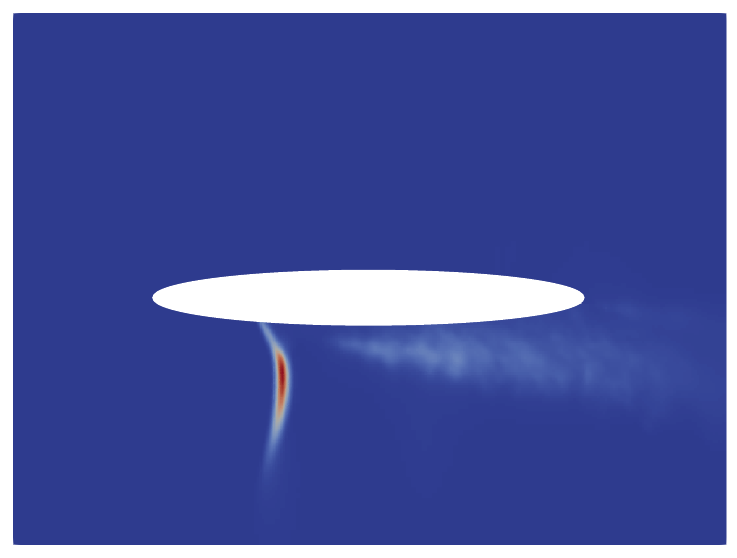}\label{fig:xrf1:metric_x}
        }
		\hspace*{1cm}
		\subfloat[interpolated metric field in the $y$-$z$-plane]{\includegraphics[width=0.45\textwidth]{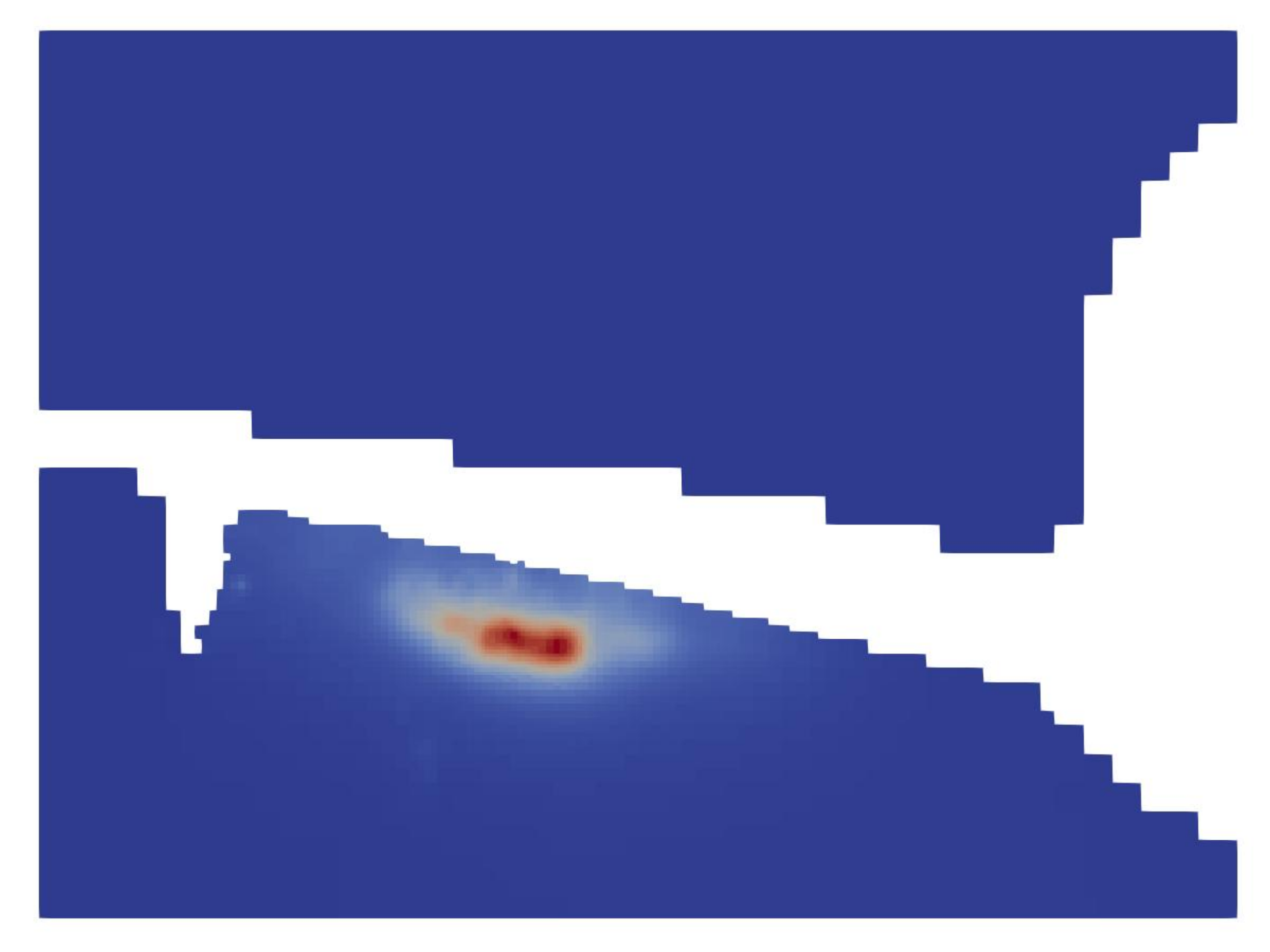}\label{fig:xrf1:metric_y}}
	\end{center}
	\caption{Grid generated by $S^3$ in the $x$-$z$-plane at $x/\mathrm{MAC} \approx 4.9$ (\ref{fig:xrf1:grid_x}) and the $y$-$z$-plane at $y/\mathrm{MAC} \approx 0.9$ (\ref{fig:xrf1:grid_y}). The lower row shows the interpolated metric field in the same planes. The airfoil geometry is proprietary to Airbus and therefore redacted in fig.~(\ref{fig:xrf1:grid_x}) and (\ref{fig:xrf1:metric_x}).}
	\label{fig:xrf1_grid_inter}
\end{figure}

We set the stopping criterion to a threshold value of $\mathcal{M}_{\min} = 0.95$.
The final grid has $N_{\ell}~=~5.167 \cdot 10^6$ cells, capturing $\mathcal{M}_{\mathrm{approx}}~=~1$ of the metric.
The cell count is reduced by $94.94\%$.\\
The resulting grid is shown in fig.~(\ref{fig:xrf1:grid_x}) and fig.~(\ref{fig:xrf1:grid_y}) in the same planes as for the original grid, along with the corresponding metric field in fig.~(\ref{fig:xrf1:metric_x}) and fig.~(\ref{fig:xrf1:metric_y}).
It is clearly visible that the grid is refined in regions where the spatial change of the metric is high, as presented in section (\ref{subsec:scube}).
The approximation of the geometry is rather poor compared to the previously shown test cases, since
no subsequent geometry refinement was employed to keep the number of cells to a minimum.
Since the first modes typically capture large, dominant flow structures, refining the grid near the aircraft model has no significant impact on the SVD results and was therefore omitted here.

\begin{figure}[htbp]
	\begin{center}
		\subfloat[original]{\includegraphics[width=0.45\textwidth]{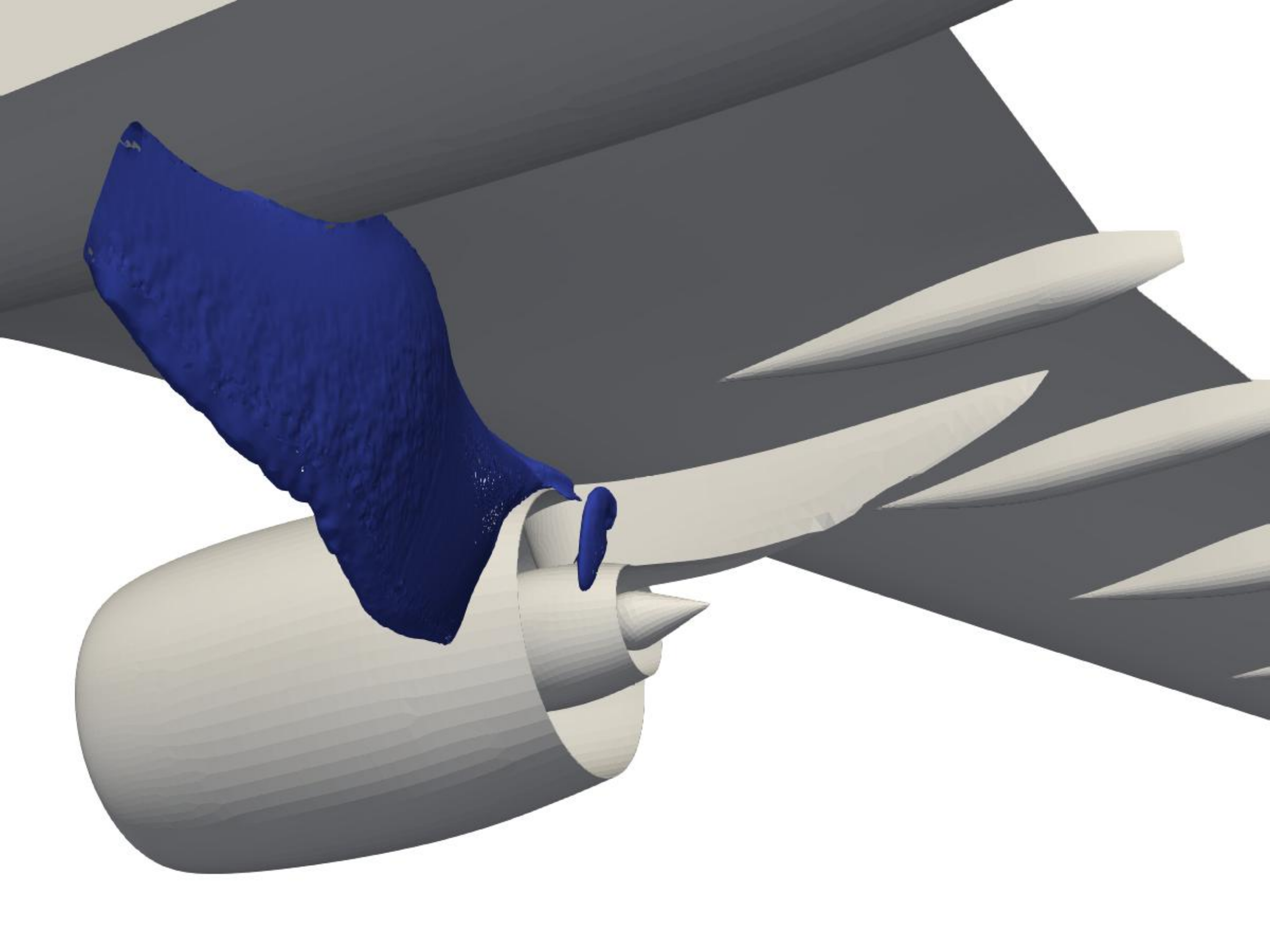}}
		\hspace*{1cm}
		\subfloat[$S^3$]{\includegraphics[width=0.45\textwidth]{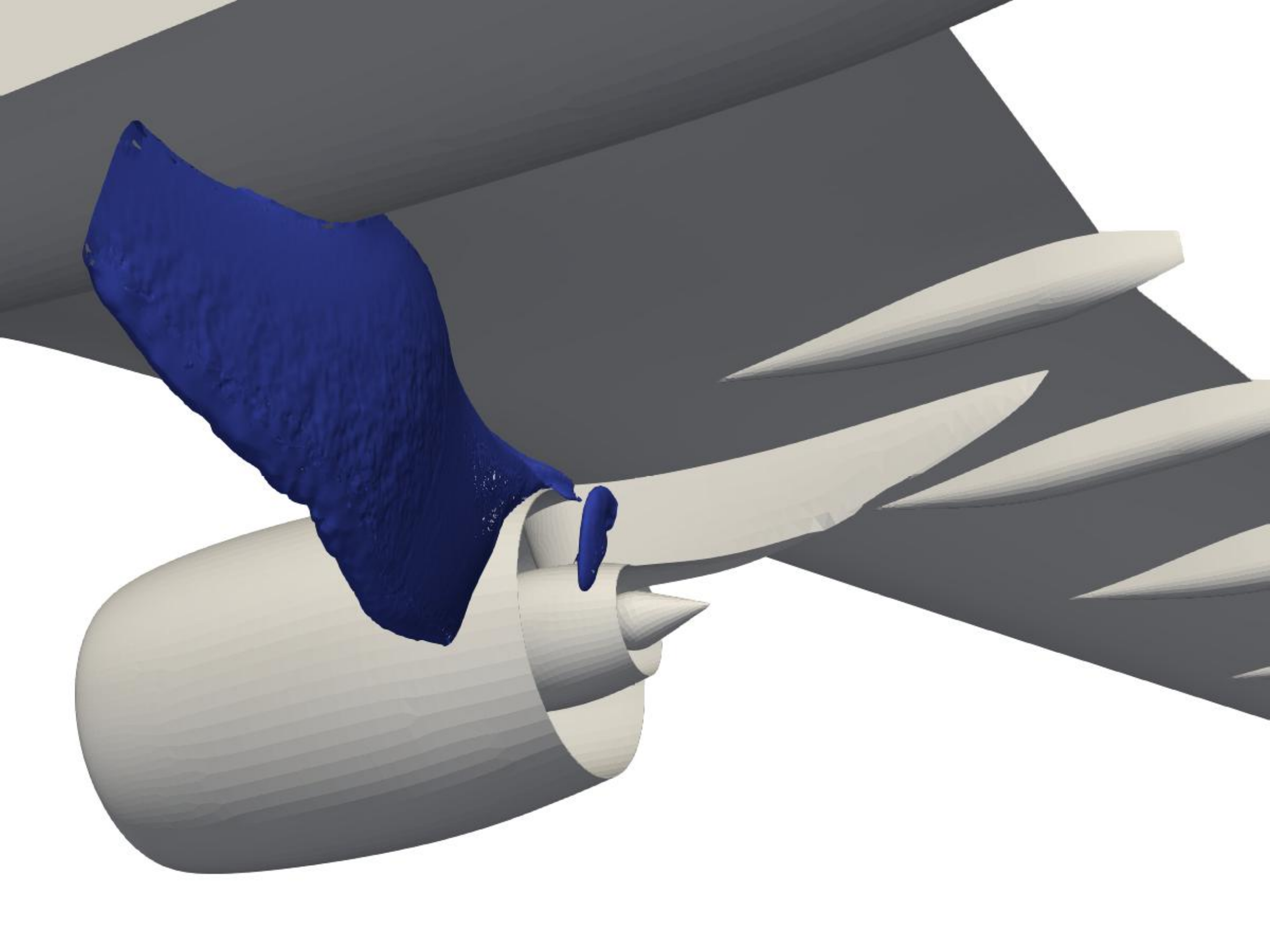}}
	\end{center}
	\caption{Isocontours of the first POD mode based on the static pressure field for the original grid, and the grid generated by $S^3$. The threshold values for the isocountours are~$-30$.}
	\label{fig:xrf1_iso_cont}
\end{figure}
We performed an SVD of the local static pressure field for both the original and the reduced flow fields.
Fig.~(\ref{fig:xrf1_iso_cont}) shows the isocontour plots of the first POD mode using a threshold value of $-30$.
Both the shape and position of the isocontours are in good agreement.
We present a comparison of the first singular values along with the temporal error of the mode coefficients in the appendix, figures (\ref{fig:eigenvalues_xrf1}) and (\ref{fig:v_temporal_error_xrf1}). 
The error between the SVD results of $S^3$ and the original data remains low.
However, the contribution of the higher-order modes for the $S^3$ data set is slightly underestimated.
The overall, weighted absolute temporal error remains in the order of $O(10^{-3}) \dots O(10^{-2})$ across the first $100$ mode coefficients.
While the first few mode coefficients are approximated very well, the error increases for higher-order modes, as presented in the appendix, fig. (\ref{fig:v_temporal_error_xrf1_first_few}) for the first $11$ uneven mode coefficients.
One cause of the deviation may be an approximation of the cell volumes for the original data set, since this quantity was not available for post-processing.
Further, it should be noted that the pressure field was not part of the original metric field.\\

Due to the large size of the original flow fields, this test case could not be executed on a local machine.
Instead, both the grid generation and SVD were performed on an HPC cluster using $64$ CPU cores.
Note that the current implementation is shared-memory parallel, so distribution across multiple nodes is not possible.
The overall execution time for $S^3$ is $t_{\mathrm{tot}}~=~15.75 \, \min$, which is composed of the uniform refinement ($t_{\mathrm{uni}}~=~44 \, s$), metric-based refinement ($t_{\mathrm{a}}~=~14.5 \, \min$) and renumbering of the leaf cells to create the final grid ($t_{\mathrm{rn}}~=~22 \, s$).

\subsection{Benchmark and comparison of the test cases}
\label{subsec:timings}
In this section, we present detailed timing for individual steps of the mesh generation process and how these timings depend on the metric threshold value.
Due to the size of the half-model simulation, we conducted these tests only for the tandem configuration and the cylinder flow.
All benchmarks are performed on a laptop equipped with an \emph{Intel~\textsuperscript{\textregistered}Core\textsuperscript{\texttrademark} i7-11800H} with $8$ physical cores and $32$ GB of RAM.
We did not repeat the tests multiple times, but made sure that the impact of other processes remained marginal.

\begin{figure}[htbp]
	\begin{center}
		\includegraphics[width=0.8\textwidth]{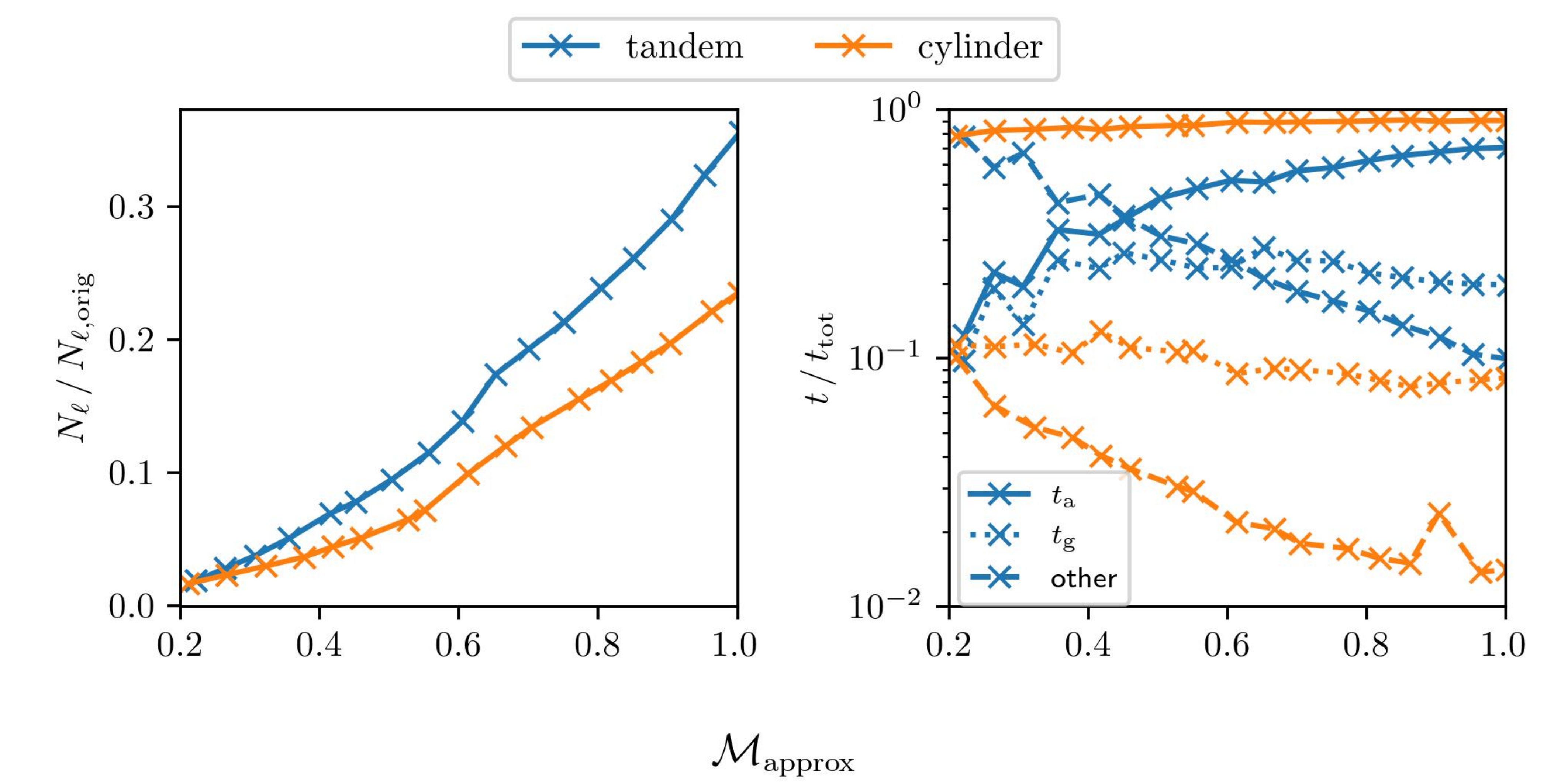}
		\caption{Cell count progression with respect to the captured metric (left) and composition of the meshing times (right) for the tandem configuration and the cylinder flow. The timing distinguishes between adaptive refinement ($t_\mathrm{a}$), geometry refinement ($t_\mathrm{g}$), and the remaining steps, i.e., uniform refinement and renumbering (\emph{other}).}
		\label{fig:n_cells_t_exec}
	\end{center}
\end{figure}

Fig.~(\ref{fig:n_cells_t_exec}) depicts the number of cells as well as the composition of the execution times for varying $\mathcal{M}_\mathrm{approx}$.
Although the number of cells increases with an improved approximation of the metric, reduction rates of $65\%$ for the tandem configuration and $76\%$ for the cylinder are achievable even with $\mathcal{M}_{\mathrm{approx}}~=~1$.
It should be noted that the choice of the metric significantly influences the overall mesh reduction and must be done carefully.

The total execution time ($t_{\mathrm{tot}}$) is composed of the uniform refinement ($t_\mathrm{uni}$), metric-based refinement ($t_\mathrm{a}$), geometry refinement ($t_\mathrm{g}$), and the final mesh renumbering ($t_{\mathrm{rn}}$).
From fig.~(\ref{fig:n_cells_t_exec}), it can be seen that adaptive refinement significantly dominates the overall execution time for larger mesh sizes.
The geometry refinement is the second most expensive part of the algorithm and depends on the geometry itself and the maximum refinement level.
For the cylinder flow, the relative contribution to the overall meshing time is relatively constant.
For the more complex airfoil geometry, the relative contribution varies in the range $10-30\%$.
We note once more that the geometry refinement is optional and not required in many cases.
The mesh renumbering at the end of the refinement process scales approximately linearly with the number of leaf cells, while the influence of the uniform refinement is negligible.
Since the latter quantities contribute only a minor part to the overall execution time, they are summarized in fig.~(\ref{fig:n_cells_t_exec}) as \emph{other}.
The physical execution times for $\mathcal{M}_{\mathrm{approx}}~=~1$ remained in the order of one minute for the tandem test case, whereas the grid generation of the cylinder test required up to $23$ minutes on the local workstation.
Both overall execution times refer to benchmarks employing a subsequent geometry refinement.
The mapping between original and new meshes is not included in the timing.
The interpolation and export of the flow fields highly depend on the available RAM, which limits the maximum number of snapshots that can be processed in parallel.
However, loading, interpolation onto the generated grid, and export to HDF5 require approximately the same time as the mesh generation on a local machine for the investigated cases.
These timings emphasize that incorporating $S^3$ into an existing post-processing pipeline results in minimal additional computational overhead relative to the overall execution time of the CFD simulation.
Using coarse meshes from $S^3$ enables additional post-processing steps to be executed on a local machine, thus avoiding the need for HPC resources in some cases.

\section{Conclusion}
\label{sec:conclusion}
The present article introduced an improved version of $S^3$ for efficiently downsampling CFD data for post-processing.
This new version was tested successfully on the scale-resolving simulation of the flow past a tandem configuration of airfoils in the transonic regime, the incompressible turbulent flow past a circular cylinder, and the flow around an aircraft half-model.
$S^3$ efficiently decreased the mesh size for all cases, while preserving a user-provided metric.
The SVD performed to investigate the information loss caused by $S^3$ revealed negligible differences for both test cases.
Parameter studies further showed achievable overall mesh reductions between $35 \% ... 98\%$, depending on the metric and stopping criteria, providing the possibility to execute post-processing steps on a local machine rather than using HPC resources.\\
Although we presented $S^3$ as a post-processing tool, it can also be applied directly during the runtime of a simulation.
One possible approach would be to generate the coarse grid after the initial transient or warm-up phase of the simulation.
Once the $S^3$ mesh is available, new snapshots created during the productive simulation run can be directly interpolated onto the coarse grid, enabling, e.g., smaller write intervals.\\

There are still some limitations for $S^3$ when dealing with large amounts of data.
The most significant limitation is the lack of support for distributed parallelism across computing nodes on an HPC system.
The limited memory per node can become an issue when generating large meshes in the order of $\mathcal{O}\left( 10^8 \right)$ or larger.
One possible solution to this problem is to remove unused nodes from the tree when generating the grid.
Currently, the unused nodes are only discarded when exporting the final leaf cells.
As presented in section (\ref{subsec:scube}), $S^3$ utilizes a recursive tree to store and address the mesh cells internally.
An intermediate pruning of the tree would decrease the memory requirements to the point where support for distributed parallelism may be redundant.

Since $S^3$ is designed to compress large amounts of volume data, it is unsuited for surface data applications.
In this case, the octree mesh cannot accurately represent the surface contour, leading to high interpolation errors when exporting the flow fields onto the grid created by $S^3$.
These errors are caused by the KNN interpolation and may be pronounced near thin geometries, such as the trailing edge of an airfoil.
However, for the test cases shown here, the error remains within acceptable bounds.

\section*{Acknowledgment}
The authors gratefully acknowledge the Deutsche Forschungsgemeinschaft DFG (German Research Foundation) for funding this work in the framework of the research unit FOR 2895 under the grant number 406435057. The authors also gratefully acknowledge the computing time made available to them on the high-performance computer at the NHR Center of TU Dresden. This center is jointly supported by the Federal Ministry of Education and Research and the state governments participating in the NHR (www.nhr-verein.de/unsere-partner).

\printbibliography[title=References, heading=bibliography]
\clearpage

\appendix
\section{Appendix}
\subsection{Singular value decomposition}\label{subsec:svd}
\noindent
This section summarizes the fundamentals of the singular value decomposition, which is used in section \ref{sec:applications} to compare the results produced by $S^3$ to the original data.
First, the state vectors $\mathbf{x}_n$ at time steps $n=1, ..., N$ are organized in a data matrix $\mathbf{X} \in \mathbb{R}^{\tilde{M}N_\mathrm{var} \times N}$
\begin{center}
	\begin{equation}
		\mathbf{X} = 
		\begin{bmatrix} 
			\mathbf{|} & \mathbf{|} & & \mathbf{|} \\
			\mathbf{x}_1 & \mathbf{x}_2 & ... & \mathbf{x}_n \\
			\mathbf{|} & \mathbf{|} & & \mathbf{|} 
		\end{bmatrix}.
	\end{equation}
\end{center}
The state vector is typically composed of $N_\mathrm{var}$ flow variables known at $\tilde{M}$ discrete locations, where $\tilde{M}=M$ for the original mesh and $\tilde{M}=N_\ell$ for the $S^3$ mesh.
To reduce the mesh dependency of the SVD computation, it is common practice to weigh each entry in the state vector with the square root of the corresponding cell volume \cite{weiner_robust_2023}.
Moreover, the temporal (rowwise) mean is subtracted from each column.
The economy SVD of $\mathbf{X}$ reads
\begin{center}
	\begin{equation}
		\mathbf{X} = \mathbf{U}\mathbf{\Sigma}\mathbf{V}^T,
	\end{equation}
\end{center}
with the matrix of left-singular vectors $\mathbf{U} \in \mathbb{R}^{\tilde{M}N_\mathrm{var} \times N}$, the matrix of right-singular vectors $\mathbf{V} \in \mathbb{R}^{N \times N}$, and the diagonal matrix containing the singular values $\mathbf{\Sigma} \in \mathbb{R}^{N \times N}$.
All triples of left-singular vector, singular value, and right-singular vector are ordered according to the singular values in descending order.
The left-singular vectors contain spatial patterns (modes), while the right-singular vectors denote the contribution of each mode over time.
For the mode visualization, the volume weighting is reversed.
For a more comprehensive overview of SVD fundamentals, we refer to \cite{weiss_tutorial_2019, goodfellow_deep_2016}.
The SVD is computed with the open-source flow analysis package \emph{flowTorch} \cite{weiner_flowtorch_2021}.

\subsection{OAT tandem configuration}\label{sec:appendix:oat}
\begin{figure}[htbp]
	\begin{center}
		\includegraphics[width=0.75\textwidth]{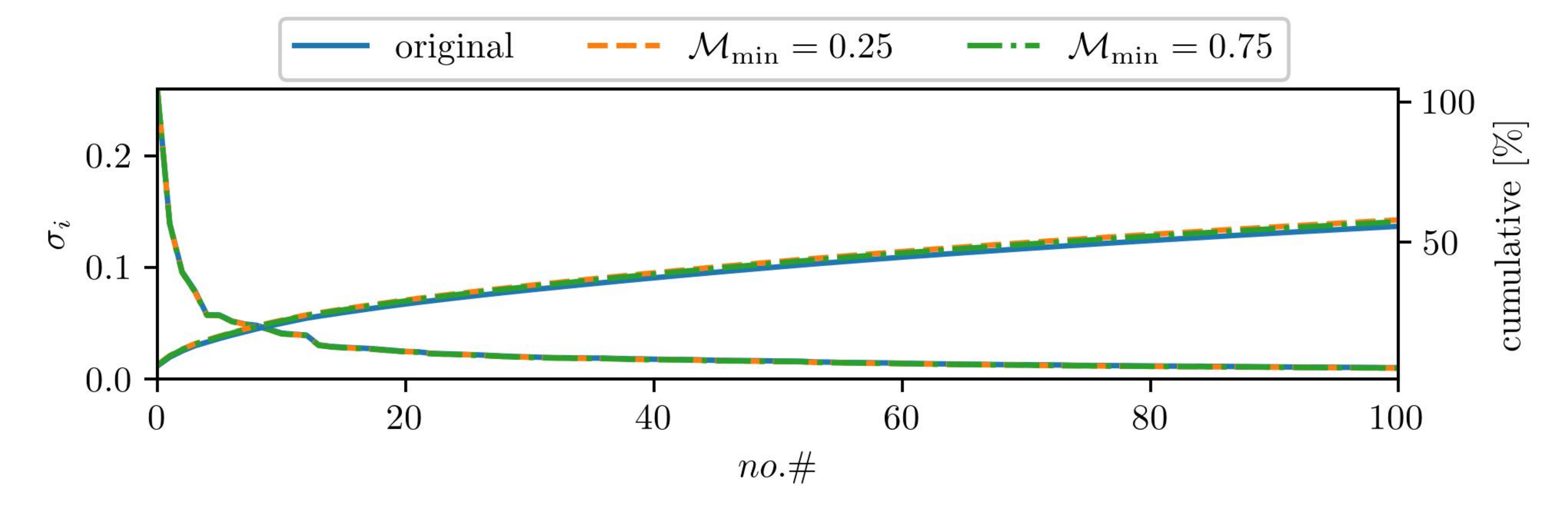}
		\caption{Comparison of the first $100$ singular values for $\boldsymbol{\mathcal{M}}_1~=~\mathrm{std}(\mathbf{Ma})$ and their cumulative contribution.}
		\label{fig:eigenvalues_oat}
	\end{center}
\end{figure}

Fig. (\ref{fig:eigenvalues_oat}) depicts the first $100$ singular values for the tandem configuration.
Both threshold values of $\mathcal{M}_{\min}~=~0.25$ and $0.75$ yield excellent agreement with the singular values of the original dataset.\\

\begin{figure}[htbp]
	\begin{center}
		\includegraphics[width=0.75\textwidth]{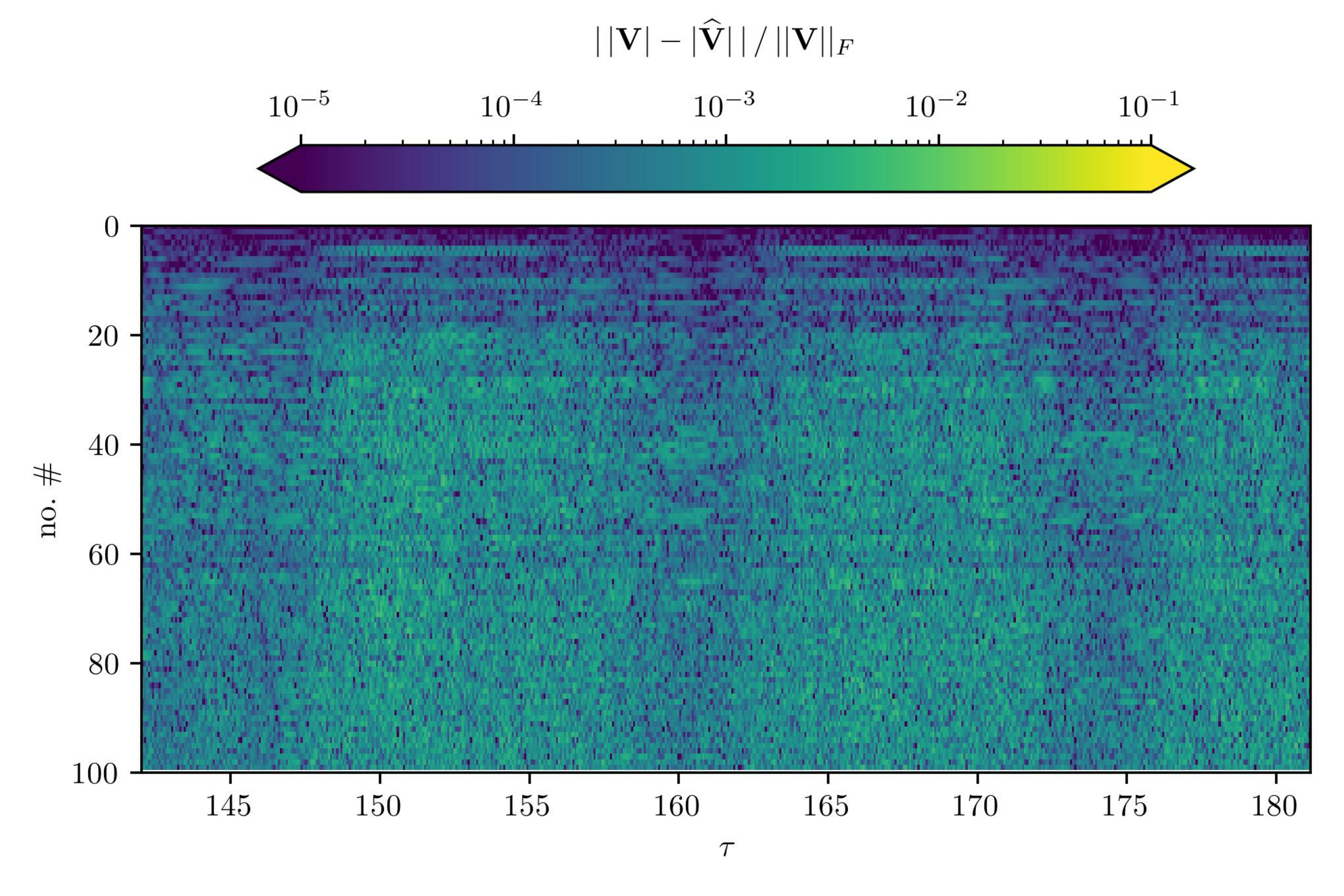}
		\caption{Temporal evolution of the error within the first $100$ right singular vectors for $\boldsymbol{\mathcal{M}}_1~=~\mathrm{std}(\mathbf{Ma})$ and $\mathcal{M}_{\min}~=~0.25$.}
		\label{fig:v_temporal_error_oat}
	\end{center}
\end{figure}

Fig. (\ref{fig:v_temporal_error_oat}) shows the absolute temporal error $|\,|\mathbf{V}| - |\mathbf{\widehat{V}}|\,|$ of the first $100$ right-singular vectors, scaled with the Frobenius norm $||\mathbf{V}||_F$.
Here, only the error for a threshold value of $\mathcal{M}_{\min}~=~0.25$ is presented.
Even though the error is increasing for higher mode coefficients, the overall error remains small, $O(10^{-4}) \dots O(10^{-3})$.\\

\begin{figure}[htbp]
	\begin{center}
		\subfloat[Absolute spatial error for the \emph{unseen} Mach number field]{\includegraphics[width=0.9\textwidth]{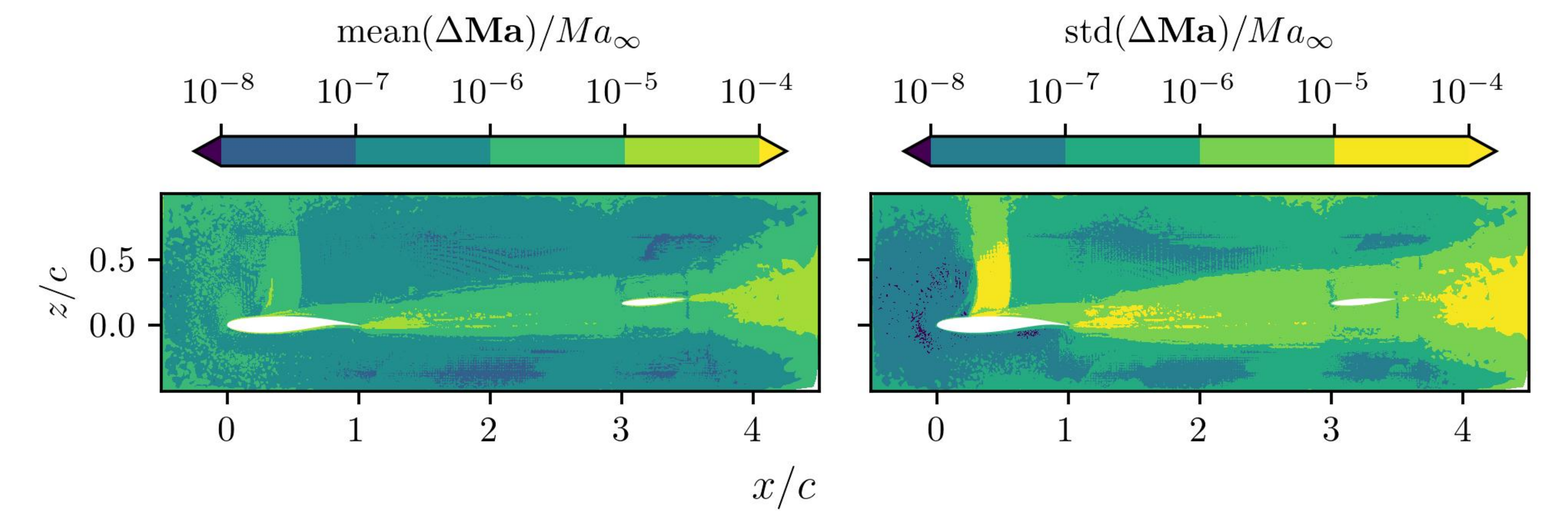}\label{fig:spatial_error_mu_only_ma}}\\
		\subfloat[Absolute spatial error for the velocity component $u_3$]{\includegraphics[width=0.9\textwidth]{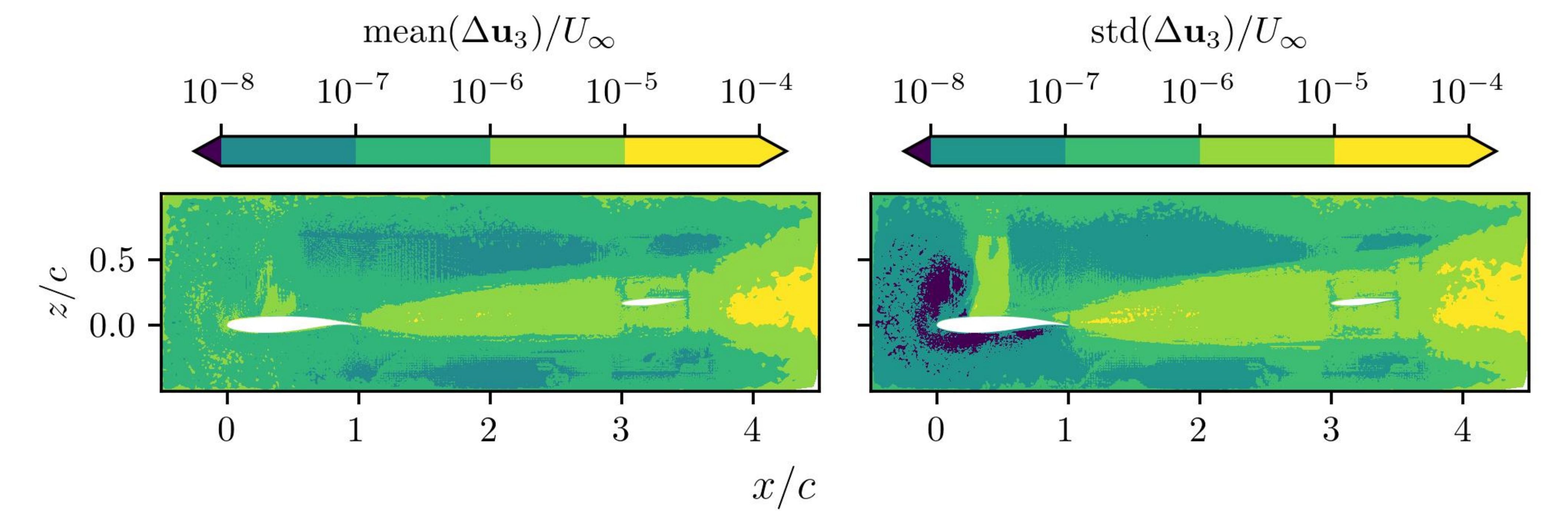}\label{fig:spatial_error_mu_only_u3}}
		\caption{Temporal mean (left column) and standard deviation (right column) of the absolute spatial error $\Delta \mathbf{f}_n~=~|\mathbf{f}^\ast_n - \mathbf{f}_n|$, weighted with the square-root of the cell volume and scaled with the free stream quantity $f_{\infty}$ for $\mathcal{M}_{\mathrm{approx}}~=~0.75$ using $\boldsymbol{\mathcal{M}}_2$.}
		\label{fig:spatial_error_mu_only}
	\end{center}
\end{figure}

Fig. (\ref{fig:spatial_error_mu_only}) presents the temporal mean spatial error of the local Mach number field (fig. (\ref{fig:spatial_error_mu_only_ma})) and $u_3$-field (fig. (\ref{fig:spatial_error_mu_only_u3})), respectively.
Additionally, the temporal standard deviation of the spatial error is shown.
All errors are weighted with the cell volumes and include both the interpolation error made by $S^3$ as well as the error from interpolation back to the original grid (required for computing the spatial error) for $\mathcal{M}_{\min}~=~0.75$.
Note that the metric $\boldsymbol{\mathcal{M}_2}$ does not include the local Mach number field.
However, it can be seen in fig. (\ref{fig:spatial_error_mu_only_ma}) that the mean error remains low, in the order of $O(10^{-6})$, in most parts of the domain (left column), while slightly increasing towards the wake region of the rear airfoil.
The temporal standard deviation of the error (right column) shows an additional increase in the error in the shock region of the leading airfoil, but remains overall below $O(10^{-4})$.

As discussed in sec. (\ref{subsec:oat}), the $u_3$-component of the velocity field consistently yields the highest temporal interpolation error of all fields.
We therefore present the mean spatial error and the temporal standard deviation of the absolute spatial error in fig. (\ref{fig:spatial_error_mu_only_u3}) for $\boldsymbol{\mathcal{M}}_2$ as well.
Even though the absolute temporal error is increased compared to all other fields, the overall spatial error is comparable to the other fields.
Solely the wake region aft of the rear airfoil shows a higher mean interpolation error, while the temporal standard deviation exhibits no significant changes compared to other fields.
It should be noted that the spatial error fields for the other metric fields tested show a similar trend and are therefore not included here.\\

\begin{figure}[htbp]
	\begin{center}
		\includegraphics[width=0.6\textwidth]{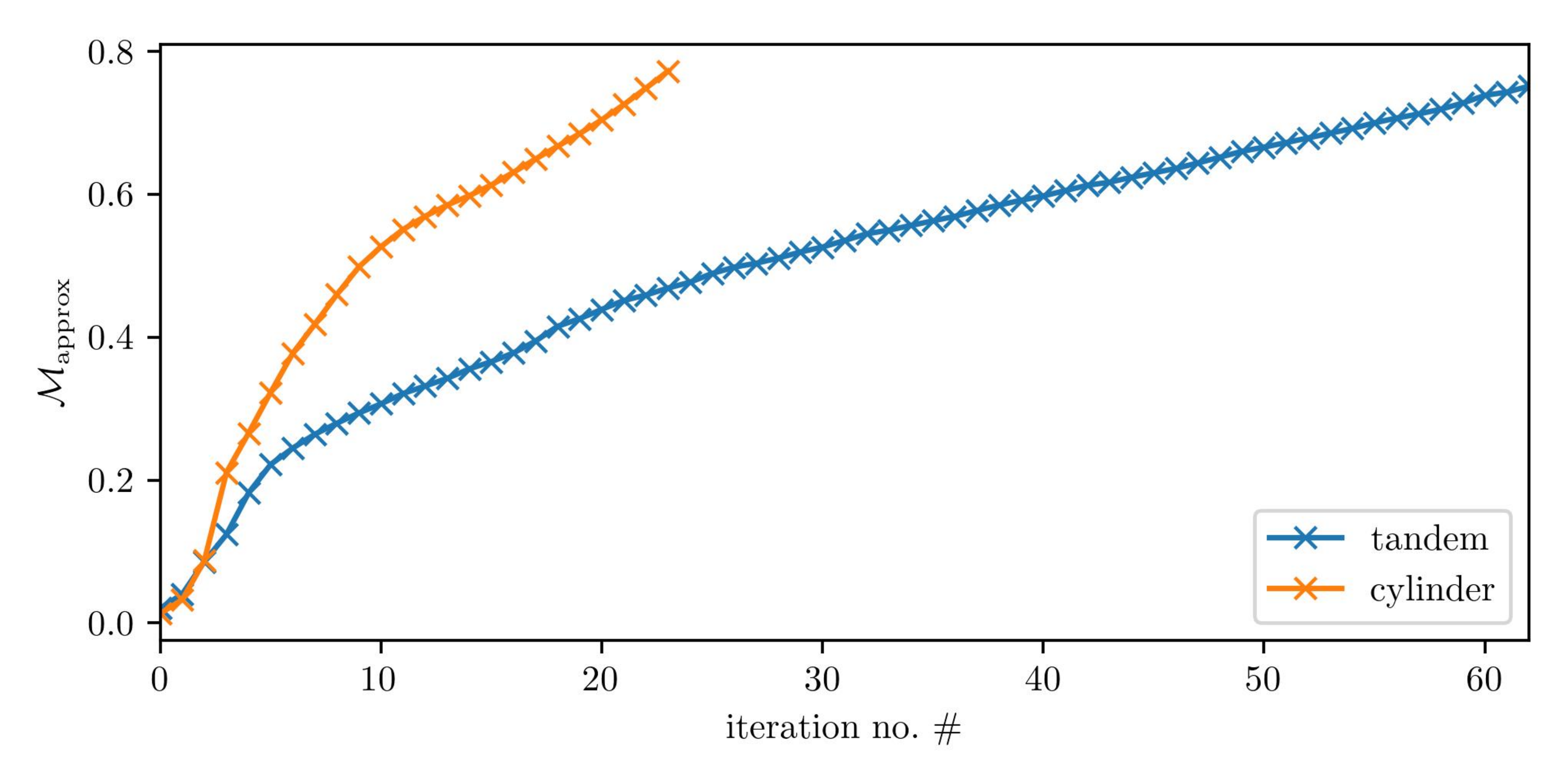}
	\end{center}
	\caption{Convergence behavior of the metric-based refinement for the tandem configuration and the cylinder flow using $\mathcal{M}_{\mathrm{min}} = 0.75$.}
	\label{fig:progress_refinement}
\end{figure}

Fig.~(\ref{fig:progress_refinement}) shows the approximation of $\mathcal{M}_{\mathrm{approx}}$ over the course of the metric-based refinement for both the tandem configuration and the cylinder flow.
The approximation improves significantly within the first few iterations, since the dominant large-scale structures present in both test cases can be represented with relatively few cells.
When the refinement process continues, the improvement with respect to the approximation of $\mathcal{M}_{\mathrm{approx}}$ converges to a linear behavior.
Since $M_{\mathrm{tandem}} \ll M_{\mathrm{cylinder}}$ results in a smaller $N_{\mathrm{c, start}}$ for the tandem configuration, more iterations are required to reach the same target value of $\mathcal{M}_{\mathrm{min}} = 0.75$.

\subsection{Flow past a circular cylinder}\label{sec:appendix:cylinder}
\begin{figure}[htbp]
	\begin{center}
		\subfloat[$x$-$y$-plane, $\mathcal{M}_{\mathrm{approx}} = 0.53$]{\includegraphics[width=0.42\textwidth,trim={0 7cm 0 7cm},clip]{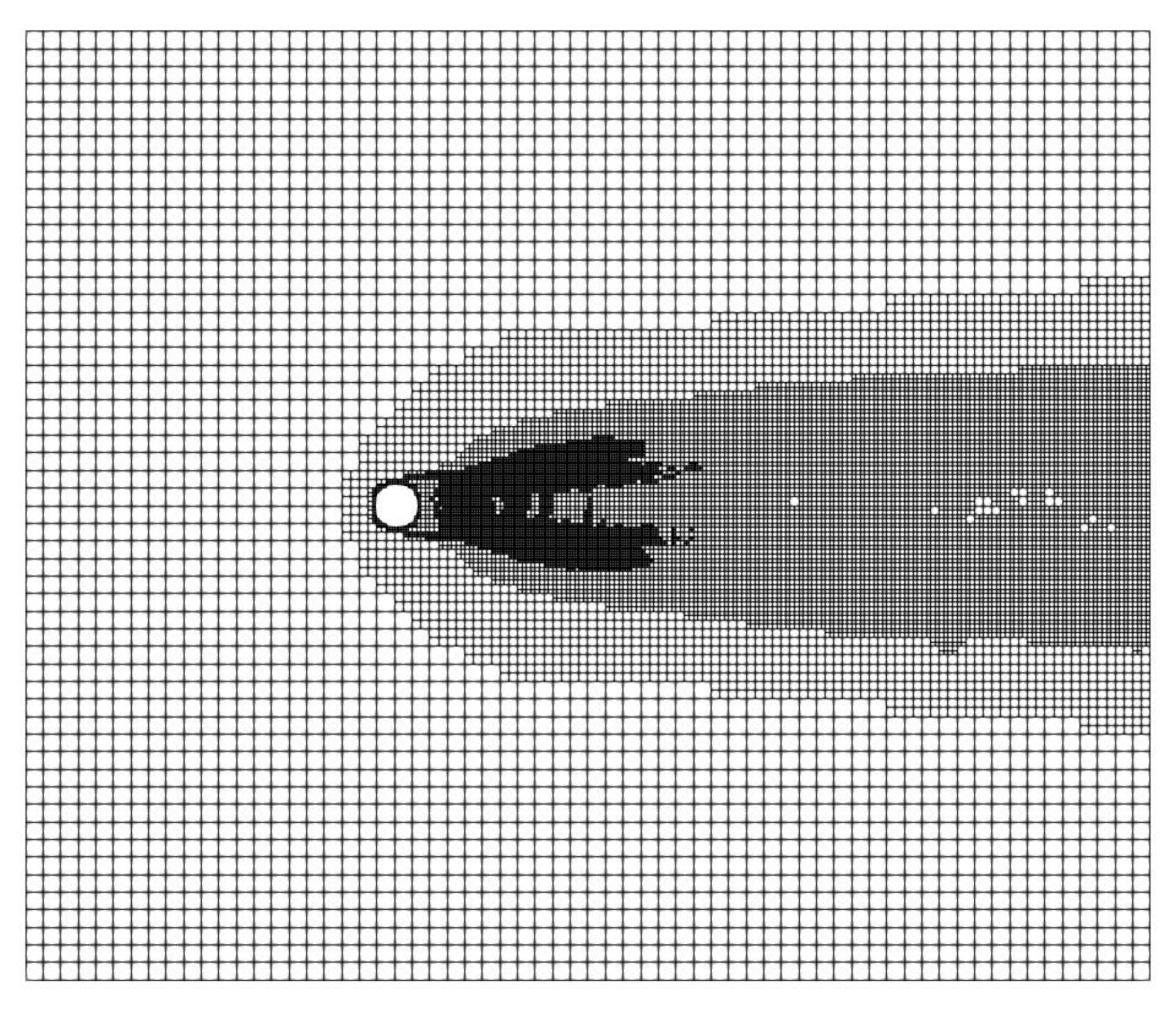}\label{fig:interpolated_grid_cylinder_xy_M50}}
		\hspace*{0.5cm}
		\subfloat[$x$-$y$-plane, $\mathcal{M}_{\mathrm{approx}} = 0.77$]{\includegraphics[width=0.42\textwidth,trim={0 7cm 0 7cm},clip]{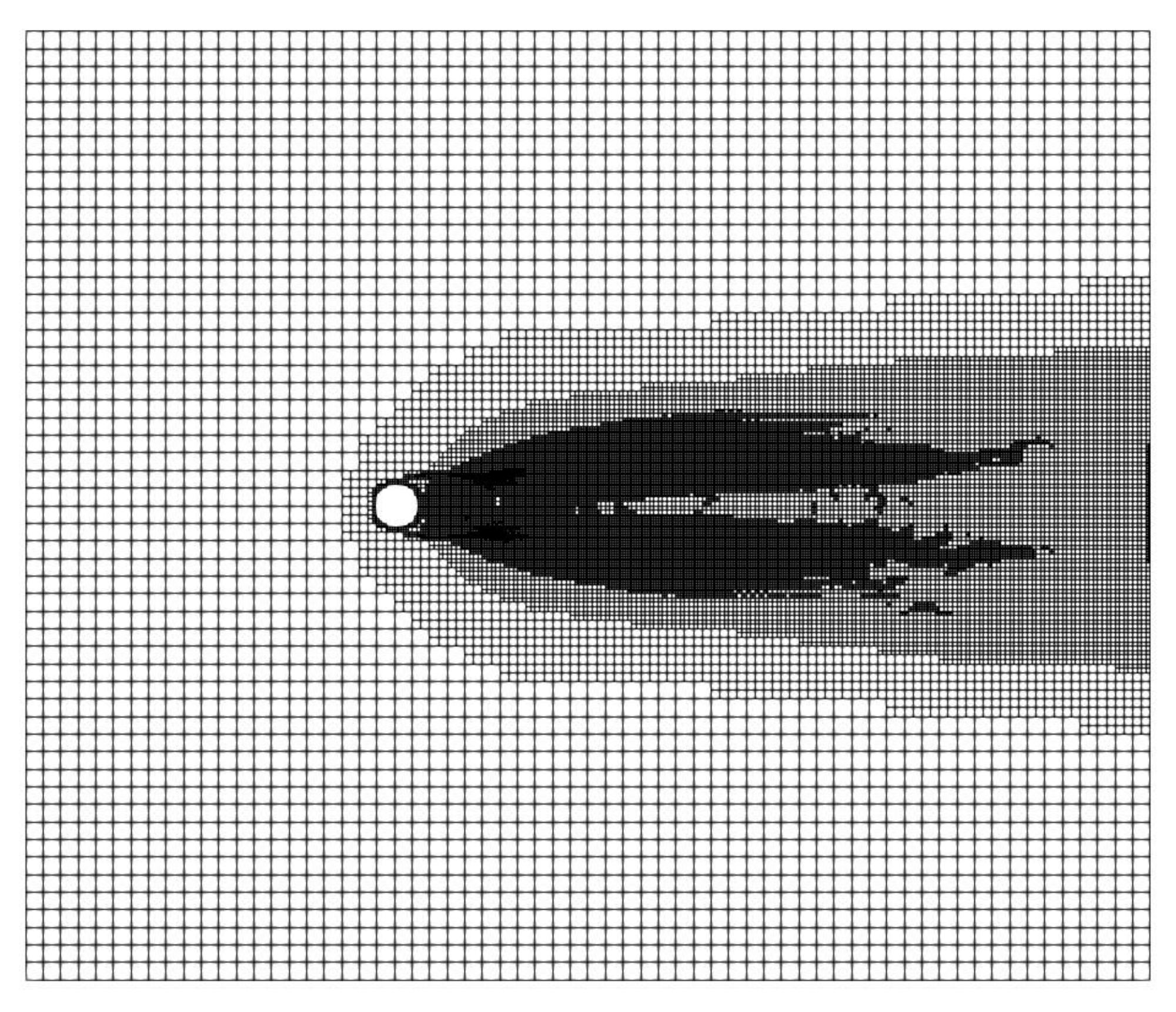}\label{fig:interpolated_grid_cylinder_xy_M75}}
		\\
		\subfloat[$x$-$z$-plane, $\mathcal{M}_{\mathrm{approx}} = 0.53$]{\includegraphics[width=0.42\textwidth]{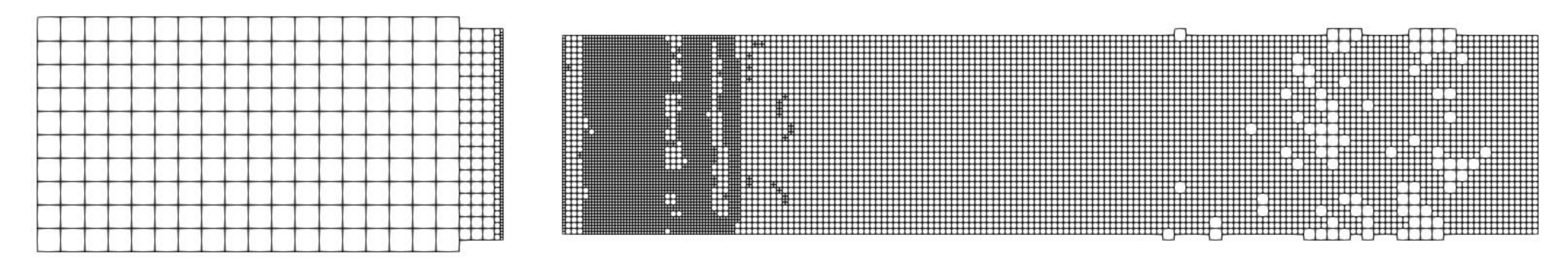}\label{fig:interpolated_grid_cylinder_xz_M50}}
		\hspace*{0.5cm}
		\subfloat[$x$-$z$-plane, $\mathcal{M}_{\mathrm{approx}} = 0.77$]{\includegraphics[width=0.42\textwidth]{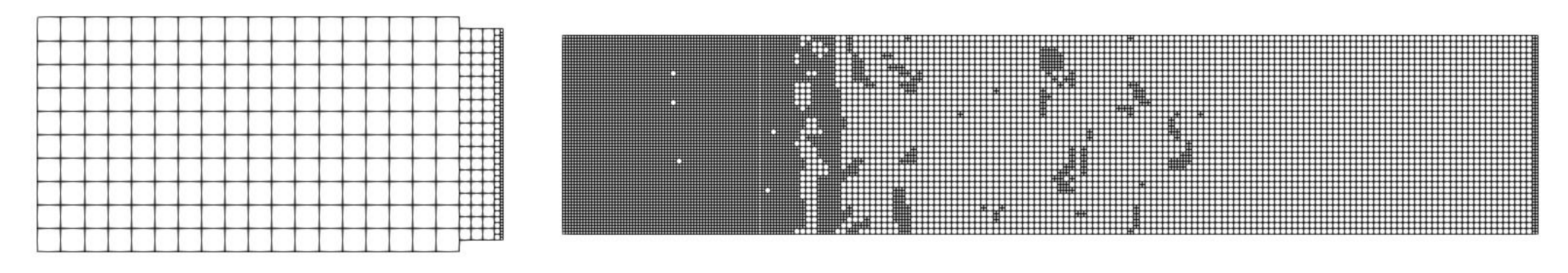}\label{fig:interpolated_grid_cylinder_xz_M75}}
	\end{center}
	\caption{The left column shows the grid generated by $S^3$ for a captured metric of $\mathcal{M}_{\mathrm{approx}} = 0.53$ in the $x$-$y$-plane at $z/d = \pi/2$ (\ref{fig:interpolated_grid_cylinder_xy_M50}) and $x$-$z$-plane (\ref{fig:interpolated_grid_cylinder_xz_M50}) at $y/d~=~8$. The right column depicts the metric interpolated onto the grid created by $S^3$ for $\mathcal{M}_{\mathrm{approx}}~=~0.77$ in the $x$-$y$-plane at $z/d~=~\pi/2$  (\ref{fig:interpolated_grid_cylinder_xy_M75}) and $x$-$z$-plane (\ref{fig:interpolated_grid_cylinder_xz_M75}) at $y/d = 8$.}
\end{figure}

Fig.~(\ref{fig:interpolated_grid_cylinder_xy_M50}) and fig.~(\ref{fig:interpolated_grid_cylinder_xy_M75}) shows slices through the grid generated by $S^3$ for $\mathcal{M}_{\mathrm{approx}} = 0.53$ and $\mathcal{M}_{\mathrm{approx}} = 0.77$, respectively.
The grid remains unchanged in the outer regions of the domain, whereas the refinement level of the wake increases with increasing approximation of the metric.
The overall symmetrical shape of the refinement regions in the wake of the cylinder is preserved for larger threshold values of $\mathcal{M}_{\mathrm{min}}$.

\begin{figure}[htbp]
	\begin{center}
		\subfloat[$x$-component]
		{\includegraphics[width=0.6\textwidth]{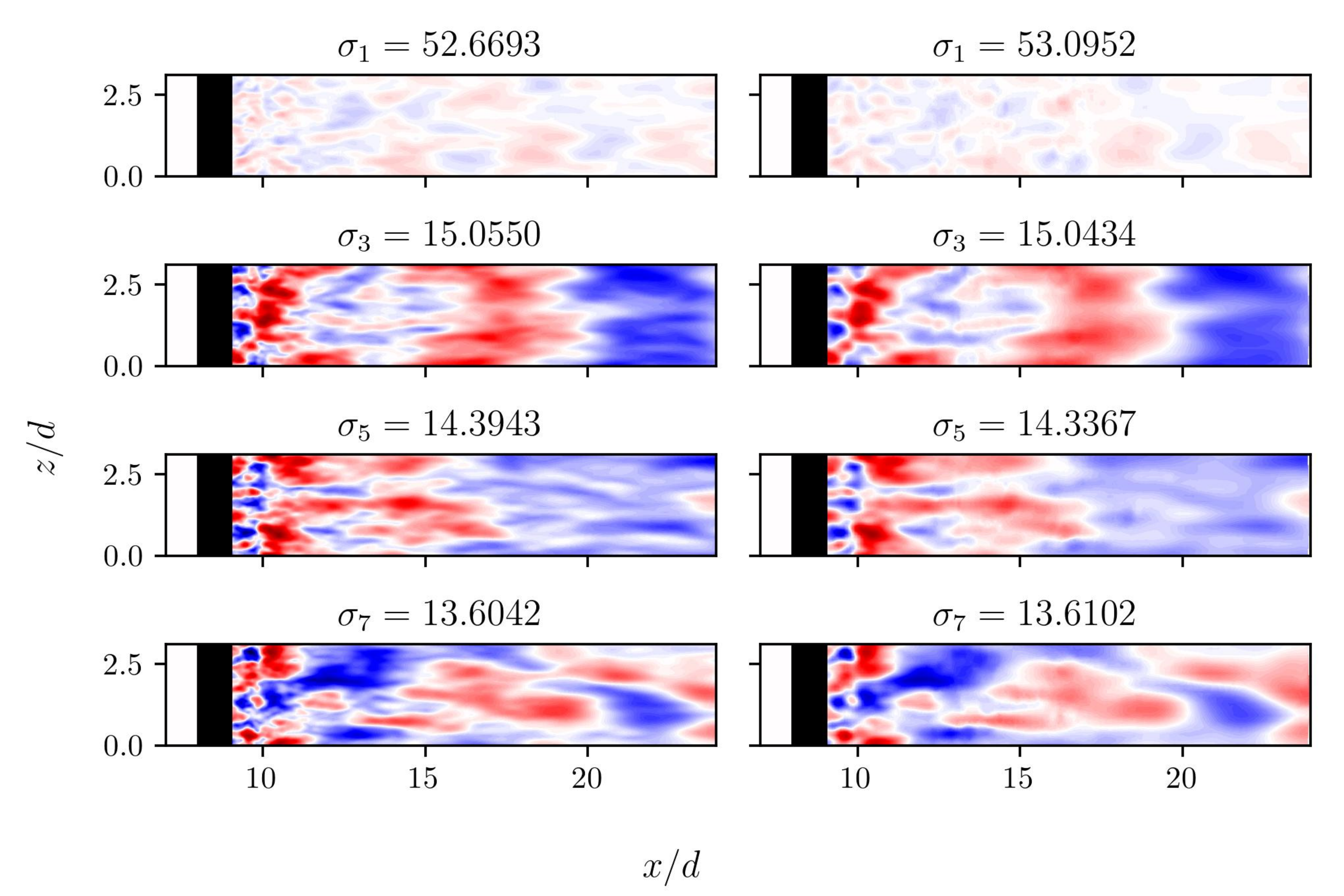}}
		\\
		\subfloat[$y$-component]{\includegraphics[width=0.6\textwidth]{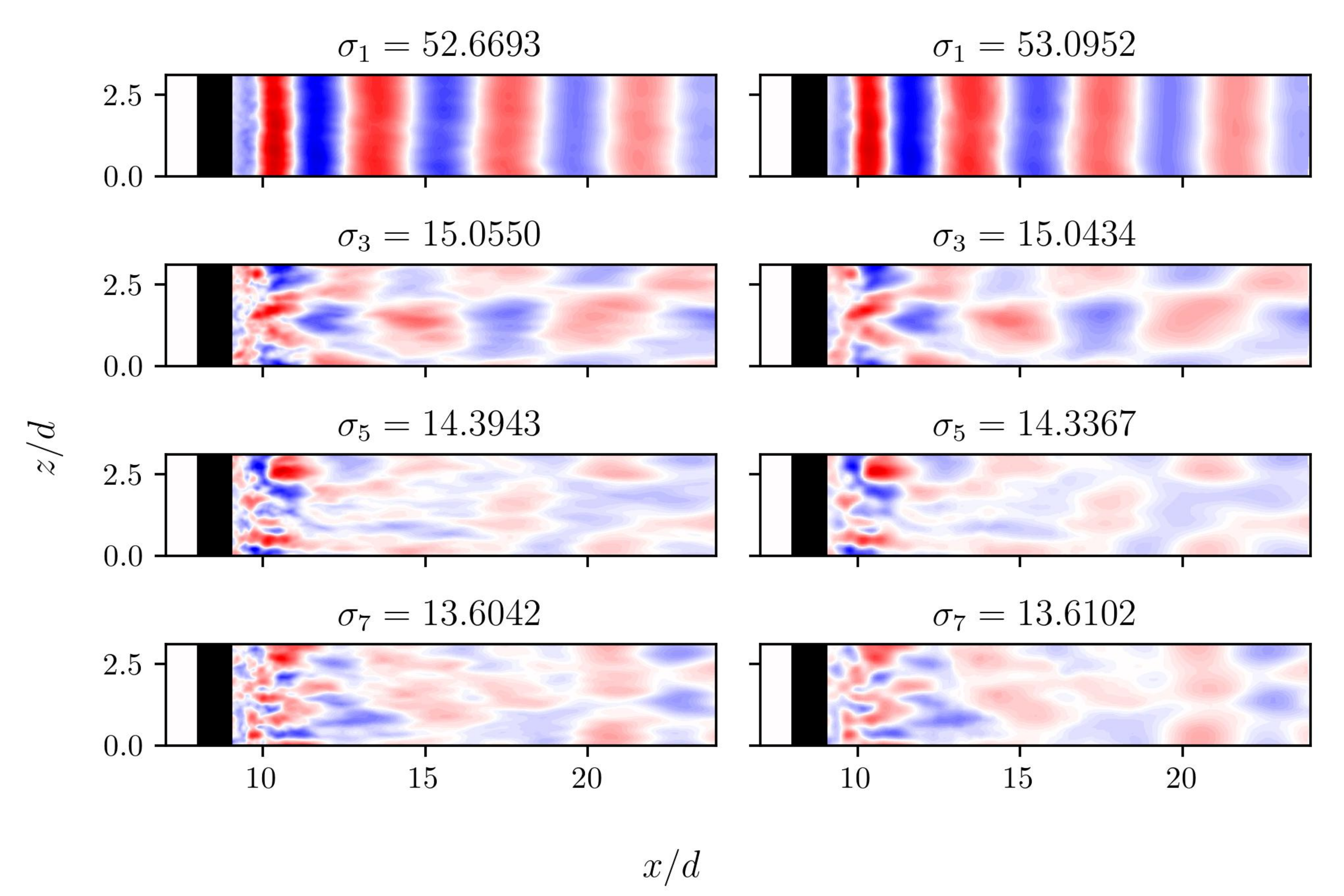}}
	\end{center}
	\caption{Comparison of the first four uneven POD modes and associated singular values for the original data (left columns) and the $S^3$-interpolated data with $\mathcal{M}_{\mathrm{approx}}~=~0.27$  (right columns) in the $x$-$z$-plane at $z/d~=~8$ for the cylinder test case. The colorscale is identical for all contours and bounded by $\pm||\mathbf{U}||_\infty$.}
	\label{fig:compoarison_pod_mode_cylinder_xz}
\end{figure}

Fig.~(\ref{fig:compoarison_pod_mode_cylinder_xz}) shows the first four uneven left-singular vectors and the associated singular values in the $x$-$z$-plane for~$\mathcal{M}_{\mathrm{approx}} = 0.27$.
As in section \ref{subsec:cylinder}, the $z$-component of the modes is omitted here for compactness.
The modes on the reduced grid show no visible difference to the modes on the original mesh from CFD, which is consistent with the results of the $x$-$y$-plane.
The results for $M_\mathrm{approx} = 0.53$ and $M_\mathrm{approx} = 0.77$ are not shown here, since the error between the SVD results of the original and reduced data is already within acceptable bounds for $M_\mathrm{approx} = 0.27$.
However, the error reduces with increasing approximation of the metric.

\clearpage
\subsection{Flow around an aircraft}\label{sec:appendix:xrf1}
\begin{figure}[htbp]
	\begin{center}
		\includegraphics[width=0.75\textwidth]{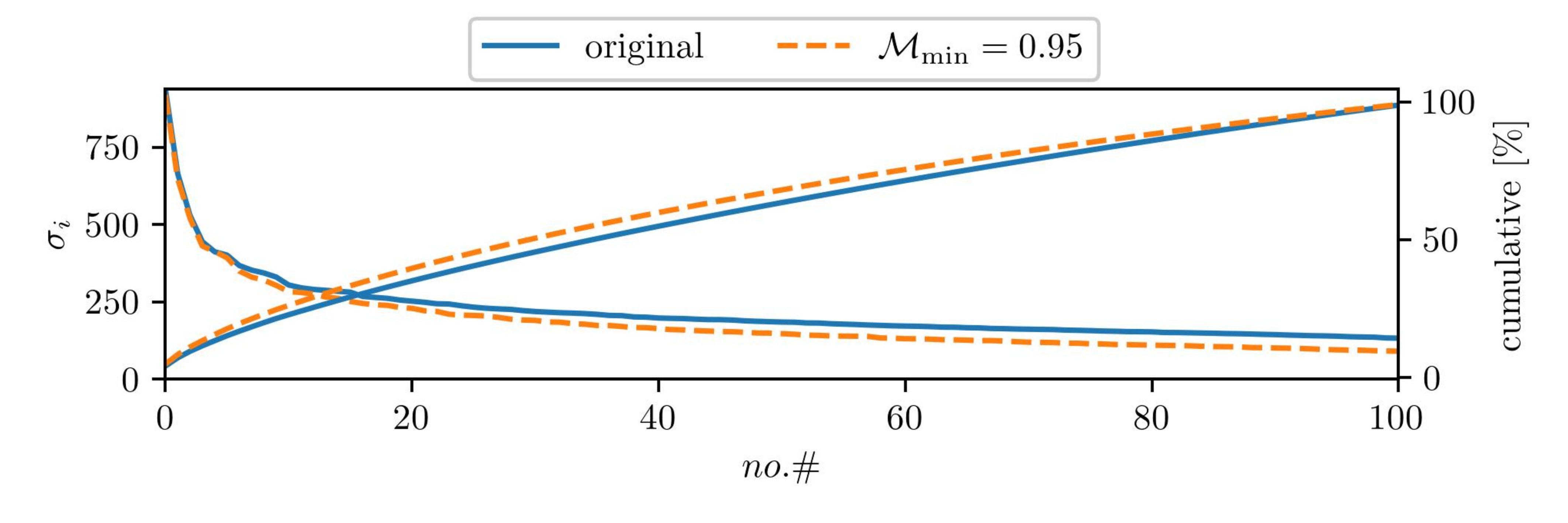}
		\caption{Temporal evolution of the error within the first $100$ right singular values for the Airbus XRF-1 aircraft.}
		\label{fig:eigenvalues_xrf1}
	\end{center}
\end{figure}

\begin{figure}[htbp]
	\begin{center}
		\includegraphics[width=0.75\textwidth]{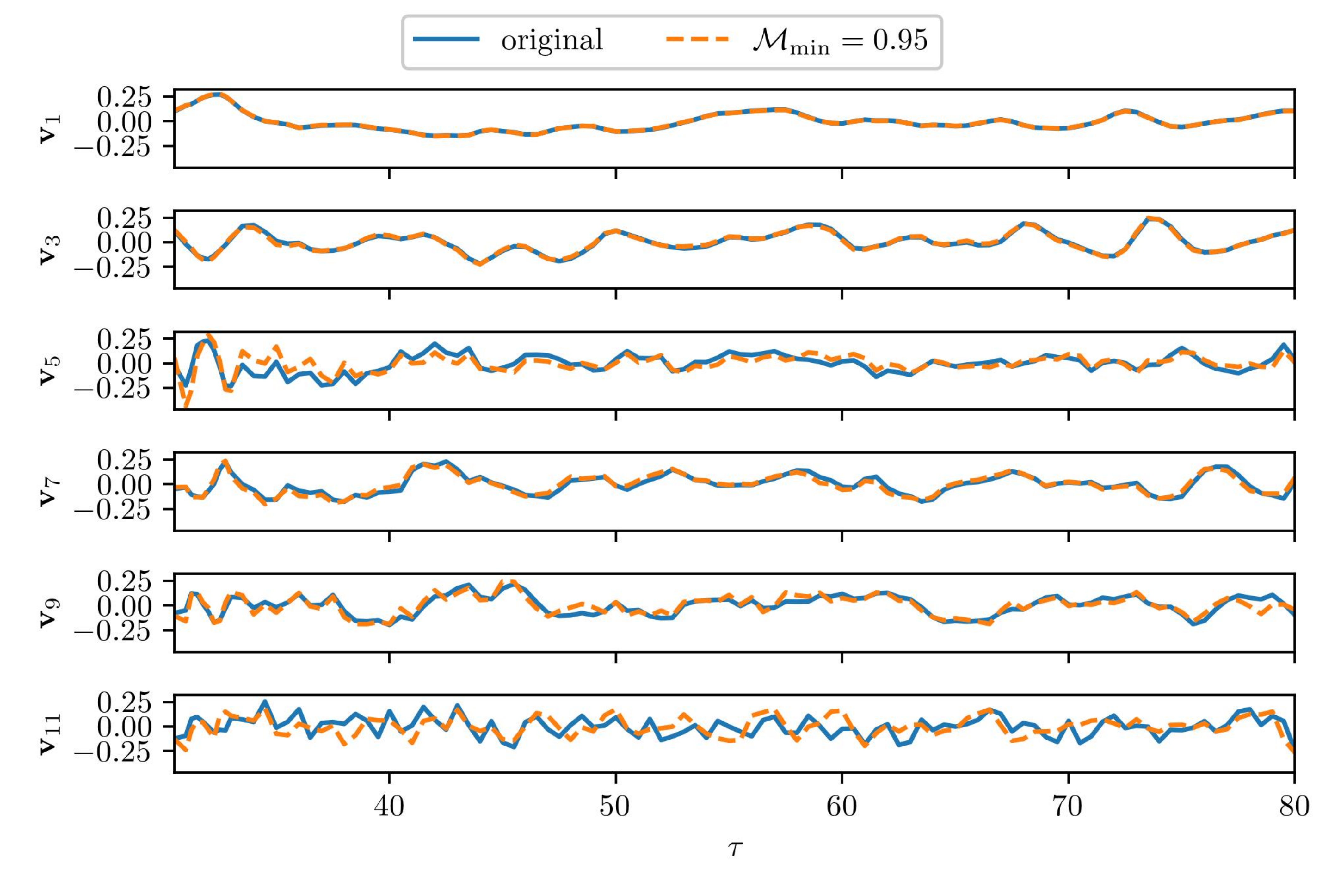}
		\caption{Comparison of the leading right-singular vectors for the Airbus XRF-1 aircraft. Every second mode coefficient is shown.}
		\label{fig:v_temporal_error_xrf1_first_few}
	\end{center}
\end{figure}

\begin{figure}[htbp]
	\begin{center}
		\includegraphics[width=0.75\textwidth]{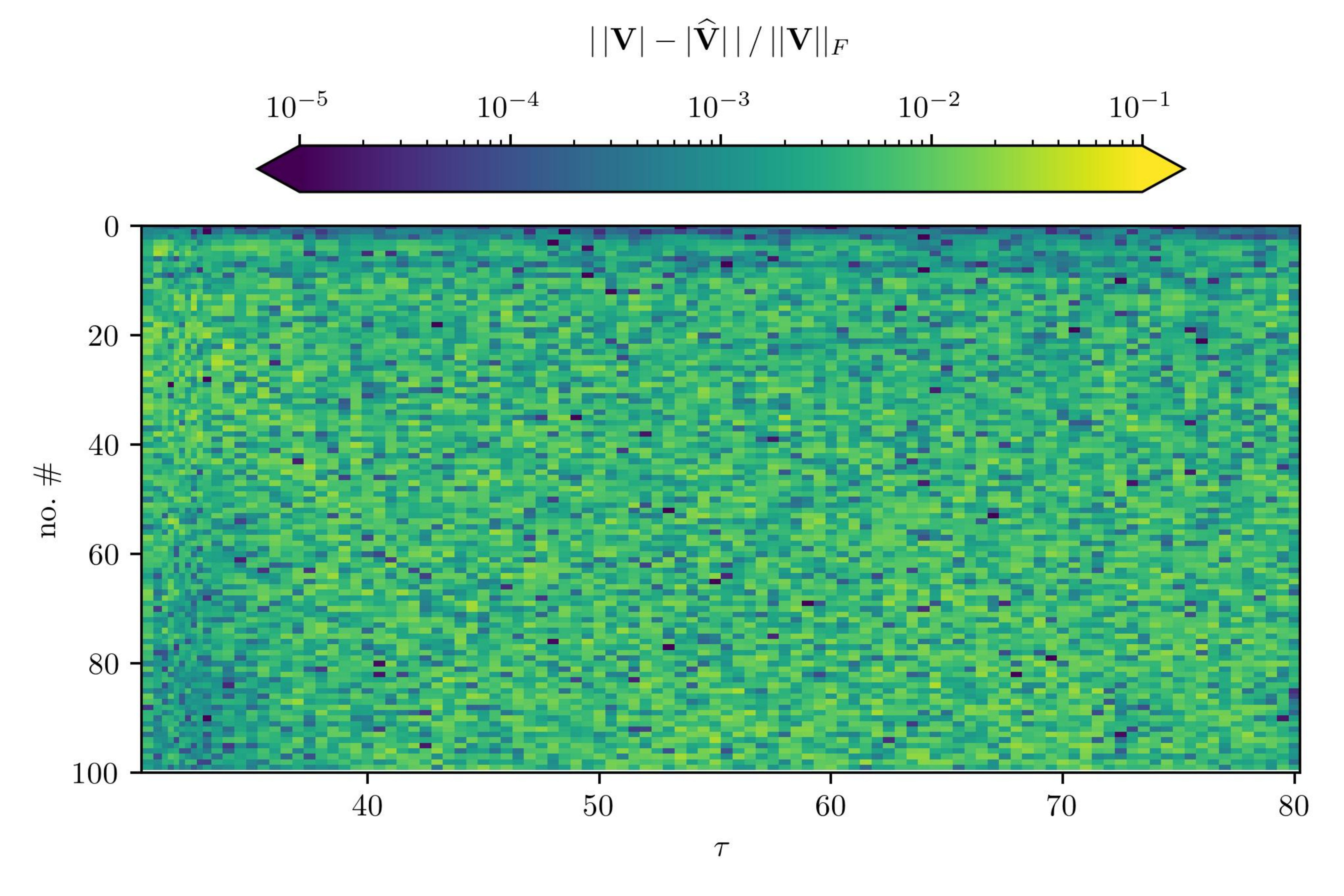}
		\caption{Temporal evolution of the error in the first $100$ right singular vectors for the Airbus XRF-1 aircraft.}
		\label{fig:v_temporal_error_xrf1}
	\end{center}
\end{figure}

It should be noted that, in contrast to the SVD results for the OAT tandem configuration, we did not have access to the exact cell volumes of the original XRF-1 mesh, required to perform the weighted SVD.
We therefore approximated the cell volumes of the CFD grid using a KNN algorithm (the implementation is open-source and linked in the accompanying code repository).
Even though this procedure yields a low approximation error, there is a small influence on the SVD results.
The results shown in Fig. (\ref{fig:eigenvalues_xrf1}) to (\ref{fig:v_temporal_error_xrf1}) not only incorporate the interpolation error made by $S^3$, but also the error made when approximating the cell volumes of the original grid.\\

Fig. (\ref{fig:eigenvalues_xrf1}) shows a comparison of the first $100$ singular values for the SVD of the pressure field, performed on the original grid and the grid generated by $S^3$.
While the dominant singular values are captured accurately, the magnitude of higher singular values is underpredicted by the SVD performed on the grid from $S^3$.

Fig. (\ref{fig:v_temporal_error_xrf1_first_few}) shows a comparison of the first six uneven right-singular vectors.
We denote the dimensionless time $\tau$ as $\tau = t / \mathrm{CTU}$.
The first two mode coefficients are captured perfectly, whereas higher mode coefficients show minor deviations.
However, the overall trend is still preserved accurately.

Finally, we present the absolute temporal error $|\,|\mathbf{V}| - |\mathbf{\widehat{V}}|\,|$ of the first $100$ right-singular vectors, scaled with the Frobenious norm $||\mathbf{V}||_F$.
Although the overall error is higher than for the tandem configuration, it remains fairly small across all mode coefficients and time steps.
The majority of the mode coefficients yield an error of $O(10^{-2})$ while the interpolation error for the first modes is in the order of $O(10^{-4})$.

\end{document}